\documentclass[prb,superscriptaddress,twocolumn,floatfix,showpacs]{revtex4-1}
\usepackage{graphicx,color}
\usepackage{amsmath,amssymb,bm}

\newcommand{\tn}{\textnormal}

\definecolor{darkred}{rgb}{0.90,0,0}
\definecolor{darkgreen}{rgb}{0,0.60,.2}
\definecolor{darkblue}{rgb}{0,0,1}
\definecolor{grey}{cmyk}{0,0,0,0.25}
\definecolor{orange}{cmyk}{0,0.6,0.8,0}

\begin{document}
\title{\boldmath Real-time and real-space spin- and energy dynamics in one-dimensional spin-1/2 systems induced by local quantum quenches at finite temperatures}

\author{C.\ Karrasch}
\affiliation{Department of Physics, University of California, Berkeley, California 95720, USA}
\affiliation{Materials Sciences Division, Lawrence Berkeley National Laboratory, Berkeley, CA 94720, USA}
\author{J.\ E.\ Moore}
\affiliation{Department of Physics, University of California, Berkeley, California 95720, USA}
\affiliation{Materials Sciences Division, Lawrence Berkeley National Laboratory, Berkeley, CA 94720, USA}
\author{F. Heidrich-Meisner}
\affiliation{Department of Physics and Arnold Sommerfeld Center for Theoretical Physics,
Ludwig-Maximilians-Universit\"at M\"unchen, D-80333 M\"unchen, Germany}

\begin{abstract}
We study the spin- and energy dynamics in one-dimensional spin-1/2 systems induced by local quantum quenches at finite temperatures using
a time-dependent density matrix renormalization group method. System sizes are chosen large enough to ensure that 
the time-dependent data for the accessible time scales represent the behavior in the thermodynamic limit.
As a main result, we observe a ballistic spreading of perturbations of the energy density in the integrable spin-1/2 $XXZ$ chain for all temperatures
and exchange anisotropies, related to the divergent thermal conductivity in this model and  the exact conservation of the energy current.
In contrast, the spin dynamics is ballistic in the massless phase, but shows a diffusive behavior at high temperatures in the 
easy-axis phase in the case of a vanishing background spin density. We extract a quantitative estimate for the spin diffusion constant
from the time dependence of the spatial variance of the spin density, which agrees well with values obtained from current-current correlation functions
using an Einstein relation. Interestingly, the diffusion constant approaches a constant value deep in the easy-axis regime.
As an example for  non-integrable models, we consider two-leg ladders, for which we observe indications of diffusive energy and spin dynamics.
The relevance of our results for recent experiments with quantum magnets and bosons in optical lattices is discussed.  
\end{abstract}

\maketitle


\section{Introduction}
\label{sec:intro}

The non-equilibrium dynamics of strongly interacting one-dimensional systems is receiving considerable 
attention, due to, on the one hand, the possibility of carrying out controlled experiments with ultra-cold 
quantum gases \cite{greiner02,kinoshita06,hofferberth07,trotzky12,schneider12,cheneau12,ronzheimer13,fukuhara13} as well as 
condensed matter experiments that probe the real-time domain\cite{montagnese13,wall10}
and, on the other hand, 
the substantial interest in understanding the relaxation dynamics and thermalization in quantum quenches,\cite{rigol08,polkovnikov11}
in which a parameter of the Hamiltonian is instantaneously or non-adiabatically fast changed to induce a non-equilibrium situation.
One-dimensional systems that can be solved via Bethe ansatz methods possess an infinite number of local conservation laws and are therefore
candidate systems for a lack of thermalization in the sense of standard statistical ensembles,\cite{rigol07} due to non-ergodicity.\cite{caux11} These include continuum models such as the Lieb-Liniger gas and the Yang-Gaudin model 
or lattice models such as hard-core bosons,  the Fermi-Hubbard model and the spin-1/2 XXZ chain. 

The existence of non-trivial local conservation laws can also lead to strictly ballistic finite-temperature transport in strongly interacting 1D systems \cite{zotos97,hm07,prosen11,sirker11} measured through non-zero
Drude weights within linear-response theory. The most famous example is dissipationless energy transport in the spin-1/2 $XXZ$ chain due to the exact conservation of the energy-current operator. \cite{zotos97,kluemper02} Spin transport in the same model at a vanishing average magnetization appears to be ballistic in the gapless phase\cite{zotos96,zotos99,benz05,prosen11,hm03,hm07,herbrych11,karrasch12,karrasch13}
and diffusive in the massive phase.\cite{zotos96,steinigeweg09,steinigeweg10,steinigeweg12,steinigeweg12a,znidaric11} Some studies also discussed the large-spin case $S>1/2$
and the classical limit.\cite{steinigeweg10,steinigeweg12b} 

Beyond integrable models, an important  question pertains to the role of integrability-breaking perturbations that, as a consequence, also violate most non-trivial conservation laws, thus affecting  qualitatively transport properties.\cite{zotos96,hm03,zotos04,jung07,karrasch12a,znidaric13}
Besides the usual formulation in terms of Kubo formulae, many studies have recently considered steady-state transport in open quantum systems coupled to baths where similar
questions, namely the conditions for the emergence of ballistic, diffusive or other forms of transport, are under active investigation,\cite{prosen09,znidaric11,saito03,mendoza-arenas13,mendoza-arenas13a}
including classical models.\cite{prosen13b}

A connection between quantum-quench dynamics and transport properties 
can be made in local quantum quenches that induce finite currents in the spin-, particle- or, energy density. Examples include the spreading
of density wave-packets,\cite{langer09,langer11,kim13} few-magnon excitations,\cite{ganahl12,liu13} or the coupling of subsystems that initially had different densities.\cite{gobert05,lancaster10a,karrasch13a,luca13,sabetta13}
To decide whether such density perturbations spread out ballistically or diffusively, one can resort to following the time evolution of  the spatial variance $\sigma^2$ of the density distribution
as a convenient measure, $\sigma \sim t$ corresponding to ballistic and $\sigma\sim \sqrt{t}$ implying diffusive dynamics.
The prevailing picture, based on mostly numerical simulations using the time-dependent density matrix renormalization group method\cite{vidal04,daley04,white04} at \textit{zero temperature},\cite{langer09,langer11}
is that the linear-response behavior carries over to the real-time and real-space dynamics in these local quenches: Energy transport is always ballistic in the spin-1/2 $XXZ$ chain,\cite{langer11}
whereas spin dynamics is ballistic in the massless regime.\cite{langer09} Deviations from ballistic spin dynamics  were observed in the massive regime,\cite{langer09} namely, perturbations in the spin density spread diffusively on attainable time scales. The quantitative analysis of the numerical data in this regime 
was, however, hampered by finite-size effects (see also the discussion in Ref.~\onlinecite{jesenko11}).
 In non-integrable models, examples for a non-ballistic spreading of perturbations in the spin density were found as well, most notably in Heisenberg ladders.\cite{langer09}
 
On the experimental side, there are exciting recent developments from both condensed matter physics and ultra-cold atomic gases 
that realize such local quenches focussing on the spin-, energy-, or particle dynamics.
Low-dimensional quantum systems are known to feature a significant contribution from magnetic excitations to 
the thermal conductivity \cite{sologubenko07,hess07}, most notably in the spin-chain materials SrCuO$_2$, Sr$_2$OCu$_3$ (Refs.~\onlinecite{sologubenko00a,sologubenko01,hlubek10}) and in the spin ladder
system (Sr,Ca,La)$_{14}$Cu$_{24}$O$_{41}$.\cite{sologubenko00,hess01}
More recently, experimentalists succeeded in  studying the heat dynamics in these strongly correlated materials in real time using two setups.
First, the spreading of heat in the surface of spin-ladder materials was monitored \cite{otter09,otter12} and second, the propagation of 
a heat pulse through the bulk of quasi one-dimensional materials was studied over macroscopically large distances.\cite{montagnese13}
The analysis of the spatial variance for the first approach suggests diffusive dynamics.\cite{otter12} 

While it is tempting to relate
the experimental observation of large thermal conductivities in certain spin-chain materials\cite{sologubenko00a,sologubenko01,hlubek10}  to conserved currents
for the underlying spin Hamiltonians, many other effects need to be taken into account to obtain a full description, including
phonons,\cite{rozhkov05,chernychev05,shimshoni03,boulat07} impurities,\cite{metavitsiadis10,barisic09,karahalios09} or spin-drag effects.\cite{boulat07,bartsch13} 
It is at present still an open question whether  anomalous transport due to  exact conservation laws of one-dimensional spin models survives 
 these perturbations in such a way  that the proximity of a realistic system to integrable models can be viewed as the core reason for the large one-dimensional heat conductivities.

\begin{figure}[t]
\includegraphics[width=\linewidth,clip]{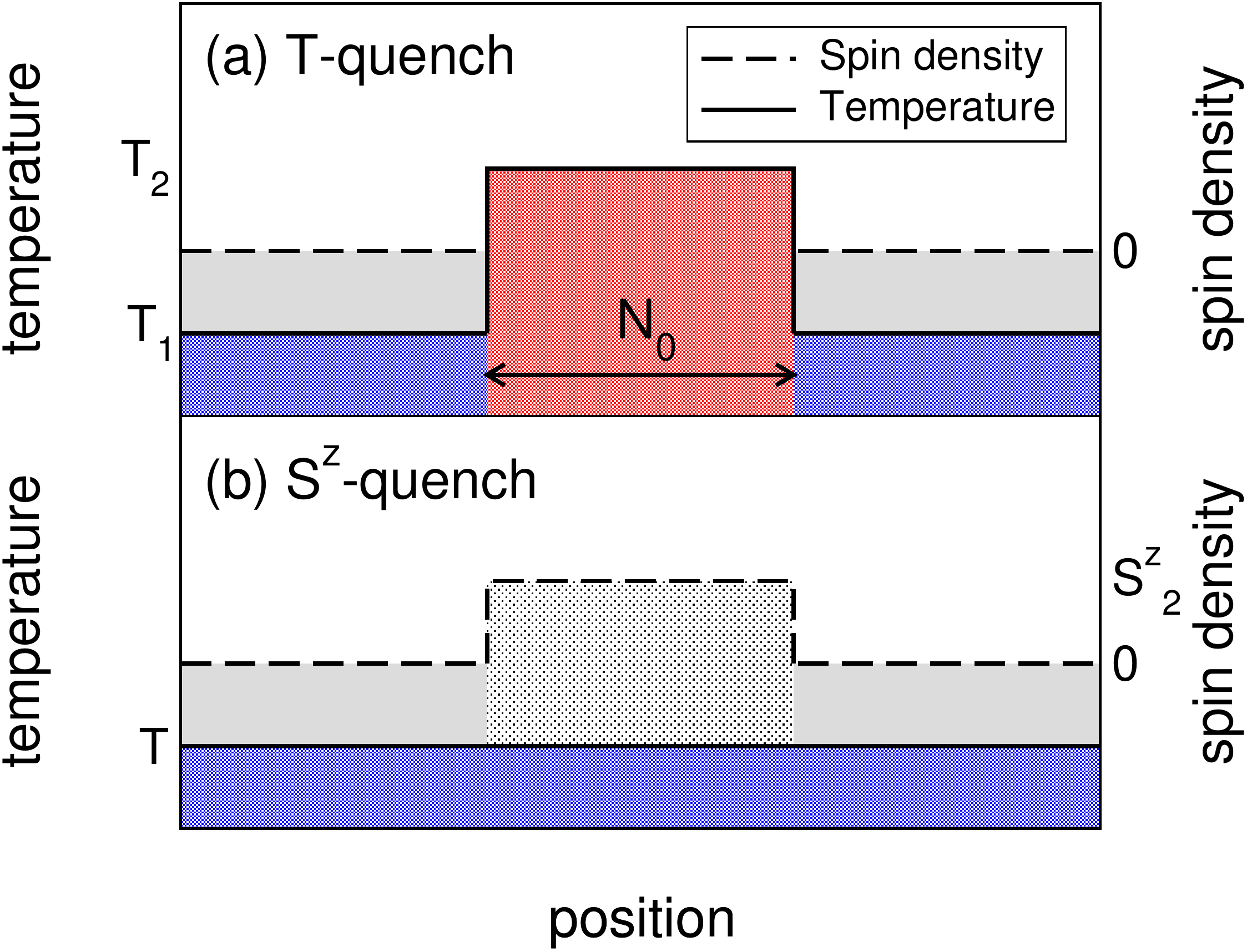}
\caption{(Color online) {\it Sketch of the initial states.} 
(a) $T$-quench.
The system has a constant spin density $\langle S^z_n\rangle=0$ (dashed line). The central $N_0$ sites are prepared at a temperature $T_2>T_1$,
where $T_1$ is the temperature of the bulk of the system ($T$-profile: solid line).
(b) $S^z$-quench. The system is at a constant background density,
but $N_0$ sites in the center have a finite spin density $S^z_2$, while in the bulk, $S_1^z=\langle S^z_n\rangle=0$. 
}
\label{fig:sketck}
\end{figure}

In experiments with ultra-cold atomic gases, pure Hubbard-type models can be realized,\cite{greiner02,joerdens08,schneider08} without a coupling to the lattice and
ideally,
without impurities. As a trade-off one deals with typically inhomogeneous densities and much smaller particle numbers
than in condensed matter physics.\cite{bloch08}
Bosons in  a single band with an onsite repulsion, realize, in the limit of infinitely strong interactions,
the spin-$1/2$ XX model.\cite{cazalilla11}
These hard-core bosons map exactly to non-interacting fermions and therefore, this system is integrable and
possesses a conserved particle current. As a consequence, one expects ballistic dynamics. In cold-gas experiments
this can, for instance, be probed 
by removing the trapping potential, allowing the gas to  expand in the optical lattice in the so-called 
sudden expansion.\cite{schneider12,ronzheimer13, reinhard13} 
In a recent experiment\cite{ronzheimer13},  ballistic dynamics
of 1D hard-core bosons was convincingly demonstrated, providing a clear realization of
anomalous transport due to the existence of non-trivial conservation laws in an integrable quantum model. 
Coupling one-dimensional systems of hard-core bosons to either two-dimensional systems\cite{ronzheimer13} or two-leg ladders\cite{vidmar13} 
results in deviations from ballistic dynamics, which in the 2D case has been argued to be related to diffusion\cite{ronzheimer13,schneider12}
(see also Refs.~\onlinecite{jreissaty13,schonmeier13}).

\begin{figure}[t]
\includegraphics[width=0.49\linewidth,clip]{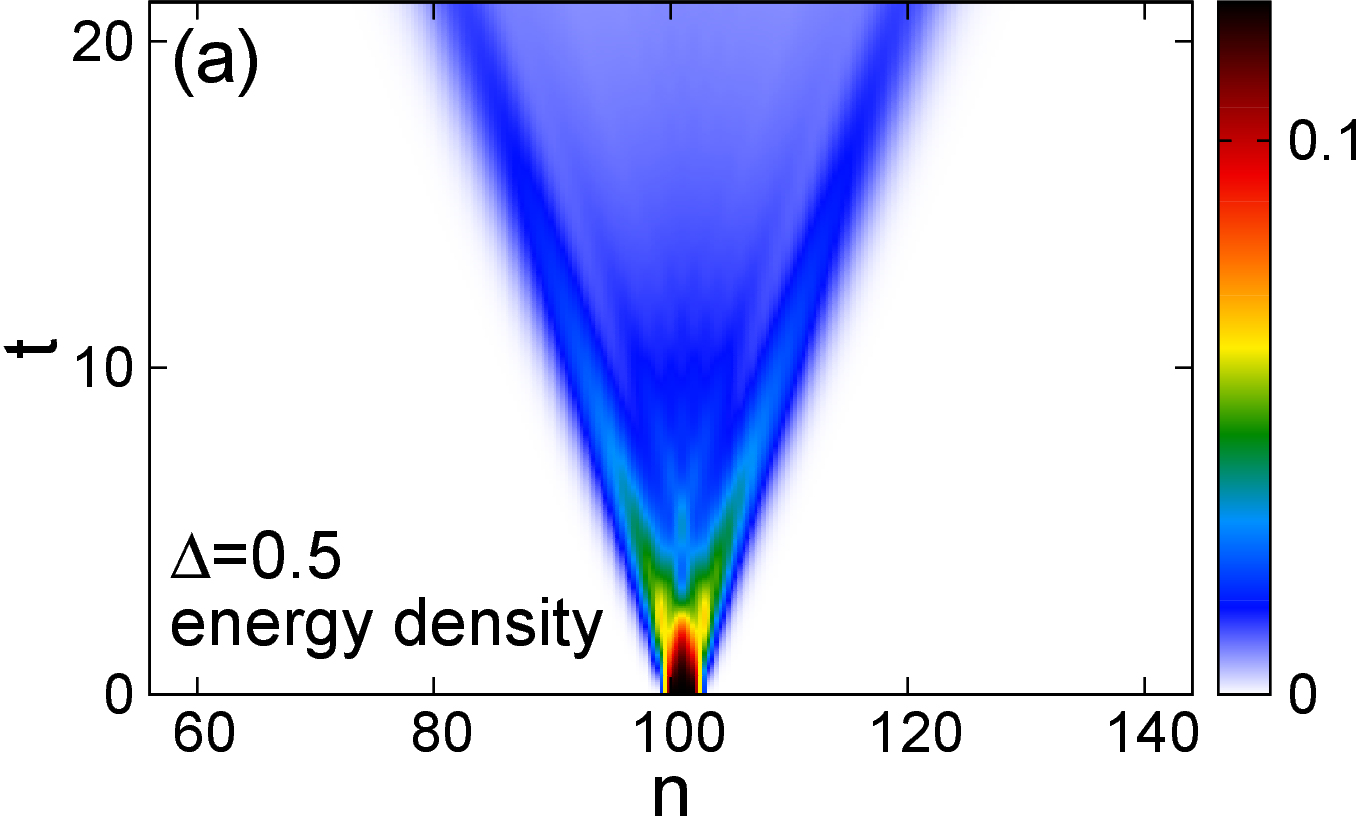}\hspace*{0.01\linewidth}
\includegraphics[width=0.49\linewidth,clip]{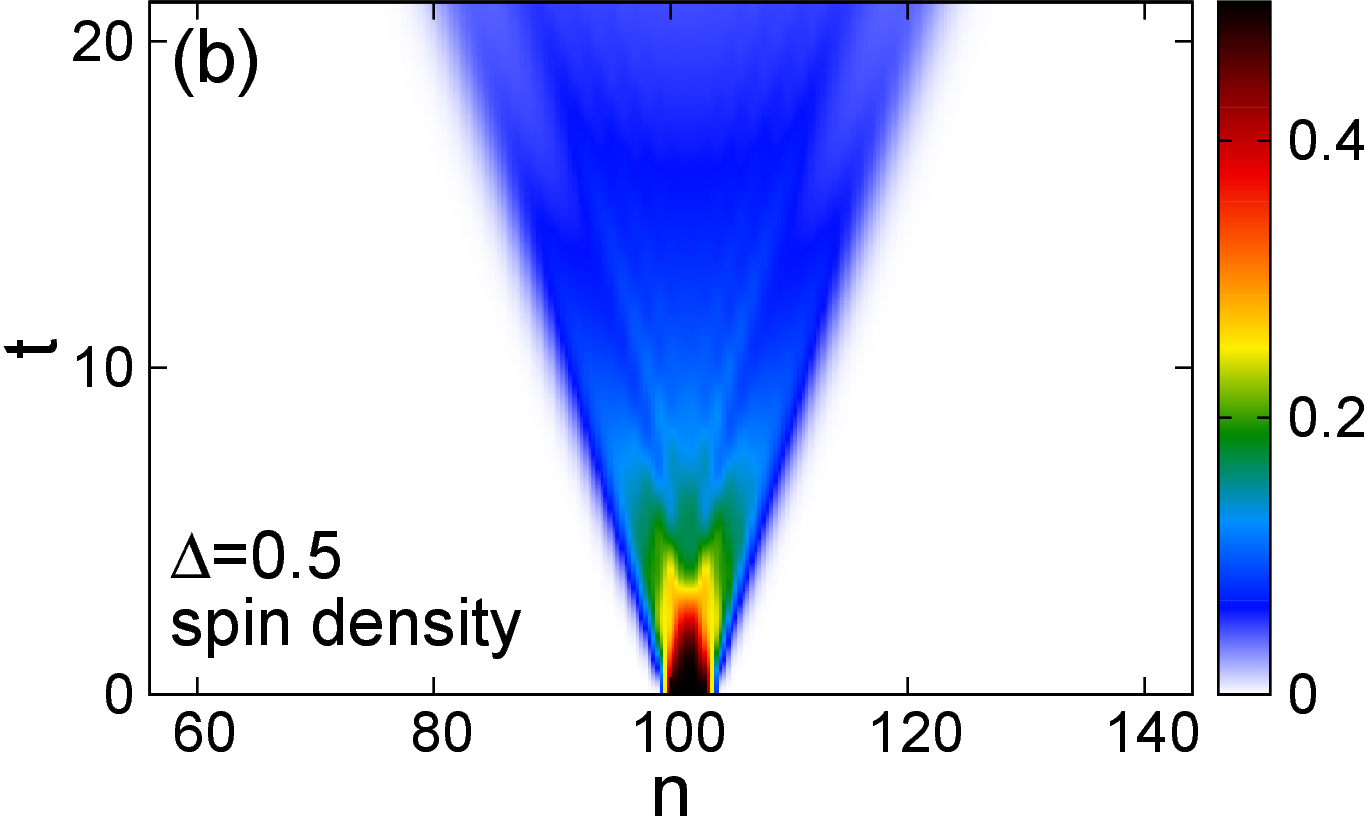} \\[0.2cm]
\includegraphics[width=0.49\linewidth,clip]{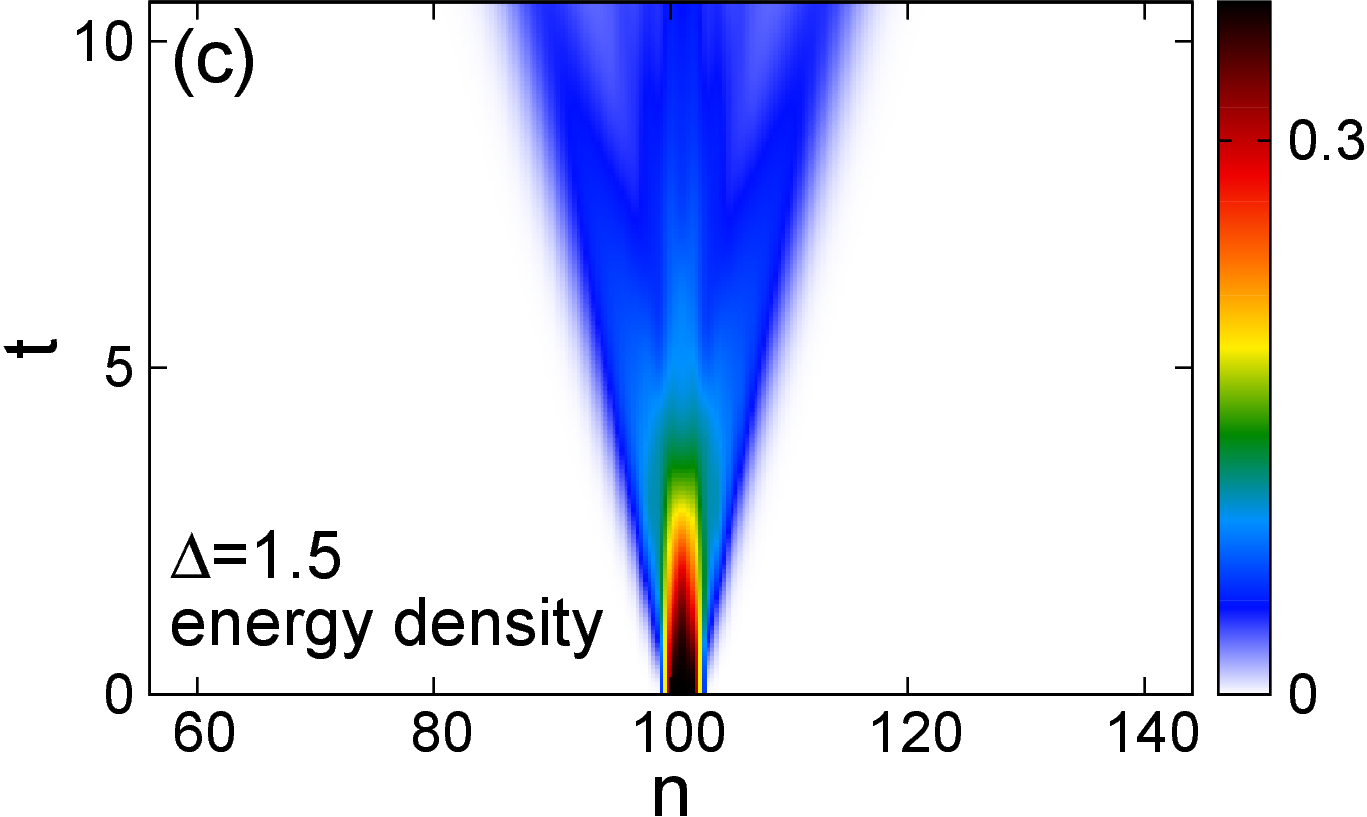}\hspace*{0.01\linewidth}
\includegraphics[width=0.49\linewidth,clip]{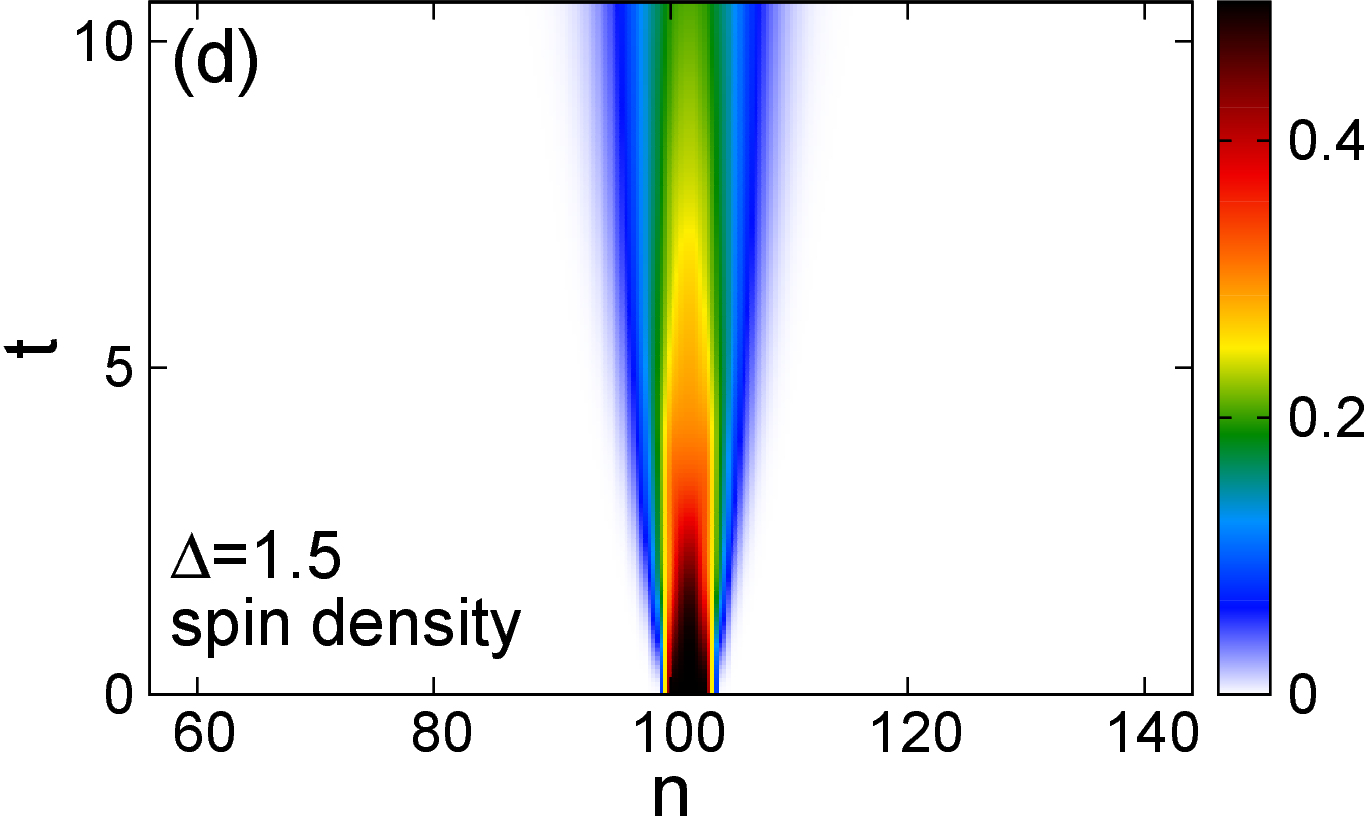}
\caption{(Color online) {\it $S^z$-quench in the integrable model}. Expansion of the spin and energy density after preparing a local spin-wave packet of size $N_0=4$ at time $t=0$ in the center of a (a,b) gapless and (c,d) gapped $XXZ$ spin chain of total size $N=204$ at infinite background temperature. (a-c) and (d) exhibit ballistic and diffusive behavior, respectively.}
\label{fig:spin}
\end{figure}

\begin{figure}[t]
\includegraphics[width=0.49\linewidth,clip]{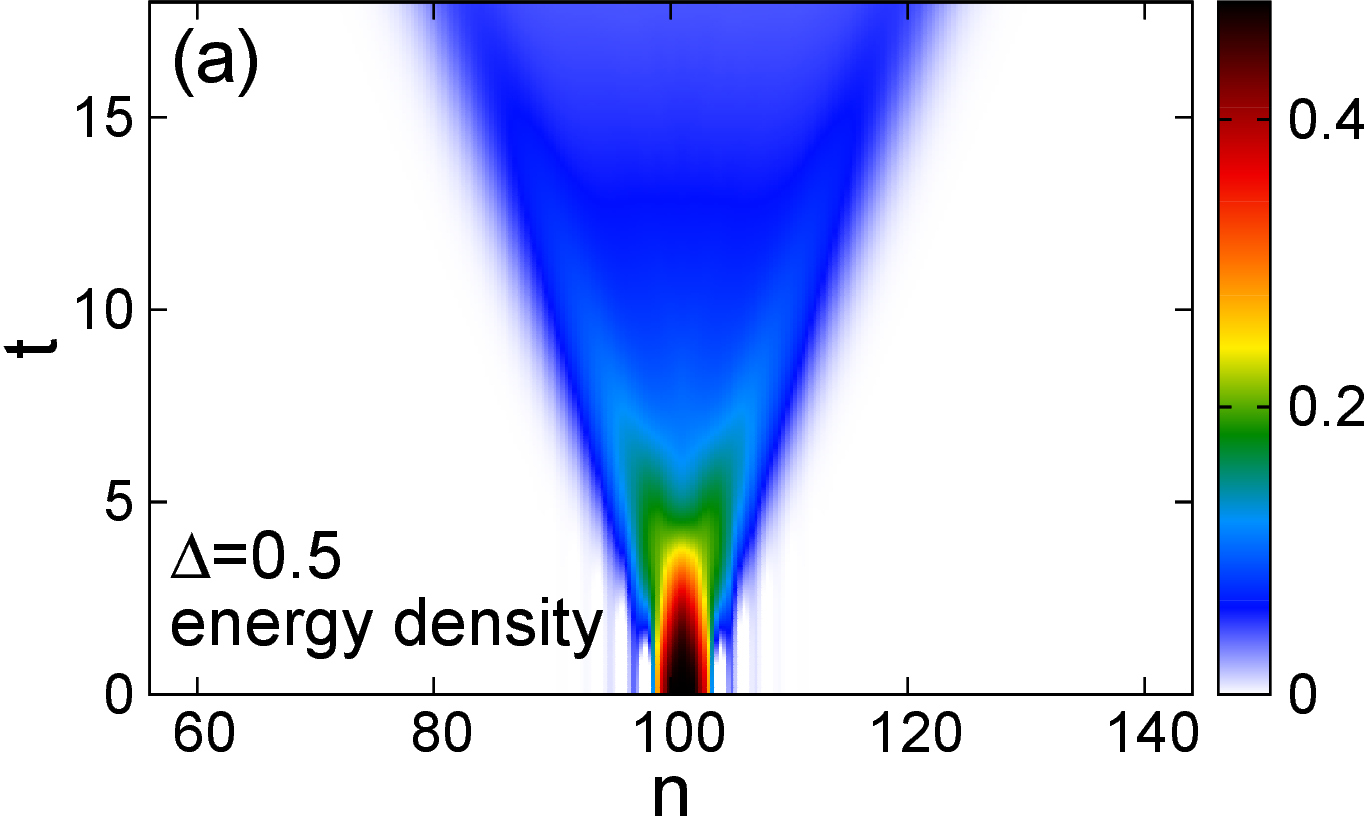}\hspace*{0.01\linewidth}
\includegraphics[width=0.49\linewidth,clip]{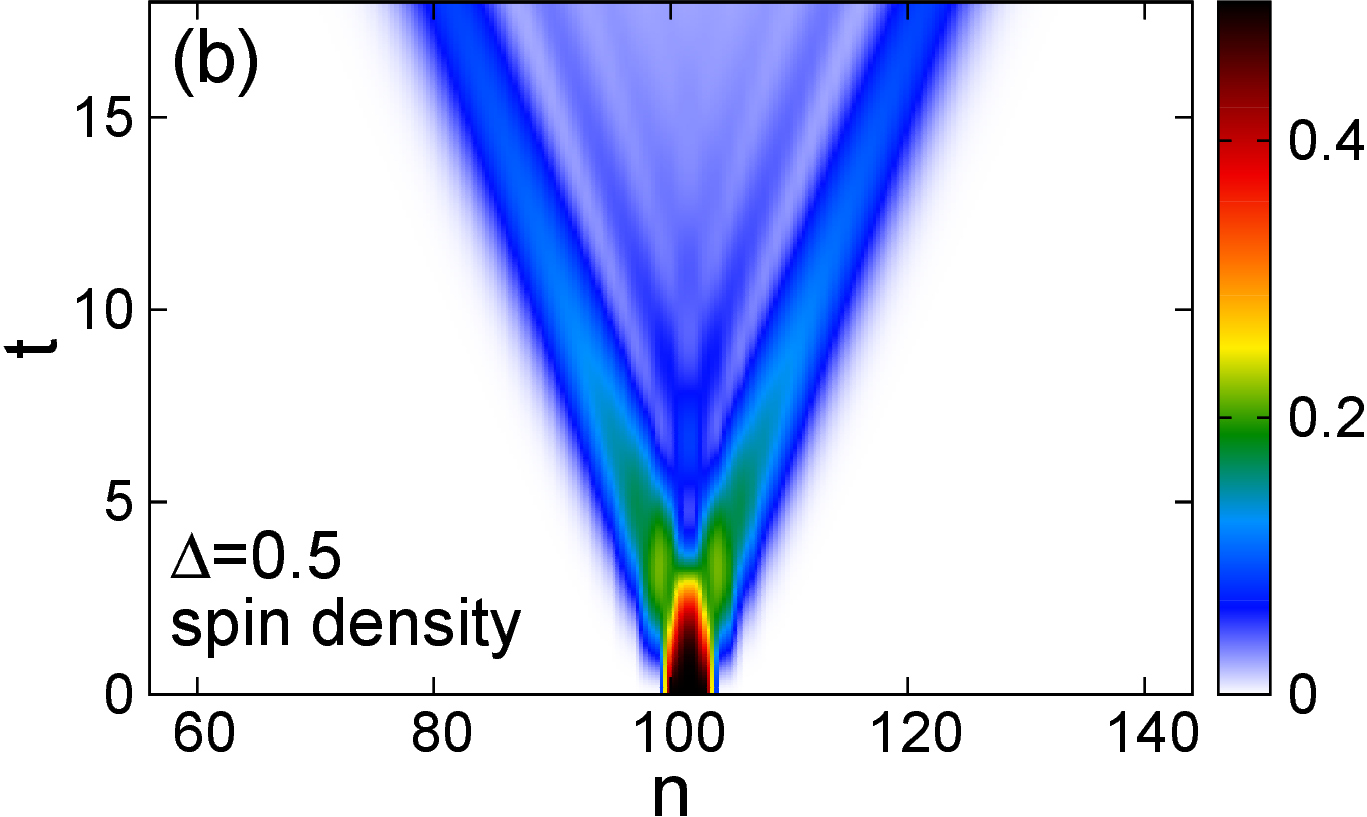} \\[0.2cm]
\includegraphics[width=0.49\linewidth,clip]{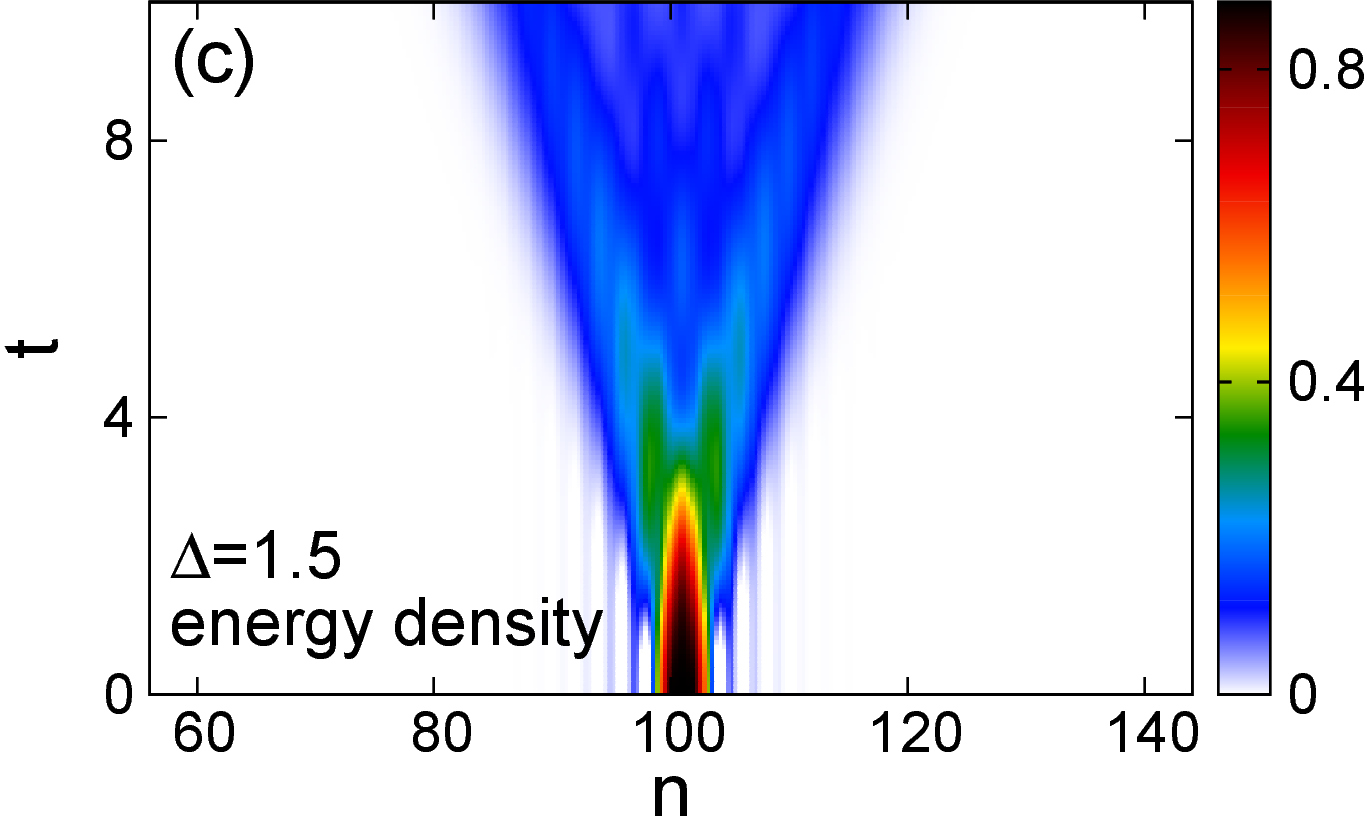}\hspace*{0.01\linewidth}
\includegraphics[width=0.49\linewidth,clip]{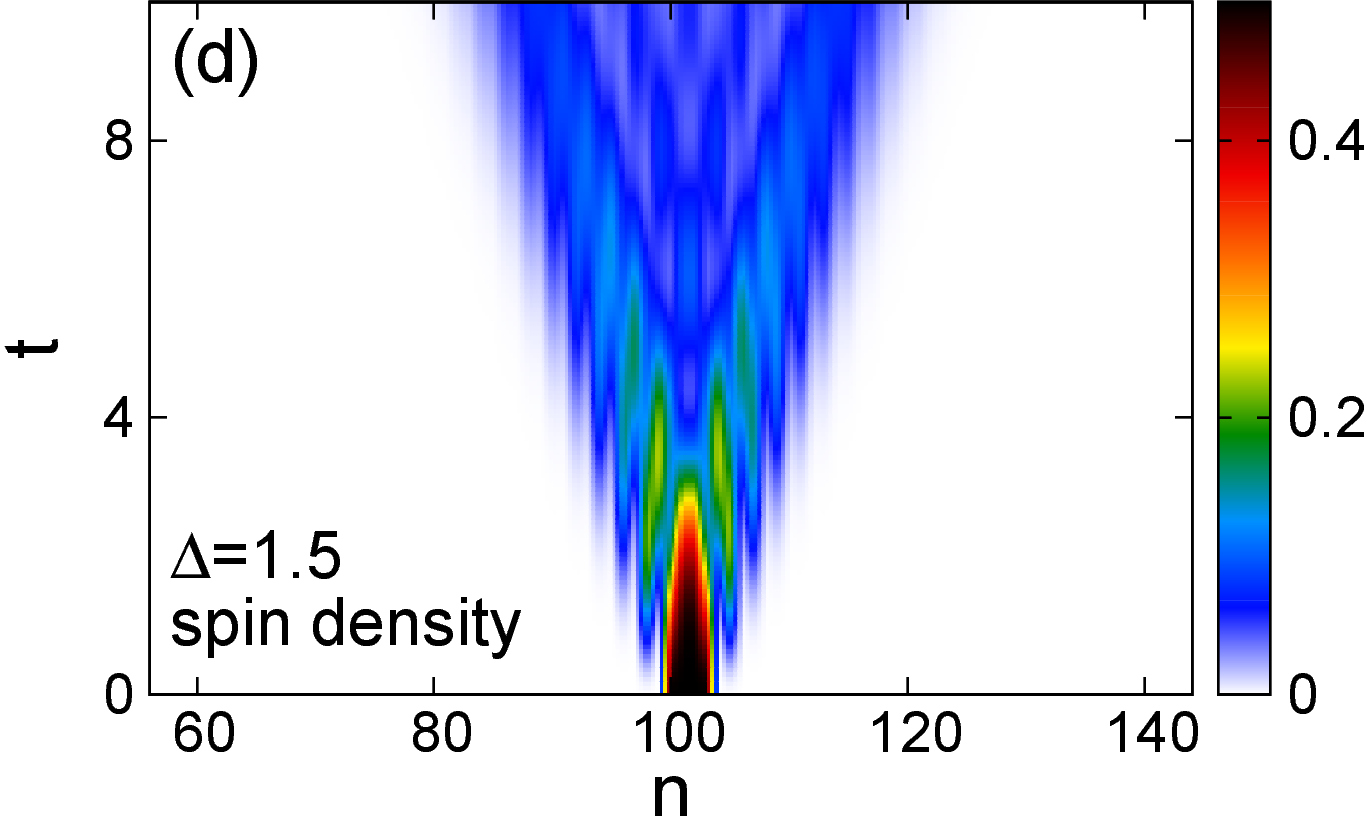}
\caption{(Color online) The same as in Fig.~\ref{fig:spin} but for $T=0.2$. Note the structure
in the density plots for the spin. }
\label{fig:spin2}
\end{figure}

\begin{figure*}[t]
\includegraphics[width=0.44\linewidth,clip]{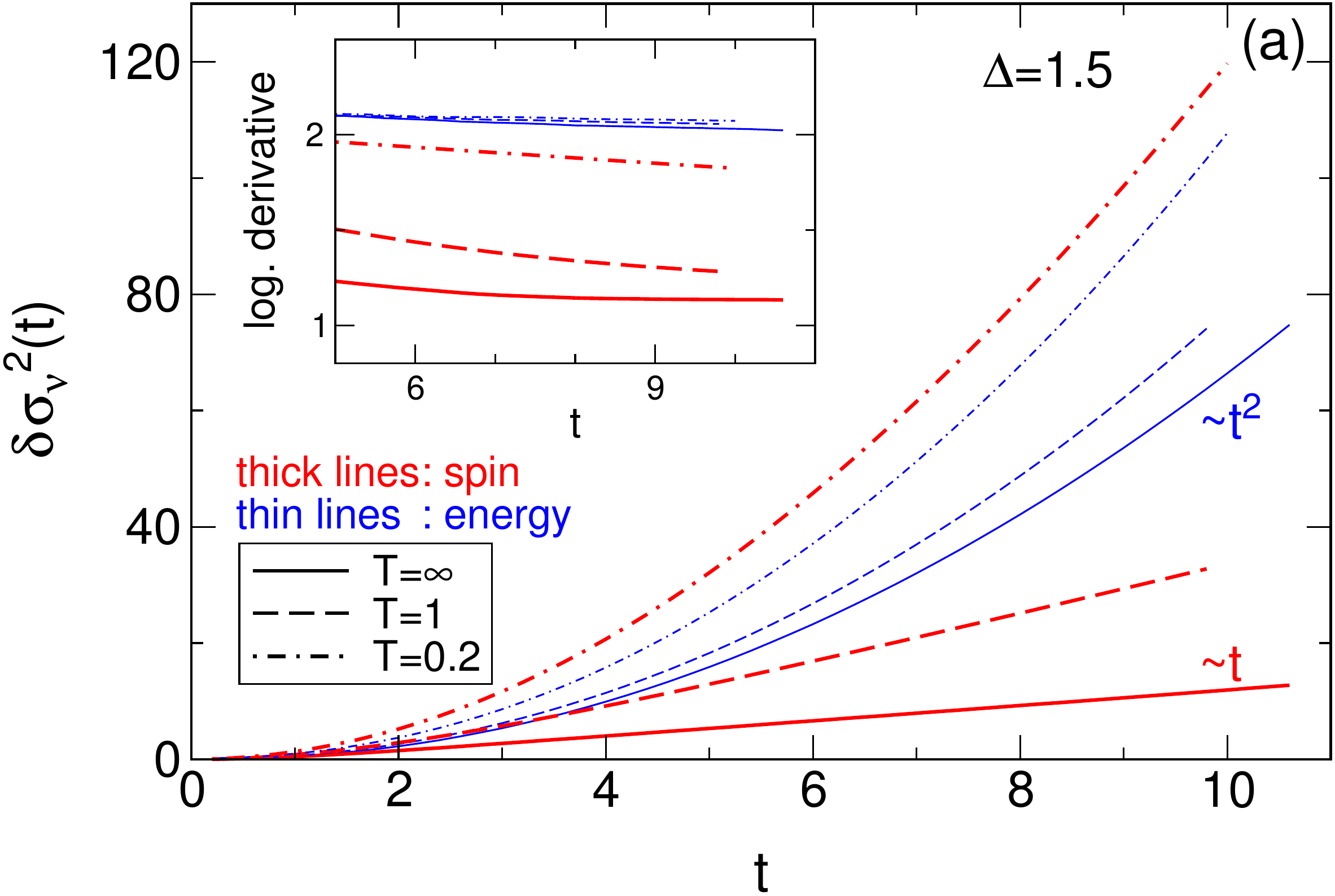} \hspace*{0.04\linewidth}
\includegraphics[width=0.44\linewidth,clip]{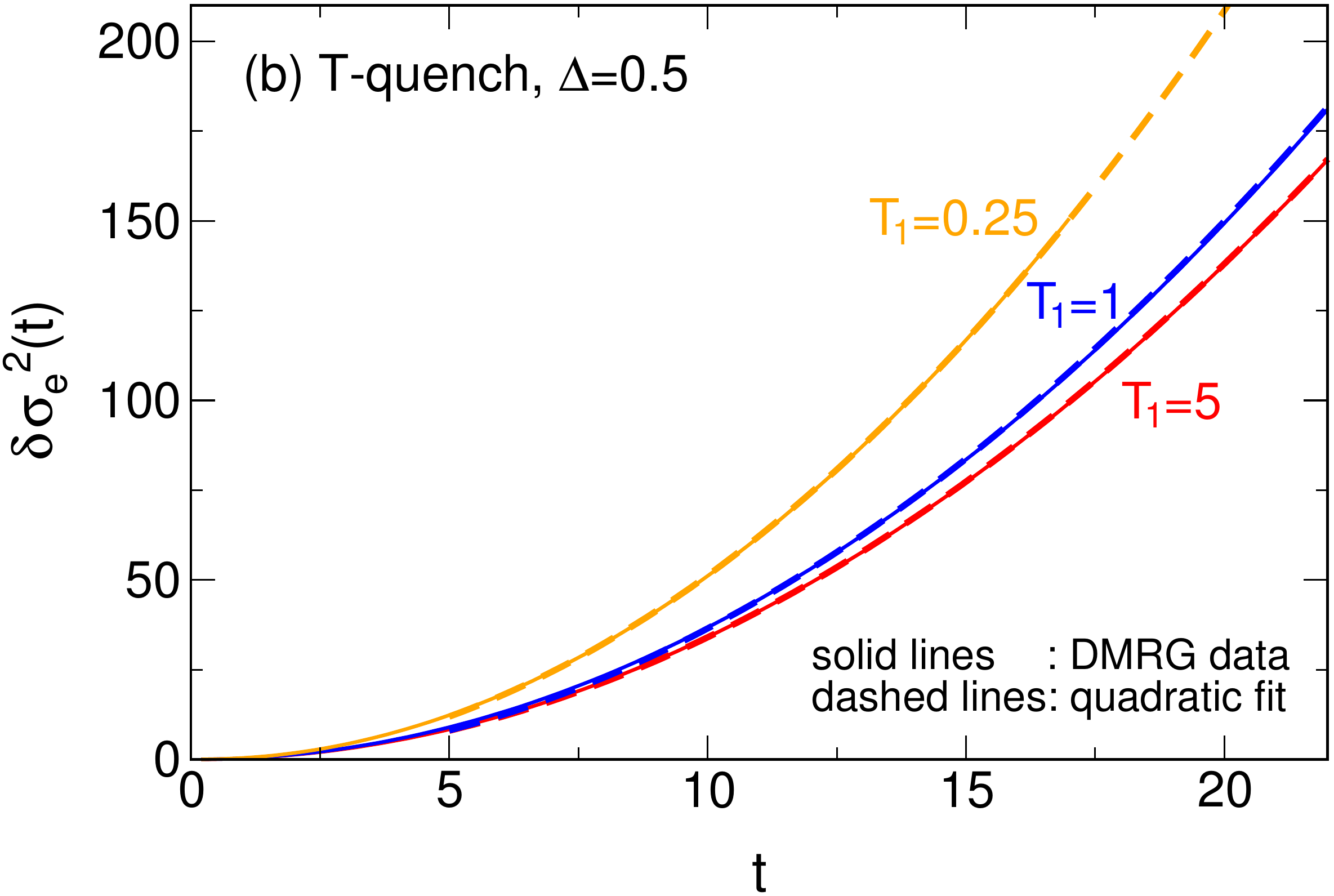}
\caption{(Color online) Time evolution of the spatial variance of the spin and energy densities shown in Figs.~\ref{fig:spin} and \ref{fig:t1t2}. Ballistic and diffusive behavior is signalled by $\delta \sigma_{\nu}^2(t)\sim t^2$ and $\delta \sigma_{\nu}^2(t)\sim t$, respectively, which can be easily identified by computing the logarithmic derivative. (a) {\it $S^z$-quench.} Expansion of a spin-wave packet in a gapped $XXZ$ chain at $\Delta=1.5$ for $T=\infty$ [Fig.~\ref{fig:spin}(c,d)] as well as for $T=1$ and $T=0.2$. (b) {\it $T$-quench.} Width of energy-wave packets of the $XXZ$ chain with $\Delta=0.5$ for $T_1=5,1,0.25$ and $T_2=\infty$  [compare  Fig.~\ref{fig:t1t2}(a) for $T_1=5$].
Dashed lines (indistinguishable from the sold lines) are fits to $a+bt^2$ for $t>5$. 
}\label{fig:sigma}
\end{figure*}

In our work we advance the study of the spreading of perturbations in the spin or energy density in spin-1/2 systems to {\it finite temperatures},
by exploiting recent progress  with finite-temperature, real-time density matrix renormalization group simulations.\cite{karrasch12,barthel13b,karrasch13a}
We consider two types of local quenches: (i) In the $T$-quench, we prepare initial states such that in the center, there is a short region
with $T_2>T_1$ [see Fig.~1(a)] embedded into a larger system with a constant background temperature $T_1$. The spin density is held at zero in this set-up.
(ii) In the second case, the $S^z$-quench, the system is at a fixed temperature $T$, but the central region has a magnetization $S^z_2>0$ whereas
the background density is kept at $S^z_1=0$ (if not stated otherwise, $S^z_2=1/2$).
We study the finite-temperature dynamics induced through these quenches in the integrable spin-1/2 $XXZ$ chain as well as non-integrable $XXZ$ two-leg ladders.
Our analysis is mainly based on the time-dependence of the spatial variance $\sigma_{\nu}^2$ of the spin and energy density ($\nu=s,e$, respectively), following Refs.~\onlinecite{langer09,langer11}.

As a main result, we observe a ballistic spreading of energy, i.e., $\sigma_e\propto t$ 
 at all temperatures and for all exchange anisotropies in the spin-1/2 $XXZ$ model in qualitative
agreement with linear-response theory, both for the $T$- and the $S^z$-quenches. 
Spin dynamics  is ballistic in the gapless regime of the 1D chain with $\sigma_s\propto t$, whereas in the massive regime our data is consistent with diffusive behavior, i.e.,  $\sigma_s\propto \sqrt{t}$.

We further present quantitative results for the spin diffusion constant of the $XXZ$ chain at infinite temperature and in the
massive regime, obtained from both $S^z$-quenches and Einstein relations, where we extract the dc-spin conductivity from the long-time limit 
of current-correlation functions. Our results from these two approaches, which are in good quantitative agreement, indicate a saturation of $\mathcal{D}_s$ at large anisotropies $\Delta$, in contrast to the 
predictions from Refs.~\onlinecite{znidaric11,steinigeweg12}. 
  Field-theoretical results\cite{sachdev97,damle05,damle98} for the low-temperature regime of gapped spin models
suggest that the diffusion constant is inversely proportional to the gap, which deep in the easy-axis phase, is given by the exchange anisotropy $\Delta$. 
 
In the case of non-integrable systems, 
we focus our attention on two-leg ladders. For both $S^z$- and $T$-quenches, we observe noticeable deviations 
from ballistic dynamics, and we extract the energy-diffusion constant at infinite temperatures.    We discuss these results in the context of recent experiments
with quantum magnets\cite{montagnese13,otter09,otter12} and bosons in optical lattices.\cite{ronzheimer13} 

The plan of this exposition is the following. Section \ref{sec:models} introduces the model Hamiltonians and definition for
the spatial variance, and provides technical details on our tDMRG simulations. 
In Sec.~\ref{sec:xxz}, we present our results for $S^z$- and $T$-quenches in the integrable $XXZ$  model.
We discuss the emergent velocities for the spreading of energy and spin in the ballistic cases and
present quantitative results for the spin-diffusion constant in the easy-axis regime. Section \ref{sec:ladder}
summarizes our results for the two-leg ladder geometry. We conclude with a summary in Sec.~\ref{sec:summary}. 

\section{Models and definitions}
\label{sec:models}

The first model of interest is the spin-1/2 $XXZ$ chain: 
\begin{equation}
H = J \sum_{n=1}^{N-1} \lbrack S^x_nS^x_{n+1} + S^y_nS^y_{n+1} + \Delta S^z_nS^z_{n+1}\rbrack,
\label{eq:ham}
\end{equation}
where $S^{\mu}_n$, $\mu=x,y,z$, are the components of a spin-1/2 operator acting on site $n$ of
a chain of length $N$. $\Delta$ parameterizes the
exchange anisotropy, $J$ sets the energy scale and will be set to unity hereafter ($\hbar=1$).
The model has a critical gapless phase for $|\Delta| \leq 1$ and  gapped  phases
with antiferromagnetic order and ferromagnetic order for $\Delta>1$ and $\Delta<-1$, respectively.\cite{kolezhuk-review}

The second model studied here is a two-leg spin ladder given by
\begin{eqnarray}
H &=& J \sum_{n=1 \atop {\lambda=1,2}}^{N-1} \lbrack S^x_{n,\lambda}S^x_{n+1,\lambda} + S^y_{n,\lambda}S^y_{n+1,\lambda} + \Delta S^z_{n,\lambda}S^z_{n+1,\lambda}\rbrack\nonumber \\ 
  && +J_2\sum_{n=1}^N \lbrack S^x_{n,1}S^x_{n,2} + S^y_{n,1}S^y_{n,2} + \Delta S^z_{n,1}S^z_{n,2}\rbrack\,. 
\end{eqnarray}
In this equation, $\lambda=1,2$ labels the upper and lower leg, $J$ is the longitudinal coupling along each leg and $J_2$
is the coupling along rungs of the ladder.
We use open boundary conditions for both models.

Our analysis will primarily focus on the time-dependence of the spatial variance of the spin- and energy-density $\rho_n^{\nu}$ ($\nu=s,e$ for spin and energy, respectively), $\rho_n^s=\langle S^z_n\rangle$ and $\rho_n^e=\langle  h_n\rangle$, respectively. We define the energy density as 
\begin{equation}
h_n =  J \lbrack  S^x_nS^x_{n+1} + S^y_nS^y_{n+1} + \Delta S^z_n S^z_{n+1} \rbrack
\end{equation}
for the chain, whereas for the ladder we assign  the term
\begin{eqnarray}
h_n &=&J \sum_{\lambda=1,2} \lbrack S^x_{n,\lambda}S^x_{n+1,\lambda} + S^y_{n,\lambda}S^y_{n+1,\lambda} + \Delta S^z_{n,\lambda}S^z_{n+1,\lambda}\rbrack \nonumber \\
  && +\frac{J_2}{2} \lbrack S^x_{n,1}S^x_{n,2} + S^y_{n,1}S^y_{n,2} + \Delta S^z_{n,1}S^z_{n,2}\rbrack  \\
  && +\frac{J_2}{2} \lbrack S^x_{n+1,1}S^x_{n+1,2} + S^y_{n+1,1}S^y_{n+1,2} + \Delta S^z_{n+1,1}S^z_{n+1,2}\rbrack\,. \nonumber
\end{eqnarray}
The  spatial variance of the density distribution is defined as
\begin{equation}
\sigma^2_{\nu}(t) = \frac{1}{{\mathcal{N}}_{\nu}}\sum_{n=n_0}^{N-n_0} (n-n_c^\nu)^2\, (\rho_n^{\nu}(t) -\rho_\tn{bg}^{\nu}) 
\end{equation}
where $\rho_\tn{bg}^{\nu}$ is the bulk background density, $n_c^\nu$ is the center of the wave packet, $n_0$ cuts off boundary effects from the left and right ends, and the normalization constant reads
\begin{equation}
\mathcal{N}_\nu =  \sum_{n=n_0}^{N-n_0}(\rho_n^{\nu}(t) -\rho_\tn{bg}^{\nu})~.
\end{equation}

We identify the dynamics as ballistic if
\begin{equation}
\delta \sigma_{\nu}=\sqrt{\sigma^2_{\nu}(t) -\sigma^2_{\nu}(t=0)} = V_{\nu} \,t
\end{equation}
where $V_{\nu}$ has units of velocity.
Diffusive dynamics in 1D should show up via a much slower growth of the variance given by
\begin{equation}
\delta \sigma_{\nu}=  \sqrt{2\mathcal{D}_{\nu}\,t} 
\end{equation}
where $\mathcal{D}_{\nu}$ is the diffusion constant.

For later comparison, we quote the usual definition of ballistic transport within linear-response theory. 
The real part of the conductivity $\sigma_{\nu}(\omega)$ can be  decomposed into the divergent zero-frequency contribution
weighted with the Drude weight $D_{\nu}$ and a regular part:
\begin{equation}
\mbox{Re}\, \sigma_{\nu}(\omega)= 2\pi D_{\nu} \delta({\omega}) + \sigma_{\rm reg,\nu}(\omega)\,.
\end{equation}
A non-zero Drude weight implies ballistic transport. 

The expansion of a local perturbation in  the spin or energy density can be computed efficiently using the real-time \cite{vidal04,daley04,white04,schmitteckert04} finite-temperature\cite{verstraete04,feiguin05} density matrix renormalization group \cite{white92,white93,schollwoeck05,schollwoeck11} (DMRG) algorithm introduced in Ref.~\onlinecite{karrasch12}. DMRG is essentially controlled by the so-called discarded weight $\epsilon$. We ensure that $\epsilon$ is chosen small enough [an example for a DMRG error analysis is shown in Fig.~\ref{fig:currcorr}(b)] and that $N$ is chosen large enough to obtain ``numerically-exact'' results in the thermodynamic limit. With $N\sim 200$, this is always ensured (see the finite-size analysis in, e.g., Refs.~\onlinecite{karrasch12,karrasch13}), while smaller systems would also be sufficient. We stop our simulation once the DMRG block Hilbert space dimension (see Ref.~\onlinecite{schollwoeck05}) has reached values of about 1000-2000.

\section{Integrable model:\\The spin-1/2 $XXZ$ chain}

\label{sec:xxz}
We will now study the $S^z$- and $T$-quench in the integrable spin-1/2 $XXZ$ chain. We first discuss the
time-dependence of spatial variances, which suggest ballistic energy dynamics for all $\Delta$, while
spin dynamics is ballistic for $0\leq \Delta<1$. Second, we discuss the dependence of the velocities $V_{\nu}$, $\nu=e,s$,
in the ballistic regimes on temperature and exchange anisotropy. 
Third, we extract the spin-diffusion constant from $S^z$-quenches at infinite temperature and analyze its dependence on $\Delta$.
Our findings for $\mathcal{D}_s$ are corroborated by computing the diffusion constant from current-correlation functions,
yielding quantitative agreement with the spreading of perturbations in the spin density.

\subsection{Time-dependence of the variance\\ in the $T$- and $S^z$-quench}
Figure~\ref{fig:spin} shows typical spin- and energy density profiles recorded in 
an $S^z$-quench at infinite background temperature for both the massless ($\Delta=0.5$) 
and massive regime ($\Delta=1.5$). In all cases, with the exception of spin dynamics at $\Delta=1.5$, we observe
a fast expansion of the initial perturbation and a splitting into two jets similar to the behavior at zero temperature.\cite{langer09,langer11}
The behavior of the spin density at $\Delta=1.5$ indicates the formation of a diffusive core that expands much slower than 
energy at the same $\Delta$. 
In a $T$-quench in the $XXZ$ model, the density profiles always exhibit a spreading into two jets [see, e.g., Fig.~\ref{fig:t1t2}(a) for the case of $\Delta=0.5$].
Figure~\ref{fig:spin2} shows that the qualitative behavior is the same at lower temperatures $T<\infty$. Note the  structure in the
density profiles [see Figs.~\ref{fig:spin2}(b) and (d)], which are particularly evident in the spin density. These can be related to single excitations split off the central block
with full polarization  in the initial state (compare Refs.~\onlinecite{gobert05,eisler13}).

Results for the spatial variances $\delta \sigma^2_{\nu}(t)$
are shown in Fig.~\ref{fig:sigma}(a) for an $S^z$-quench for three temperatures $T=0.2,1,\infty$ at $\Delta=1.5$. The inset shows the
time-dependence of the logarithmic derivative
\begin{equation}
\alpha_{\nu} =\frac{d \mbox{ln }\delta \sigma_{\nu}^2(t)}{d (\mbox{ln}\,t)}\,.
\end{equation}
Energy always spreads ballistically with  $\delta \sigma_e^2 \propto t^2$, both for $S^z$ and $T$-quenches
[see the data shown in Figs.~\ref{fig:sigma}(a) and (b), respectively]. 
For the spin density, we observe $\delta \sigma_s^2 \propto t^2$ for $\Delta=0.5$ (not shown in the figures), while
clear deviations from this ballistic behavior emerge for $\Delta=1.5$. In that case, the logarithmic derivative decreases
as a function of time, indicating a crossover to diffusive dynamics. The cleanest example is infinite temperature
$T=\infty$. In that case, the logarithmic derivative $\alpha_s$ saturates close to $\alpha_s \gtrsim 1$ 
at the longest times reached, suggestive of diffusive spin dynamics at finite temperatures. 
This result is qualitatively consistent with the conclusions of Refs.~\onlinecite{damle98,damle05,steinigeweg09,prosen09,znidaric11,jesenko11,steinigeweg12,huber12}.

For the special case of $\Delta=1$ (results not shown here),  the qualitative behavior of spin dynamics is still debated.
\cite{zotos99,hm03,benz05,karrasch12,herbrych11,karrasch13,znidaric11,znidaric11a, sirker11,prosen13a} By studying the time-dependence of the spatial 
variance in finite-temperature $S^z$-quenches, we make the interesting observation that on the accessible time scales,
$\delta \sigma_s^2(t) \propto t^{\alpha_s(T)}$, i.e., $\alpha_s$ depends on temperature.
At low $T$, $\alpha_s\approx 2$, while at infinite temperature, $\alpha_s \approx 1.5$.
This could be consistent with several scenarios. Znidaric\cite{znidaric11,znidaric11a} suggested that 
the behavior at $\Delta=1$ corresponds to anomalous diffusion at infinite temperature, implying $1<\alpha_s <2$.
His conclusions are derived from analyzing the steady-state of a spin chain coupled to baths.
The analysis of linear response functions has so far yielded no conclusive picture on the absence or presence
of a ballistic contribution, with some exact
diagonalization studies\cite{herbrych11} arguing for a vanishing Drude weight, while 
other authors argue that a finite $D_s$ cannot be ruled out.\cite{hm07,hm03}

Karrasch {\it et al.}~(Ref.~\onlinecite{karrasch13}) demonstrated that the relative weight of the Drude weight compared to finite
frequency contributions can at best be  small at infinite temperature at $\Delta=1$, using extrapolated exact diagonalization  data. This would imply
that ballistic dynamics in this case can only be seen on very long time scales in the wave-packet dynamics studied in our
work (see also the discussion in Ref.~\onlinecite{sirker11}).

\subsection{Analysis of velocities in the ballistic regime}

\begin{figure}[t]
\includegraphics[width=\columnwidth,clip]{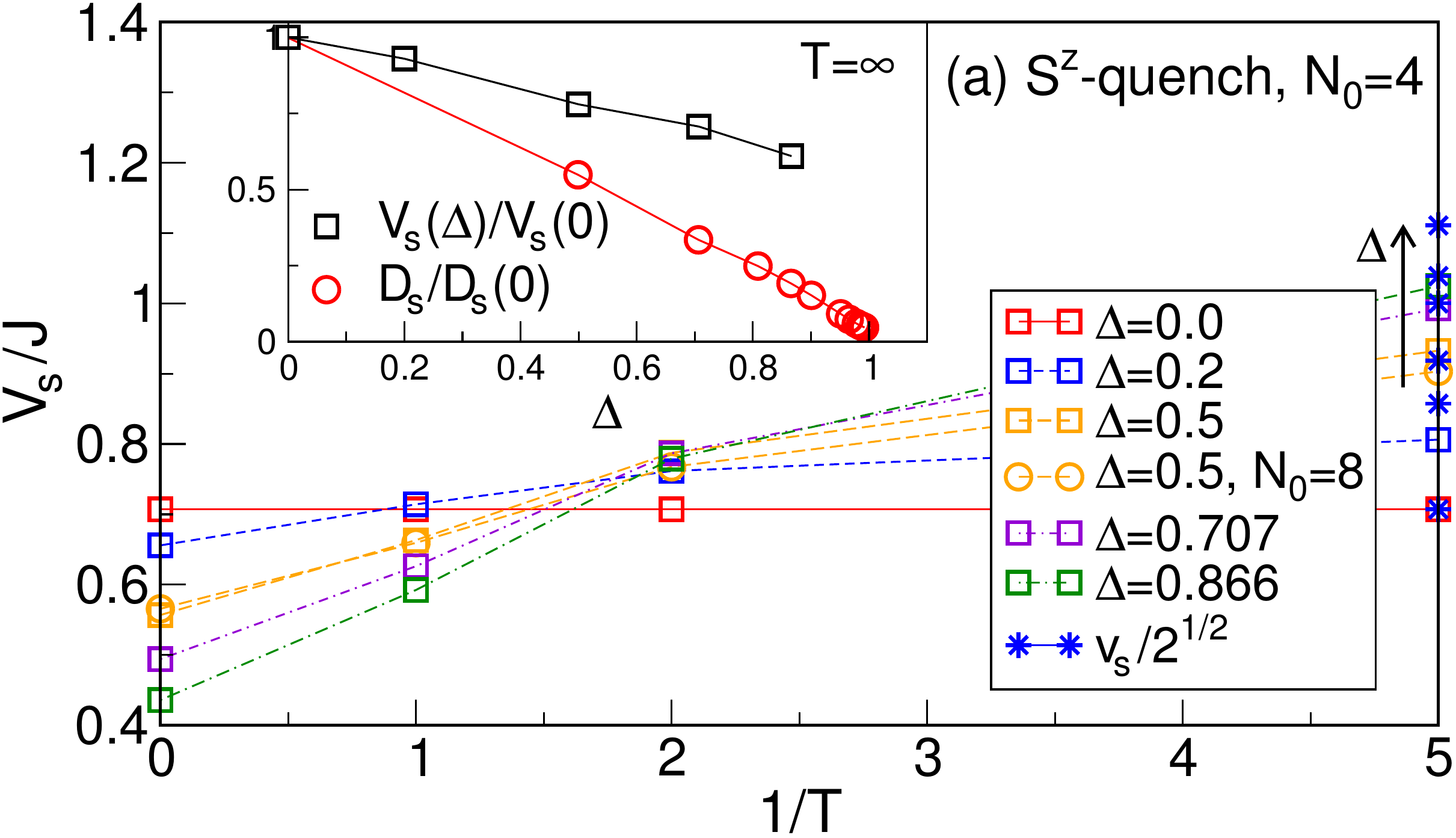}\\
\vspace{0.2cm}
\includegraphics[width=\columnwidth,clip]{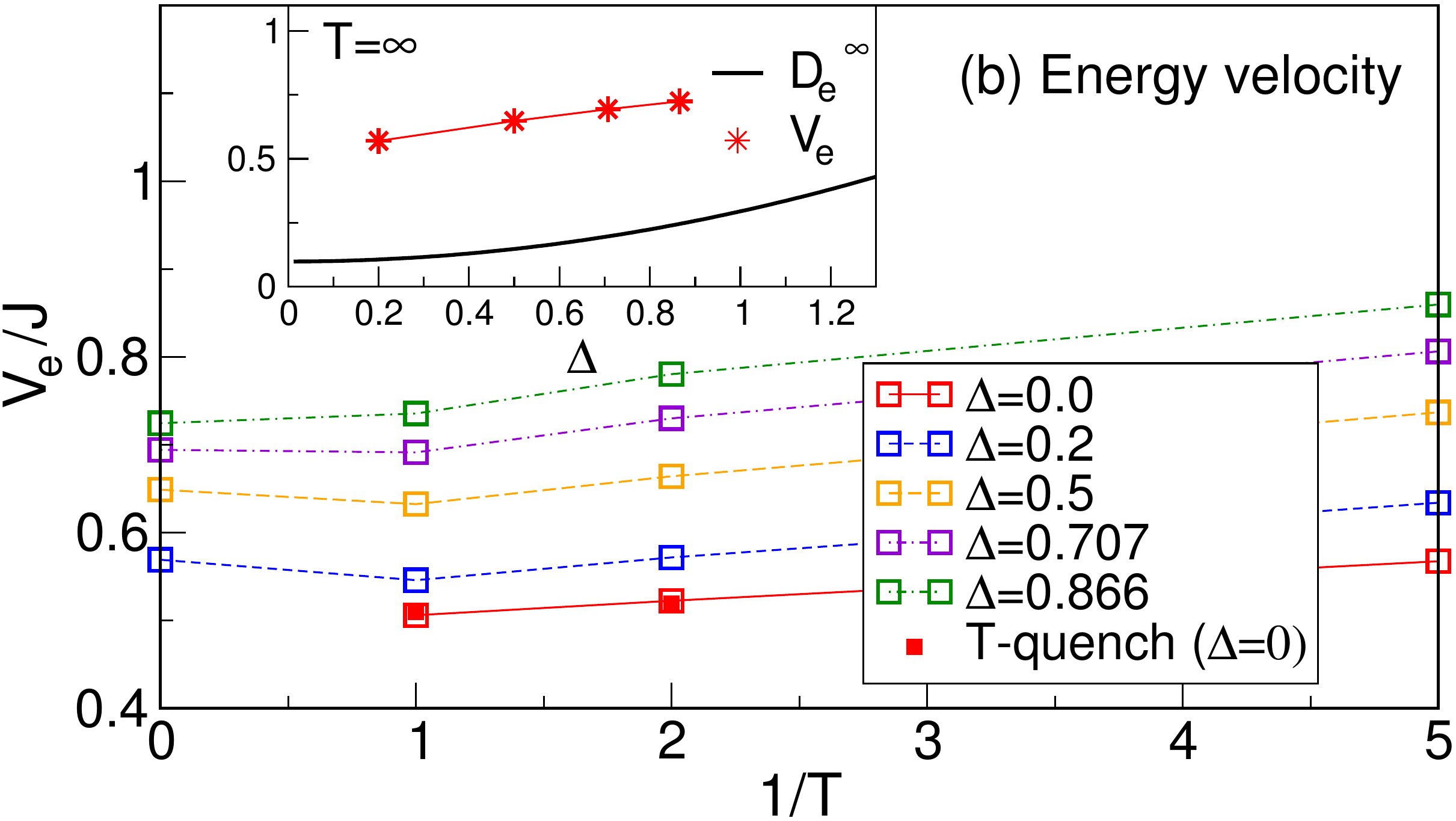}
\caption{(Color online)
(a) {\it $S^z$-quench.} Velocity $V_s$ for the ballistic spreading of perturbations in the  spin density in $S^z$-quenches with $N_0=4$ as a function of inverse temperature $\beta=1/T$.
The stars at $\beta=5$ are the spinon velocity $v_s/\sqrt{2}$ from Eq.~\eqref{eq:vs} calculated for $\Delta=0,0.2,0.5, \cos(\pi/4), \cos(\pi/5), 1$ (bottom to
top). Inset: $V_s(\Delta)/V_s(0)$ and spin-Drude weight $D_s(\Delta)/D_s(\Delta=0)$ at $T=\infty$ versus $\Delta$. The data for the Drude weight are taken
from Ref.~\onlinecite{karrasch13}.
(b) Energy velocity $V_e$ extracted from $S^z$-quenches (open symbols) and $T$-quenches (solid symbols) versus inverse temperature
$\beta=1/T$ for various values of $\Delta$. For the $T$-quench, $T=T_1$ and $T_2=\infty$. Inset: $V_e$ and Drude weight $D_{e}^{\infty}$ versus $\Delta$ at $T=\infty$. The result
for $D_e$ was derived from the Bethe ansatz (BA) in Ref.~\onlinecite{kluemper02} and $D_{e}^{\infty} = \lim_{T\to \infty} \lbrack T^2 D_e(T) \rbrack$.
}
\label{fig:velocity}
\end{figure}

\subsubsection{Spin-perturbation velocity $V_s$}

In the ballistic regime (spin transport for $0\leq \Delta<1$ and energy transport at all $\Delta$), the prefactor $V_{\nu}$
has the meaning of velocity. For small perturbations spreading out at \textit{zero} temperature and in the Luttinger liquid phase,
this velocity equals the renormalized spinon velocity.\cite{langer09,langer11}

Our results for the finite-temperature case are shown in Fig.~\ref{fig:velocity}(a). For the $S^z$-quench, we observe that (i) $V_s$ does not depend
on $\beta=1/T$ at $\Delta=0$, (ii) it increases monotonically with $\beta$ in the interacting case $0<\Delta<1$, and (iii) it only weakly depends on $N_0$. We will now try to understand $V_s$ qualitatively in the limits of high and low temperatures.

For the non-interacting system $\Delta=0$, the velocity $V_s$ is given by
\begin{eqnarray}
V_s^2 &=&\frac{1}{2\delta S^z} \sum_k v_k^2 \delta n(\epsilon_k, \beta) \label{eq:vs2n}
\end{eqnarray}
where
$\delta S^z= S^z$  for the initial state in $S^z$-quenches and $\delta n(\epsilon_k,\beta)$ is the change in the quasi-momentum distribution
function of spinless fermions at inverse temperature $\beta$ induced by the quench.\cite{langer11}
$\epsilon_k=-J\cos(k)$ is the dispersion at $\Delta=0$ with group velocity  $v_k=J\sin(k)$.

 The fact that in the initial state, the central region of width $N_0$ 
is fully polarized, implies that  the momentum distribution changes across the whole Brillouin zone.
For non-interacting particles,
we conclude from Eq.~\eqref{eq:vs2n} that $V_s$ is temperature independent  
with
\begin{eqnarray}
V_s^2 &=& J^2/2 \,. \label{eq:vs2n1}
\end{eqnarray}
While this argument cannot be strictly applied to the interacting case since there, Eq.~\eqref{eq:vs2n} 
is not valid, we still conjecture that at low temperatures and for small $N_0$ or sufficiently small $S^z_2<1/2 $, 
$$V_s(\Delta)=v_s(\Delta)/\sqrt{2}\,,$$
where 
$v_s$ is the renormalized spinon velocity:\cite{schulz96}
\begin{equation}
v_s=J \frac{\pi}{2}\frac{\sin{\gamma}}{\gamma}\,,~~\Delta=\cos(\gamma)\,.
\label{eq:vs}
\end{equation}
This behavior, i.e., a decrease of $V_s$ with increasing $\Delta$ is qualitatively consistent with our numerical results at low temperatures $\beta=5$ [see Fig.~\ref{fig:velocity}(a)].
Note that at low temperatures and in the gapless regime where conformal field-theory is valid, 
we can relate $V_s$ to the spin-Drude weight since the latter is:\cite{shastry90}
\begin{equation}
D_s = \frac{\pi}{8}\frac{\sin{\gamma}}{\gamma(\pi-\gamma)}; \enspace \Delta=\cos(\gamma)\,,
\end{equation}
i.e.,
$D_s \propto K \, v_s$, where $K$ is the Luttinger parameter.
Therefore, in the low temperature regime,
\begin{equation}
D_s \propto K \, V_s\,.
\end{equation}

At infinite temperatures, $V_s(T=\infty)$ decreases with increasing $\Delta$,  opposite to the behavior at low temperatures:
 The velocities follow a similar trend as the infinite-temperature Drude weight, which decreases monotonously
as a function of $\Delta$ for $\Delta>0$, according to some studies [see Ref.~\onlinecite{hm03,karrasch13,herbrych11}; other studies that computed only 
lower bounds to $D_s(T=\infty)$ suggest a {\it fractal}, non-monotonic dependence of $D_s(T=\infty)$ on $\Delta$ \cite{prosen11,prosen13}].
For comparison we show both quantities, $V_s(\Delta)$ and $T \cdot D_s(\Delta)$ at $T=\infty$ in the inset of Fig.~\ref{fig:velocity}(a),
which remarkably, exhibit a very similar dependence on $\Delta$. Therefore, the decrease of $V_s$ as $\Delta$ approaches $\Delta=1$
is related to the disappearance of the ballistic contribution, which is believed to be fully absent for $\Delta>1$.

Because of these opposite trends of $V_s=V_s(\Delta)$ at $T=0$, where $V_s(\Delta>0)$ is a decreasing function with $V_s(\Delta>0)>V_s(\Delta=0)$, versus $T=\infty$, where
 $V_s(\Delta>0)<V_s(\Delta=0)$, we qualitatively understand the overall temperature-dependence of $V_s=V_s(\beta)$ for $0\leq\Delta<1$, while at present
we cannot analytically predict the numbers beyond $T\approx 0$.

\subsubsection{Energy-perturbation velocity $V_e$}
 
For the velocities associated with energy spreading and for the $T$-quench,
we observe that these depend on $T_2$ at low back-ground temperatures $T_1$. 
This behavior is similar to the results of Ref.~\onlinecite{langer11}, where 
the spreading of energy at zero temperature was studied, with initial states 
that contained a central region with $E_2>E_0$, where $E_0$ isthe ground state
energy. The initial states in Ref.~\onlinecite{langer11} also have a box-like
structure, and therefore, in both cases, excitations all across
the Brillouin zone are excited, and not only those with a linear dispersion.
This can happen even for small $\delta T= T_2-T_1$ and $\delta E=E_2-E_0$, respectively, such that 
effects of a non-linear dispersion and band-structure always become relevant,
rendering the ballistic expansion velocities usually $\delta T$ or $\delta E$-dependent.
This is obvious from the expression for the ballistic velocity for 
the non-interacting case:\cite{langer11}
\begin{eqnarray}
V_e^2 &=&\frac{1}{\delta E} \sum_k \epsilon_k v_k^2 \delta n(\epsilon_k) \label{eq:vs2na} \,, \nonumber \\
\delta n(\epsilon_k) & =& n_{\rm init}(\epsilon_k)- n(\epsilon_k, \beta_1)\, ,
\end{eqnarray}
where $n_{\rm init}(\epsilon_k)$ is measured in the initial state.
Eq.~(\ref{eq:vs2na}) is in agreement with our numerical results.

In the following, we 
restrict the analysis of $V_e$ to those temperatures $T_1$ for which $V_e$ is independent of $T_2$.
The results for $V_e$ extracted from either $S^z$- or $T$-quenches agree with each other for $\Delta=0$, yet we observe
deviations for finite $\Delta>0$. 

Our results for $V_e$ versus inverse temperature are shown in Fig.~\ref{fig:velocity}(b) for various values of $0\leq \Delta <1$, 
unveiling a weak dependence on $\beta$, with a slight decrease towards higher temperatures. Moreover, at all $T$, $V_e$ increases with $\Delta$
and therefore follows  the trend of the infinite-temperature thermal Drude weight\cite{kluemper02,sakai03} [see the inset in Fig.~\ref{fig:velocity}(b)].
Note that at $\Delta=0$ and at $T=\infty$, an $S^z$-quench does not induce a change in the energy density and therefore, $V_e=0$ in that case.
The low-temperature behavior is again controlled by the renormalization of the spinon velocity $v_s$.

\subsection{Spin-diffusion constant in the easy-axis regime $\Delta>1$}

\begin{figure}[t]
\includegraphics[width=\columnwidth,clip]{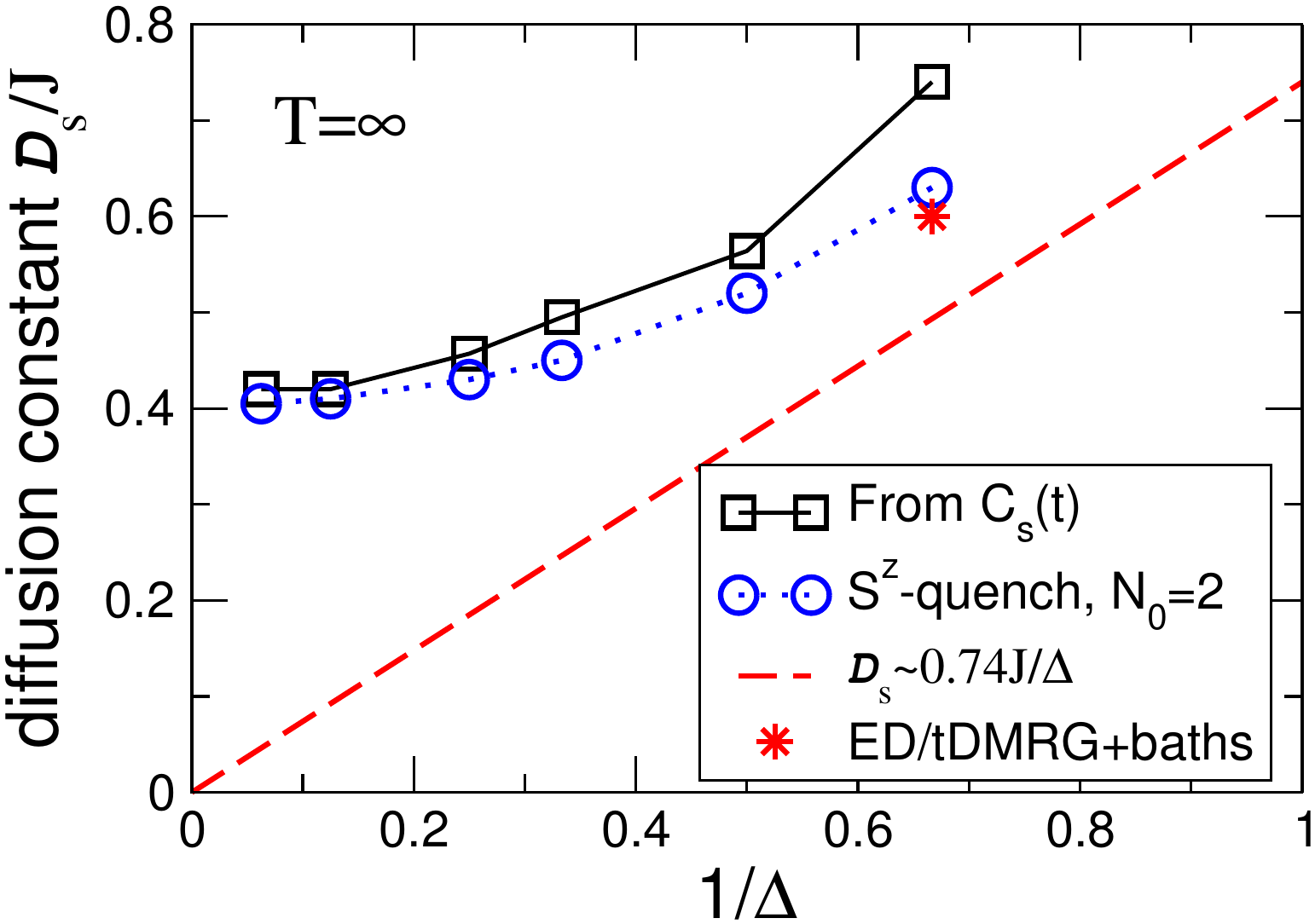}
\caption{(Color online)
{\it XXZ chain, $\Delta>1$, diffusion constant.}
Infinite-temperature
diffusion constant estimated from current-current correlation functions
(squares) compared to the results from open systems coupled to baths (Ref.~\onlinecite{znidaric11}), and exact diagonalization (star, Refs.~\onlinecite{znidaric11,steinigeweg09}).
Star: Data from Refs.~\onlinecite{znidaric11,steinigeweg09,steinigeweg12} for $\Delta=1.5$;  the dashed line $\mathcal{D}_s\propto 1/\Delta$ is the large-$\Delta$ prediction from Ref.~\onlinecite{znidaric11}.
We further include our estimates from the expansion of local spin-density perturbations in $S^z$-quenches for $N_0=2$ (circles),
which follow the same  trend as the data obtained from current correlations.}
\label{fig:comparison}
\end{figure}

We extract the infinite temperature diffusion constant $\mathcal{D}_s$ in two ways: (i) From the time dependence of the spatial 
variance $\delta \sigma_s^2$ in $S^z$-quenches and (ii) from the time dependence of current-current autocorrelations functions 
using  an Einstein relation.

\begin{figure}[t]
\includegraphics[width=\columnwidth,clip]{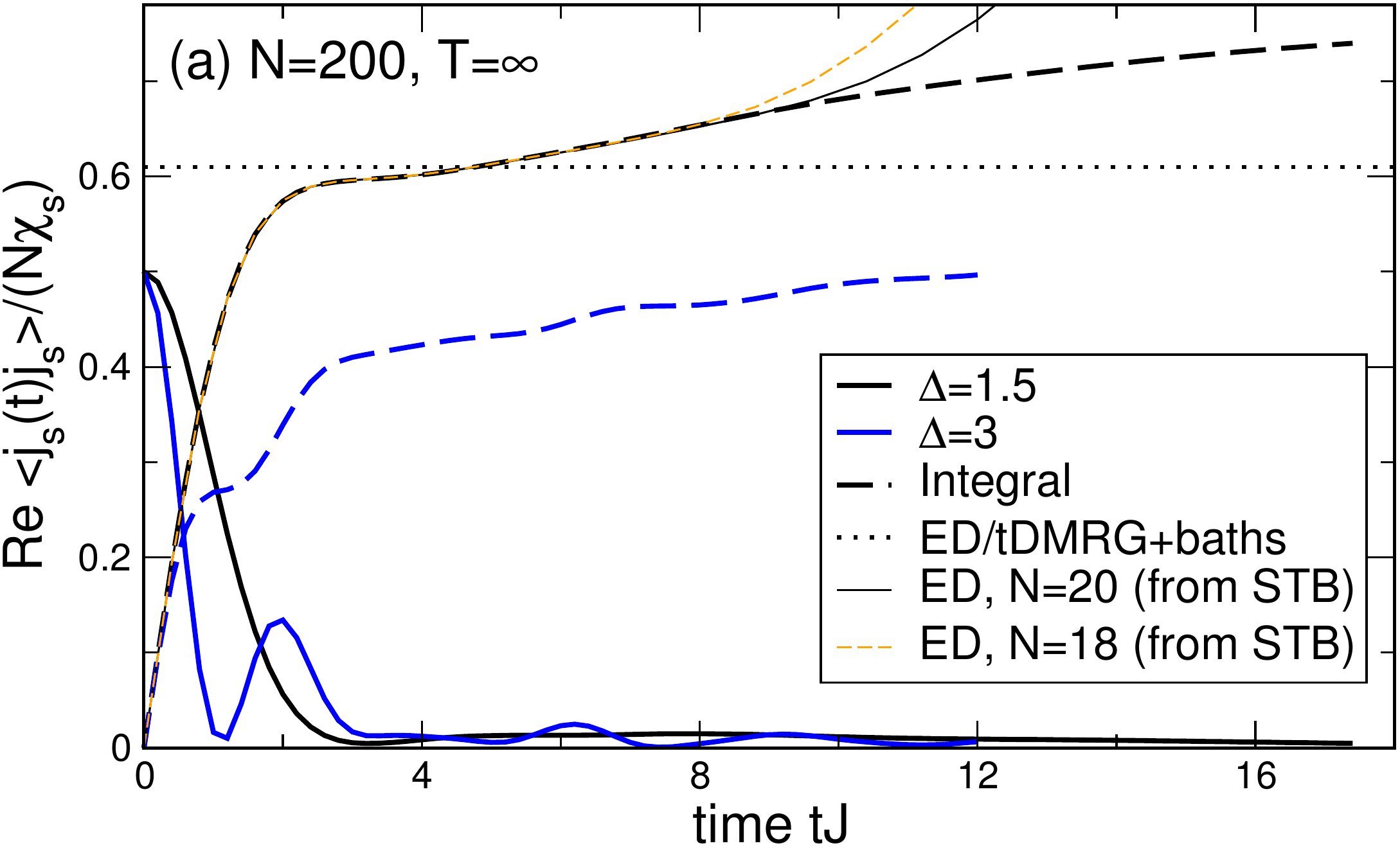}\\
\vspace{0.2cm}
\includegraphics[width=\columnwidth,clip]{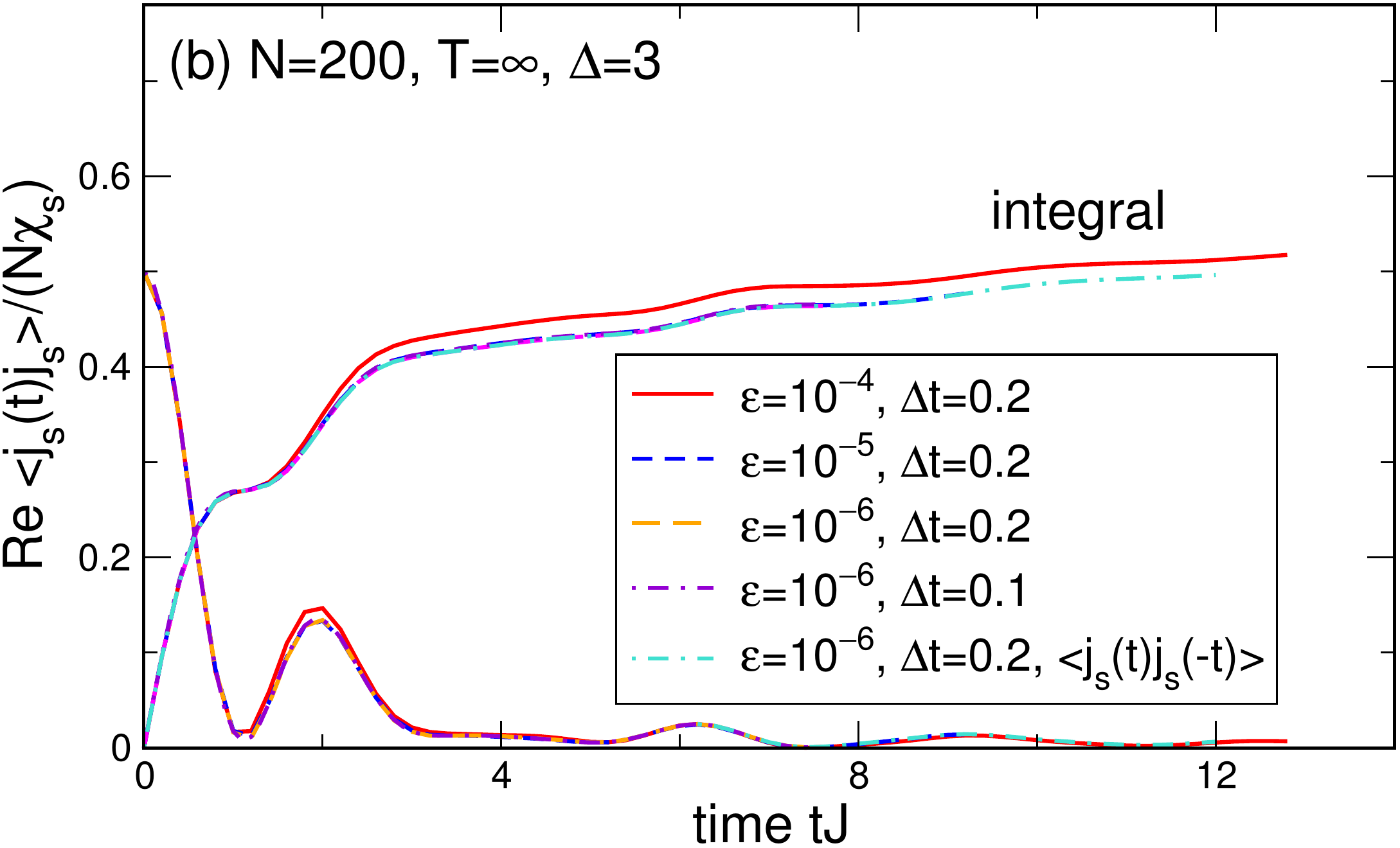}
\caption{(Color online)
{\it $XXZ$ chain, $\Delta>1$, Einstein relation.} (a) Spin-current-correlation function (solid lines) at infinite temperature for $\Delta=1.5,3$ and its integral (dashed lines, $N=200$). From this we conclude
$\mathcal{D}_s > 0.6$ in the case of $\Delta=1.5$  [compare Eq.~\eqref{eq:diff}], since $\mathcal{D}_s(t)$ keeps increasing for $t>8/J$. Since on the time scales reached,
the data are converged with respect to system size, the increase for large times is {\it not} due to a residual Drude weight (which, based on
the ED data from Ref.~\onlinecite{hm03}, would need to be a much faster increase for this $N$).
The thin lines are the ED results for $N=18$ (thin-dashed) and $N=20$ (thin-solid) for $\mathcal{D}_s$ from Ref.~\onlinecite{steinigeweg12} (STB). (b) Dependence on the discarded weight $\epsilon$ and Trotter step size $\Delta t$. By re-writing $\langle j_s(2t)j_s(0)\rangle=\langle j_s(t)j_s(-t)\rangle$, time scales twice as large can be reached.
}
\label{fig:currcorr}
\end{figure}
 
In $S^z$-quenches, $\mathcal{D}_s$ depends on $N_0$ but saturates as we decrease $N_0$. We have also investigated initial states
with $0< S^z_2 <1/2$, for which the diffusion constant agrees with our standard $S^z$-quench as $N_0$ is decreased. 
To obtain $\mathcal{D}_s$, we 
fit
\begin{equation}
\delta \sigma^2_s(t) = a_0 + 2 \mathcal{D}_s t
\end{equation}
to the spatial variance at the longest times where it increases linearly in time.
We resolve a linear increase of $\delta \sigma^2_s(t)$ 
for high temperatures whereas 
at lower temperatures, where presumably the diffusion
constant will increase from its infinite-temperature value, we usually do not reach such a saturation.
Even at $T=\infty$, we  underestimate $\mathcal{D}_s(t)$ because of the finite times reached in our tDMRG simulations.

As an estimate for the diffusion constant at $T=\infty$ and $\Delta=1.5$, we obtain $\mathcal{D}_{s}\approx 0.63 J$ from the spreading of wave-packets. This value is comparable to, but significantly larger than  results from the literature that were obtained with either exact diagonalization \cite{steinigeweg09,steinigeweg10,steinigeweg12}
or simulations of  steady-state transport in open systems coupled to baths using tDMRG methods.\cite{prosen09,znidaric11}
 
To further elucidate the $\Delta$-dependence, we plot our results for $\mathcal{D}_s$ versus $1/\Delta$, inspired by a prediction 
for the behavior of the diffusion constant at large $\Delta \gg 1$ by Znidaric\cite{znidaric11} (see also the discussion in Refs.~\onlinecite{steinigeweg10,steinigeweg12}): 
\begin{equation}
\mathcal{D}_s \approx 0.74/\Delta \label{eq:diffzni}
\end{equation}
(note that in Ref.~\onlinecite{znidaric11},  a different definition for the spin current is used and hence the  numbers are larger by a trivial factor of four).
This functional form was motivated by perturbative arguments and agrees with tDMRG simulations for steady-state transport in open systems at sufficiently large $\Delta>1$.\cite{znidaric11} A similar expression was derived from an analysis of moments of the Fourier transform of spin-density correlation function,\cite{huber12} (see also Refs.~\onlinecite{huber69,krueger71}) with a
slightly larger prefactor, ${D}_s \approx 0.89/\Delta$.
However, our results qualitatively disagree with Eq.~\eqref{eq:diffzni} for large $\Delta$, as is evident from Fig.~\ref{fig:comparison}.
Our data is rather described by
\begin{equation}
\mathcal{D}_s= \mathcal{D}_s^{\infty} + \frac{a}{\Delta^2} + \dots;
\end{equation}
where $a$ is a constant.

The diffusion constant $\mathcal{D}_{\nu}$ can also be extracted from the long-time behavior of current-correlation functions in combination with 
an Einstein relation:
\begin{equation}
\mathcal{D}_{\nu} = \frac{\sigma_{\rm dc, \nu}}{\chi_{\nu}}
\end{equation}
where $\sigma_{\rm dc, \nu}$ is the dc-conductivity and $\chi_{\nu}$ is the associated static susceptibility.

 We follow this approach to test our results for $\mathcal{D}_s$ obtained from $S^z$-quenches. 
As an improvement of the estimate of $\mathcal{D}_s(T)$ from exact diagonalization (see Refs.~\onlinecite{steinigeweg09,steinigeweg12}) we use finite-temperature DMRG to evaluate current-current
correlations functions (see Refs.~\onlinecite{karrasch12,karrasch13,karrasch13a})
\begin{equation}
C_{\nu}(t)=\langle j_{\nu}(t) j_{\nu}\rangle\,,
\end{equation}
where $j_{\nu}$ is the spin- ($\nu=s$) or energy- ($\nu=e$) current.
We exploit a recently-introduced trick,\cite{barthel13b}
\begin{equation}
C_{\nu}(2t)=\langle j_{\nu}(2t) j_{\nu}(0)\rangle = \langle j_{\nu}(t) j_{\nu}(-t)\rangle \,,
\end{equation}
which allows us to access time scales twice as large within DMRG.

Following Ref.~\onlinecite{steinigeweg09}, we introduce a time-dependent diffusion constant $\mathcal{D}_{\nu}(t)$, now related to the time dependence of $C_{\nu}(t)$
\begin{equation}
\mathcal{D}_{\nu}(t)= \frac{1}{N\, \chi_{\nu}} \int_0^tdt'\, C_{\nu}(t')\,.
\label{eq:diff}
\end{equation} 
Here,  $\chi_{\nu}$ is the static susceptibility
\begin{equation}
\chi_{\nu} =\frac{1}{N} \lbrack \langle X_{\nu }^2\rangle -\langle X_{\nu }\rangle^2 \rbrack 
\,.
\end{equation}
where $X_s=S^z$ and $X_e=H$. 
At infinite temperature,  $\chi_s=1/4$ for sufficiently large $N$.  
 
\begin{figure}[t]
\includegraphics[width=0.9\columnwidth,clip]{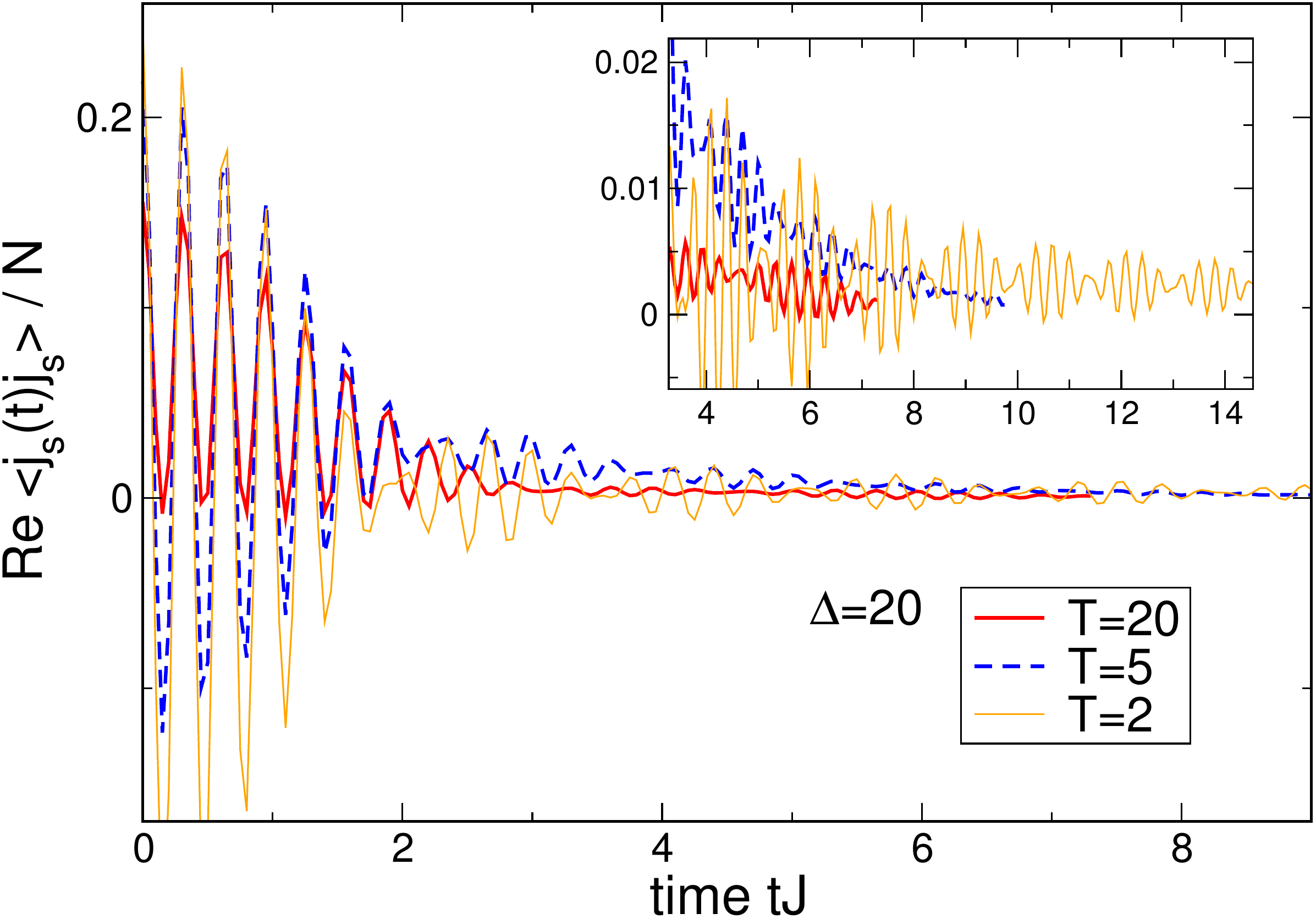}
\caption{(Color online) Spin current correlation function of the XXZ model at large anisotropies $\Delta=20$ and various temperatures.}
\label{fig:currcorrD20}
\end{figure}

The time dependence of $C_s(t)$ and $\mathcal{D}_s(t)$ for $T=\infty$ at $\Delta=1.5,3$ is shown in Fig.~\ref{fig:currcorr}.
For $\Delta=1.5$,  $\mathcal{D}_s(t)$ appears to saturate at $t\approx 4/J$, yet then starts to increase again.  Our results
agree quantitatively with the exact diagonalization data from Ref.~\onlinecite{steinigeweg10,steinigeweg12} for $t\lesssim 8/J$,
yet we are able to reach $t\lesssim 17/J$ and in this time window, there are no finite-size dependencies. Therefore, the increase
of $\mathcal{D}_s(t)$ beyond the plateau at $\mathcal{D}_s(t)\approx 0.6J$ (the value reported in  Refs.~\onlinecite{steinigeweg09,steinigeweg10,steinigeweg12})
is {\it not} related to a residual, finite-size Drude weight, which would lead to an increase at large $t$  with a slope that is larger by a factor of four for the system sizes considered here, 
compare Ref.~\onlinecite{hm03}.
Note that for small $\Delta>1$, our tDMRG simulations for  $\mathcal{D}_s(t)$ do obviously not reach time scales at which
saturation is reached and therefore (see the example of $\Delta=1.5$ in Fig.~\ref{fig:currcorr}), the results for small $\Delta$
are only a {\it lower} bound to the true diffusion constant.
Entanglement also grows the faster the larger $\Delta$ is, but fortunately, the time-dependent diffusion
constant also saturates faster deeper in the Ising regime (compare Fig.~\ref{fig:currcorrD20}).
 
The qualitative behavior of $\mathcal{D}_s(t)$ becomes more transparent by inspecting the data for larger values of $\Delta$:  $C_s(t)$
exhibits an oscillatory behavior at large times, leading to maxima. In between those maxima, $\mathcal{D}_s(t)$ takes roughly constant values, leading to 
a sequence of plateaus in time. Such a behavior was anticipated in Ref.~\onlinecite{steinigeweg12}, where, however, the residual Drude weight screened all this long-time
dynamics for the  smaller system sizes $N\leq 20$ considered there.

Most strikingly, the results from our two approaches of estimating $\mathcal{D}_s$ (i.e., from the time dependence of the spatial variance in $S^z$-quenches
and Einstein relations), agree fairly well with each other (see Fig.~\ref{fig:comparison}) but are inconsistent with Eq.~\eqref{eq:diffzni} at large $\Delta$. 
Rather, $\mathcal{D}_s$ appears to saturate at large $\Delta$. 
We attribute the quantitatve discrepancies at small $\Delta\gtrsim 1$ between our two ways of extracting of $\mathcal{D}_s$ to the different
time scales reached in computing time-evolving density profiles versus current correlation functions.

To get a picture of the dynamics of the $XXZ$ Hamiltonian at large $\Delta \gg1= J_{xy}$ [with $J_{xy}=1$ being the prefactor of the kinetic energy formed by the first two terms in Eq.~(\ref{eq:ham})], let us divide the states into ``sectors'' of constant Ising energies (i.e., with the same value of the $\Delta$ term in the Hamiltonian).  Then the action generated by the $J_{xy}$ term is only within a sector, up to terms of size $(J_{xy})^2 / \Delta$, which we ignore.  At low temperature, we need to analyze the low-lying sectors, while at infinite temperature, all Ising sectors are weighted equally.  The numerical observation that the diffusion constant is finite in the $T \gg \Delta \gg J_{xy}$ limit suggests that a finite fraction of sectors have nonzero diffusion constant and that at most a set of measure zero are ballistic.
The fact that $\mathcal{D}_s$ does not depend on $\Delta$ implies that  scattering within each sector  (controlled by $J_{xy}$) dominates over intra-band processes.

An interesting case of nearly ballistic transport, but with zero Drude weight, appears as follows.  The physics at large antiferromagnetic $\Delta \gg T \gg J_{xy}$ can be analyzed analytically.  The dynamics in this limit breaks up into decoupled sectors as above with the same Ising model energy, and the ground states are the two antiferromagnetic Ising states, which are static in this limit.  The lowest excited states have a single domain wall of the antiferromagnetism (i.e., a single point where two up spins or two down spins are adjacent).  This domain wall is hopped one site in either direction by the $J_{xy}$ term, so it will move ballistically if there is a single such wall.  However, the number of such walls is exponentially small once $\Delta/T$ is large.  This leads to the following picture: as $T \rightarrow 0$, the distance traveled by a domain wall before it interacts with others and ceases to be ballistic diverges.  At the same time the number of such domain walls, or equivalently diffusing spins, goes to zero.

This increasingly ballistic behavior is consistent with our numerical observations at low temperature, where we cease to see the current decay (see Fig.~\ref{fig:currcorrD20}), and with the idea that the Drude weight is zero at any nonzero temperature (and at $T=0$ where only the two ground states are present).  A similar tradeoff between exponential factors in the density of excitations and the distance between them occurs in the study of transport in massive nonlinear sigma models by Damle and Sachdev.\cite{damle98,damle05} For gapped one-dimensional models with triplet excitations, they predict a universal low-temperature behavior for temperatures smaller than the spin gap $\Delta_{\rm gap}$ of the form:
\begin{equation}
\mathcal{D}_s \propto \frac{1}{\Delta_{\rm gap}} e^{\Delta_{\rm gap}/T}\,.
\end{equation}

Finally, our results for the diffusion constant could be put to a quantitative experimental test by measuring the diffusion constant in quantum gas experiments that realize spin-1/2 $XXZ$ Hamiltonians. For instance the dynamics
of few-magnon excitations has recently been studied\cite{fukuhara13,fukuhara13a} in two-component Bose gases, 
which for strong repulsive interactions maps to an $XXZ$-type Hamiltonians with anisotropies ($\Delta \approx 1$) 
and a ferromagnetic exchange.

\section{Non-integrable models: The two-leg ladder}
\label{sec:ladder}

\begin{figure}[t]
\includegraphics[width=0.49\linewidth,clip]{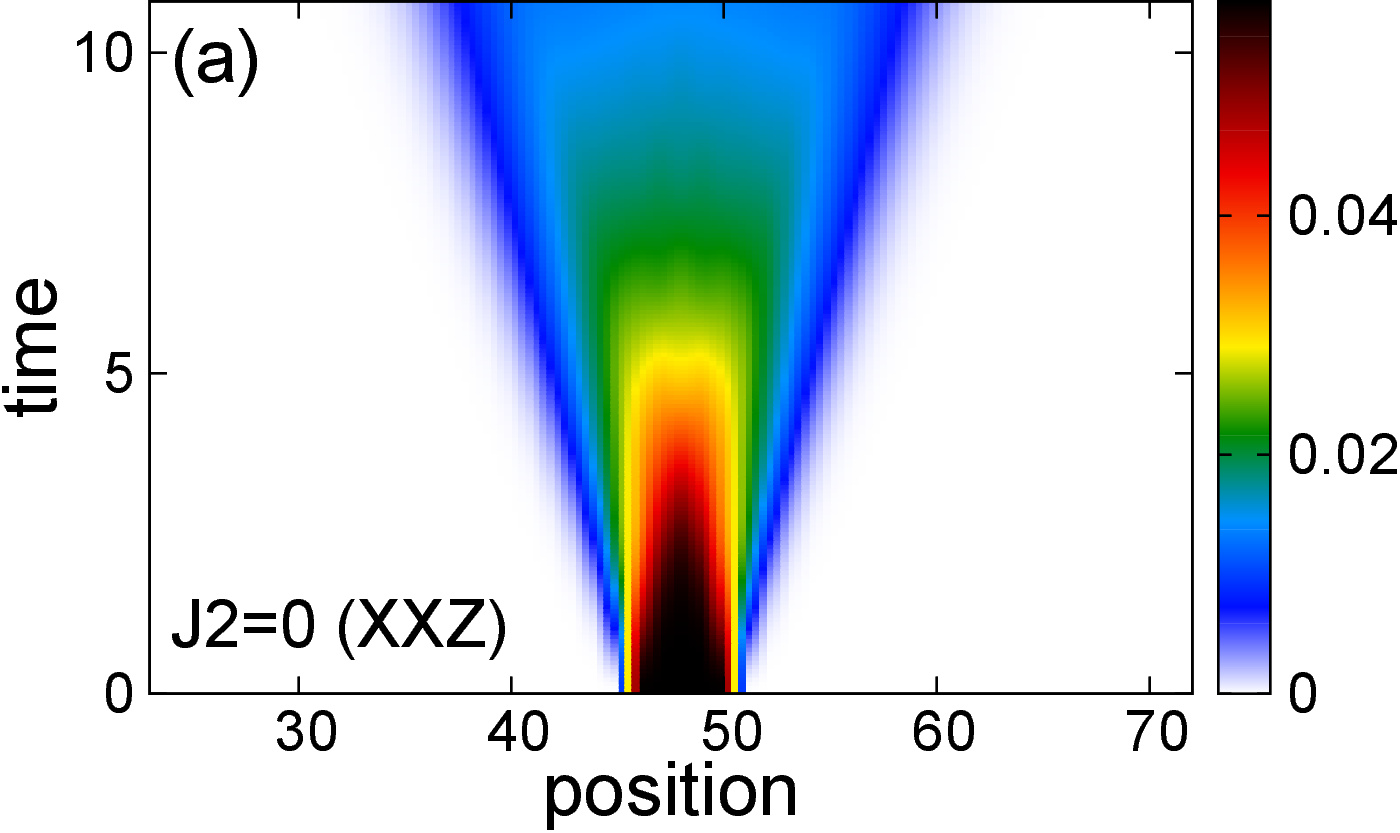}\hspace*{0.01\linewidth}
\includegraphics[width=0.49\linewidth,clip]{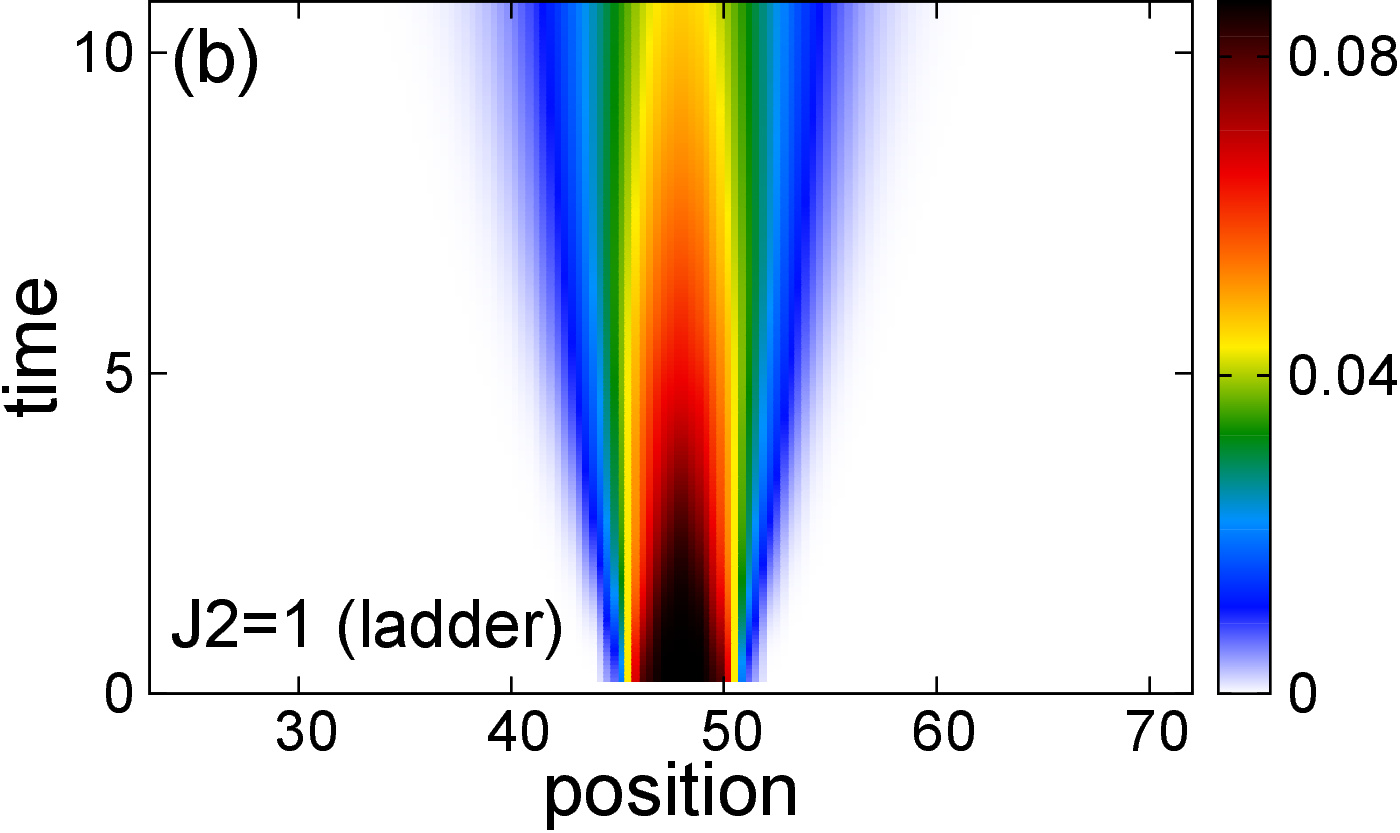} \\[0.1cm]
\includegraphics[width=0.49\linewidth,clip]{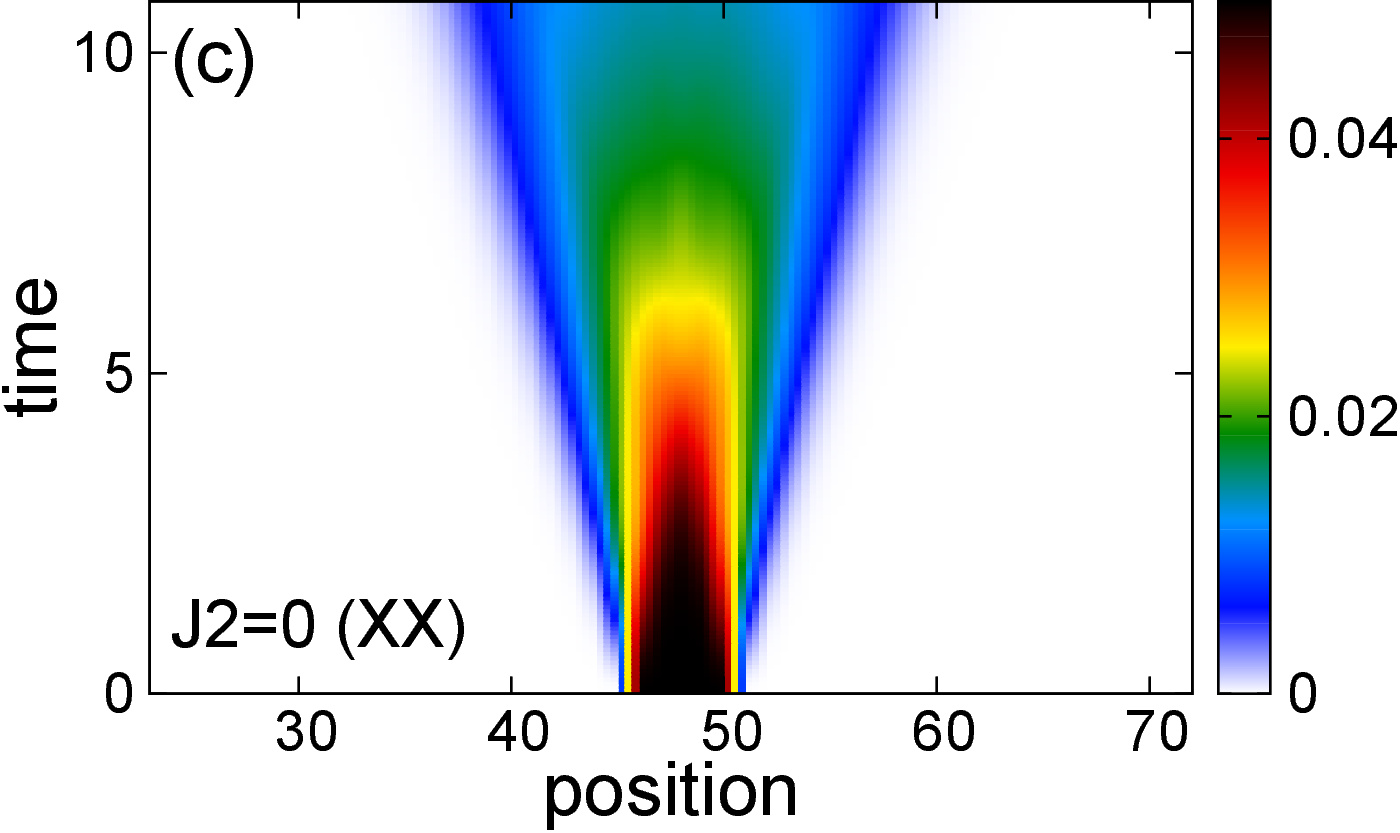}\hspace*{0.01\linewidth}
\includegraphics[width=0.49\linewidth,clip]{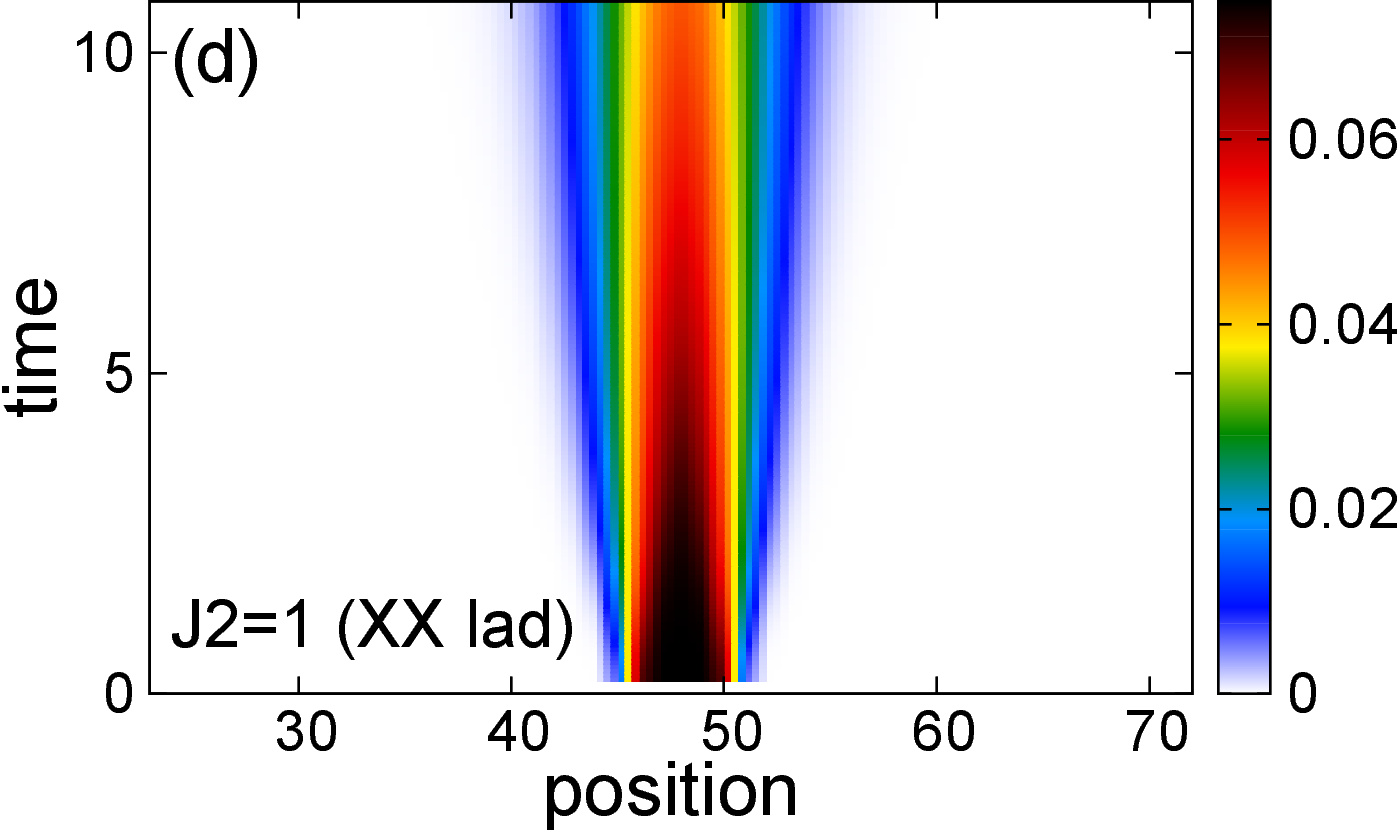} \\[0.2cm]
\caption{(Color online) {\it $T$-quench.} Expansion of a local energy wave packet defined by $T_1=5$, $T_2=\infty$, $N=100$, $N_0=4$  within (a) the $XXZ$ chain ($J_2=0$) at $\Delta=0.5$, (b) a two-leg ladder with $\Delta=0.5$, $J_2=1$, (c) the $XXZ$ chain ($J_2=0$) at $\Delta=0$, and (d) a two-leg XX ladder with $\Delta=0$, $J_2=1$.}
\label{fig:t1t2}
\end{figure}

\begin{figure}[t]
\includegraphics[width=\columnwidth,clip]{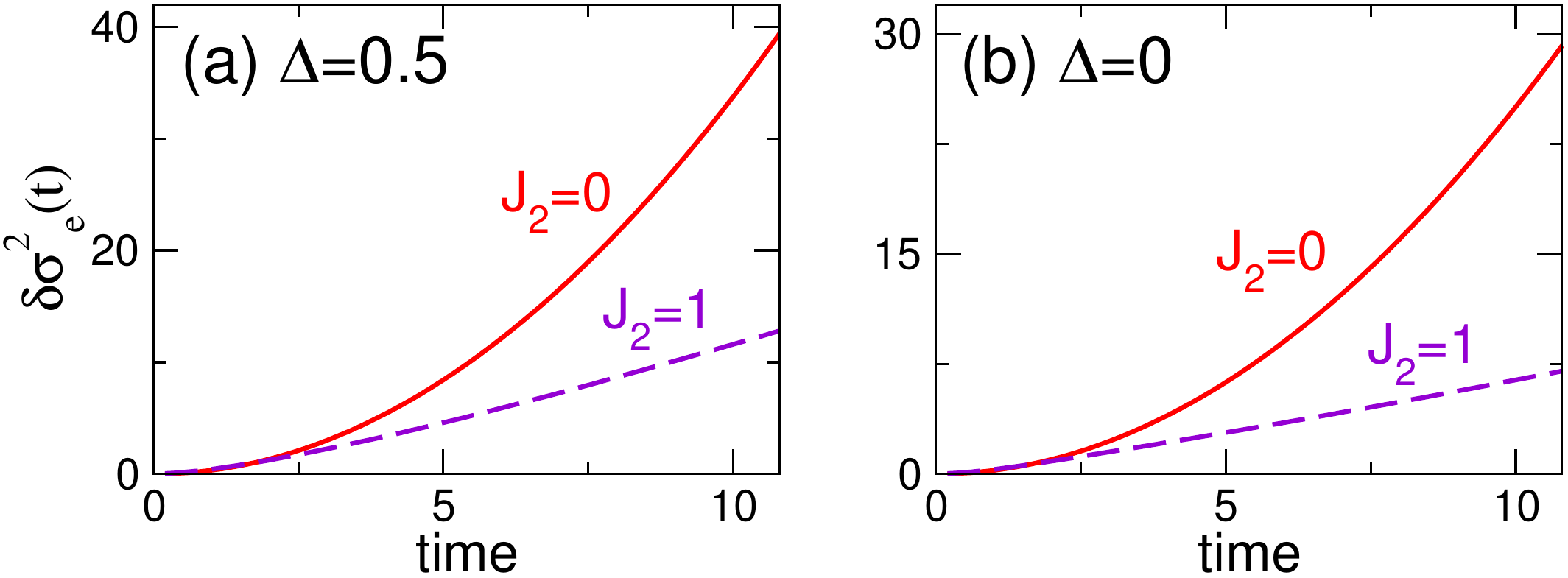} 
\caption{(Color online) Variance of the energy density corresponding to the data shown in Fig.~\ref{fig:t1t2}.}\label{fig:sigma_ladder}
\end{figure}

We next consider a non-integrable system that is relevant in the context of recent experiments:
Spin-1/2  $XXZ$ two-leg ladders ($J_2>0$). We focus on the cases $\Delta>0$ in Eq.~\eqref{eq:ham} and the specific example 
of $XX$ ladders
[$J_2>0$ and $\Delta=0$ in Eq.~\eqref{eq:ham}].
The former system is interesting for the interpretation of time-resolved heat transport measurement on 
(La,Ca,Sr)$_{14}$Cu$_{24}$O$_{41}$ (Refs.~\onlinecite{otter09,montagnese13,otter12}), while the latter can be
realized with hard-core bosons in optical lattices (see the discussion in Ref.~\onlinecite{vidmar13}).

Unfortunately, the time scales for which we obtain reliable data are  shorter for the ladders
than for chains. Moreover, the numerical simulations get expensive in general. We therefore mostly focus on the cases of $\Delta=0$ and $\Delta=0.5$, while we expect the qualitative
picture to carry over to  Heisenberg ladders as well ($\Delta=1$), which are relevant in the context of condensed matter experiments.

Figure~\ref{fig:t1t2} shows contour plots of the energy density in a $T$-quench for a chain ($J_2=0$)
and a ladder with $J_2=1$, both with $\Delta=0.5$.
The qualitative differences are evident: The spreading of energy for $J_2>0$ is much slower with a well defined high-density core. 
The corresponding variances (for $\Delta=0$ and 0.5) are displayed in  Fig.~\ref{fig:sigma_ladder}: The behavior of both ladder systems
is consistent with energy diffusion at the longest times, i.e., a linear increase of $\delta \sigma^2_e$.
The difference with the integrable cases ($\Delta=0.5$ and $\Delta=0$ at $J_2=0$) is evident.

By following the time dependence of the spatial variance $\delta \sigma_e^2(t)$ shown in  Fig.~\ref{fig:sigma_ladder}, we  
obtain a lower bound for the energy diffusion constant of ladders at infinite temperature, namely $\mathcal{D}_e\approx 0.88 J$ for $\Delta=0.5$
and $\mathcal{D}_e\approx 0.29 J$ for $\Delta=0$.

Qualitatively similar observations for the time dependence of the spatial variance can be made for $S^z$-quenches (results not shown here): For a sufficiently
large $J_2$ we are able to resolve clear deviations from  ballistic dynamics.
A more quantitative analysis of spin- and energy diffusion constants as a function of temperature and $J_2$ is left for future work.

The behavior of the energy spreading in ladders with $J_2=1$ is similar to the analysis of experimental results with complex metal oxides,\cite{otter12} where a diffusive
spreading of heat was observed by measuring temperature profiles in the surface of oxide materials (the initial 
spatially inhomogeneous temperature profile was induced by laser light).
However, a direct and quantitative connection with these experiments can at present not be made since the time scales reached in tDMRG simulations
are too short to estimate the temperature dependence of $\mathcal{D}_e(T)$. In the future, one would also be interested in including phonons into
numerical simulations.  

The results for the $XX$ ladder support the picture suggested in Ref.~\onlinecite{vidmar13}, where the sudden expansion
of hard-core bosons on a ladder was analyzed, equivalent to spin-transport in our model. The sudden expansion is the release
of particles from a region with a finite density into an empty lattice, realized in recent experiments with quantum gases in optical lattices.\cite{ronzheimer13, schneider12} In the limit
of $\Delta=0$, a 1D system of either hard-core bosons or spin-1/2 moments maps to free spinless fermions and consequently, both the energy and spin current are conserved. 
Turning on $J_2$ renders these fermions interacting, breaking the conservation laws of the free-fermion point, which are
the occupation of fermionic quasi-momenta. 
DMRG simulations\cite{vidmar13} for the sudden expansion of hard-core bosons on ladders find clear deviations from ballistic dynamics, however, the interpretation of
the sudden expansion is more complicated since the density overall decreases during the expansion in the   lattice. Therefore, the density-dependence of the diffusion constant (in the diffusive regime)
becomes relevant, rendering the diffusion equation non-linear.
In our present set-up, we probe directly the usual linear diffusion equation, clearly showing diffusive high-temperature dynamics for
hard-core bosons on a ladder.

An interesting possible experiment would be the analysis of the time-dependent spreading of 
a perturbation in the particle density of hard-core bosons in optical lattices at finite temperatures (or with non-equilibrium initial conditions), which could be realized using
so-called single-site resolution and manipulation techniques.\cite{bakr09,sherson10,weitenberg11,fukuhara13,fukuhara13a} For a 1D system, the spreading must 
be ballistic (see also Ref.~\onlinecite{ronzheimer13}), yet for two-ladders, our results suggest a slow and diffusive expansion. For hard-core bosons 2D, experimental results for sudden expansions\cite{ronzheimer13} are consistent with diffusive
dynamics. 

Finally, we mention that other non-integrable spin models that are obtained by perturbing the integrable $XXZ$ chain show a similar behavior in $S^z$-quenches
and $T$-quenches, including, e.g.,   chains with a staggered field.


\section{Summary and Discussion}
\label{sec:summary}
In this work we studied the real-time and real-space dynamics of spin and energy in one-dimensional quantum spin systems at finite temperatures. The
time-evolution was induced through local quenches in which initial states were prepared that have either an inhomogeneous temperature ($T$-quench) or spin-density profile ($S^z$-quench).
Using the purification trick,\cite{verstraete04,karrasch12,barthel13b} we applied the time-dependent DMRG method to follow the time-evolution of the spin and energy density. The background spin density
is kept at zero, equivalent to half-filling in the language of spinless fermions.
As a main qualitative result, we observe ballistic energy dynamics in the spin-1/2 $XXZ$ for all exchange anisotropies studied.  
Spin dynamics is ballistic in the gapless phase, whereas at high temperatures and in the massive regime, we clearly observe diffusive dynamics
on the accessible time scales. This allowed us to quantitatively estimate the infinite-temperature diffusion constant $\mathcal{D}_s$, which for large $\Delta/J <\infty$  becomes
independent of $\Delta$, in contrast to earlier numerical results for open systems.\cite{znidaric11}
The results obtained for the diffusion constant from the $S^z$-quenches are corroborated by extracting the diffusion constant from current correlation functions using an Einstein relation.

We further considered two-leg ladders as an example of non-integrable models, for which we find indications of diffusive dynamics both for spin and energy, allowing
us to extract a lower bound to the infinite-temperature energy diffusion constant. We discussed our results in view of recent experiments with quantum magnets\cite{otter12,montagnese13}
and bosons in optical lattices.\cite{ronzheimer13} Future experiments with either quantum gases or quantum magnets could provide  a quantitative test of our predictions for energy and spin
diffusion constants. 

\emph{Note} --- During completion of our work we became aware of related work by Steinigeweg, Gemmer and Brenig.\cite{steinigeweg13} 

\emph{Acknowledgments} ---  We are indebted to P. van Loosdrecht, M. Montagnese, and R. Steinigeweg for fruitful discussions, and we thank R. Steinigeweg further for sending us exact diagonalization data from Ref.~\onlinecite{steinigeweg12} for comparison. We thank T. Prosen and M. Znidaric for their comments on a previous version of the manuscript. We gratefully  acknowledge  support from to the Deutsche Forschungsgemeinschaft through  grant-no. 
KA3360-1/1 (C.K.) and through  Research unit FOR 912 (grant no. HE-5242/2-2 (F.H.-M.)) as well as from the Nanostructured Thermoelectrics program of LBNL (C.K.).


\bibliographystyle{apsrev}
\bibliography{references}

\begin{thebibliography}{107}
\expandafter\ifx\csname natexlab\endcsname\relax\def\natexlab#1{#1}\fi
\expandafter\ifx\csname bibnamefont\endcsname\relax
  \def\bibnamefont#1{#1}\fi
\expandafter\ifx\csname bibfnamefont\endcsname\relax
  \def\bibfnamefont#1{#1}\fi
\expandafter\ifx\csname citenamefont\endcsname\relax
  \def\citenamefont#1{#1}\fi
\expandafter\ifx\csname url\endcsname\relax
  \def\url#1{\texttt{#1}}\fi
\expandafter\ifx\csname urlprefix\endcsname\relax\def\urlprefix{URL }\fi
\providecommand{\bibinfo}[2]{#2}
\providecommand{\eprint}[2][]{\url{#2}}

\bibitem[{\citenamefont{Greiner et~al.}(2002)\citenamefont{Greiner, Mandel,
  H\"ansch, and Bloch}}]{greiner02}
\bibinfo{author}{\bibfnamefont{M.}~\bibnamefont{Greiner}},
  \bibinfo{author}{\bibfnamefont{O.}~\bibnamefont{Mandel}},
  \bibinfo{author}{\bibfnamefont{T.}~\bibnamefont{H\"ansch}}, \bibnamefont{and}
  \bibinfo{author}{\bibfnamefont{I.}~\bibnamefont{Bloch}},
  \bibinfo{journal}{Nature (London)} \textbf{\bibinfo{volume}{419}},
  \bibinfo{pages}{51} (\bibinfo{year}{2002}).

\bibitem[{\citenamefont{Kinoshita et~al.}(2006)\citenamefont{Kinoshita, Wenger,
  and Weiss}}]{kinoshita06}
\bibinfo{author}{\bibfnamefont{T.}~\bibnamefont{Kinoshita}},
  \bibinfo{author}{\bibfnamefont{T.}~\bibnamefont{Wenger}}, \bibnamefont{and}
  \bibinfo{author}{\bibfnamefont{D.~S.} \bibnamefont{Weiss}},
  \bibinfo{journal}{Nature} \textbf{\bibinfo{volume}{440}},
  \bibinfo{pages}{900} (\bibinfo{year}{2006}).

\bibitem[{\citenamefont{Hofferberth et~al.}(2007)\citenamefont{Hofferberth,
  Lesanovsky, Fisher, Schumm, and Schmiedmayer}}]{hofferberth07}
\bibinfo{author}{\bibfnamefont{S.}~\bibnamefont{Hofferberth}},
  \bibinfo{author}{\bibfnamefont{I.}~\bibnamefont{Lesanovsky}},
  \bibinfo{author}{\bibfnamefont{B.}~\bibnamefont{Fisher}},
  \bibinfo{author}{\bibfnamefont{T.}~\bibnamefont{Schumm}}, \bibnamefont{and}
  \bibinfo{author}{\bibfnamefont{J.}~\bibnamefont{Schmiedmayer}},
  \bibinfo{journal}{Nature} \textbf{\bibinfo{volume}{449}},
  \bibinfo{pages}{324} (\bibinfo{year}{2007}).

\bibitem[{\citenamefont{Trotzky et~al.}(2012)\citenamefont{Trotzky, Chen,
  Flesch, McCulloch, Schollw\"ock, Eisert, and Bloch}}]{trotzky12}
\bibinfo{author}{\bibfnamefont{S.}~\bibnamefont{Trotzky}},
  \bibinfo{author}{\bibfnamefont{Y.-A.} \bibnamefont{Chen}},
  \bibinfo{author}{\bibfnamefont{A.}~\bibnamefont{Flesch}},
  \bibinfo{author}{\bibfnamefont{I.~P.} \bibnamefont{McCulloch}},
  \bibinfo{author}{\bibfnamefont{U.}~\bibnamefont{Schollw\"ock}},
  \bibinfo{author}{\bibfnamefont{J.}~\bibnamefont{Eisert}}, \bibnamefont{and}
  \bibinfo{author}{\bibfnamefont{I.}~\bibnamefont{Bloch}},
  \bibinfo{journal}{Nature Phys.} \textbf{\bibinfo{volume}{8}},
  \bibinfo{pages}{325} (\bibinfo{year}{2012}).

\bibitem[{\citenamefont{Schneider et~al.}(2012)\citenamefont{Schneider,
  Hackerm\"uller, Ronzheimer, Will, Braun, Best, Bloch, Demler, Mandt, Rasch
  et~al.}}]{schneider12}
\bibinfo{author}{\bibfnamefont{U.}~\bibnamefont{Schneider}},
  \bibinfo{author}{\bibfnamefont{L.}~\bibnamefont{Hackerm\"uller}},
  \bibinfo{author}{\bibfnamefont{J.~P.} \bibnamefont{Ronzheimer}},
  \bibinfo{author}{\bibfnamefont{S.}~\bibnamefont{Will}},
  \bibinfo{author}{\bibfnamefont{S.}~\bibnamefont{Braun}},
  \bibinfo{author}{\bibfnamefont{T.}~\bibnamefont{Best}},
  \bibinfo{author}{\bibfnamefont{I.}~\bibnamefont{Bloch}},
  \bibinfo{author}{\bibfnamefont{E.}~\bibnamefont{Demler}},
  \bibinfo{author}{\bibfnamefont{S.}~\bibnamefont{Mandt}},
  \bibinfo{author}{\bibfnamefont{D.}~\bibnamefont{Rasch}},
  \bibnamefont{et~al.}, \bibinfo{journal}{Nature Phys.}
  \textbf{\bibinfo{volume}{8}}, \bibinfo{pages}{213} (\bibinfo{year}{2012}).

\bibitem[{\citenamefont{Cheneau et~al.}(2012)\citenamefont{Cheneau, Barmettler,
  Poletti, Endres, Schau\ss, Fukuhara, Gross, Bloch, Kollath, and
  Kuhr}}]{cheneau12}
\bibinfo{author}{\bibfnamefont{M.}~\bibnamefont{Cheneau}},
  \bibinfo{author}{\bibfnamefont{P.}~\bibnamefont{Barmettler}},
  \bibinfo{author}{\bibfnamefont{D.}~\bibnamefont{Poletti}},
  \bibinfo{author}{\bibfnamefont{M.}~\bibnamefont{Endres}},
  \bibinfo{author}{\bibfnamefont{P.}~\bibnamefont{Schau\ss}},
  \bibinfo{author}{\bibfnamefont{T.}~\bibnamefont{Fukuhara}},
  \bibinfo{author}{\bibfnamefont{C.}~\bibnamefont{Gross}},
  \bibinfo{author}{\bibfnamefont{I.}~\bibnamefont{Bloch}},
  \bibinfo{author}{\bibfnamefont{C.}~\bibnamefont{Kollath}}, \bibnamefont{and}
  \bibinfo{author}{\bibfnamefont{S.}~\bibnamefont{Kuhr}},
  \bibinfo{journal}{Nature (London)} \textbf{\bibinfo{volume}{481}},
  \bibinfo{pages}{484} (\bibinfo{year}{2012}).

\bibitem[{\citenamefont{Ronzheimer et~al.}(2013)\citenamefont{Ronzheimer,
  Schreiber, Braun, Hodgman, Langer, McCulloch, Heidrich-Meisner, Bloch, and
  Schneider}}]{ronzheimer13}
\bibinfo{author}{\bibfnamefont{J.}~\bibnamefont{Ronzheimer}},
  \bibinfo{author}{\bibfnamefont{M.}~\bibnamefont{Schreiber}},
  \bibinfo{author}{\bibfnamefont{S.}~\bibnamefont{Braun}},
  \bibinfo{author}{\bibfnamefont{S.}~\bibnamefont{Hodgman}},
  \bibinfo{author}{\bibfnamefont{S.}~\bibnamefont{Langer}},
  \bibinfo{author}{\bibfnamefont{I.}~\bibnamefont{McCulloch}},
  \bibinfo{author}{\bibfnamefont{F.}~\bibnamefont{Heidrich-Meisner}},
  \bibinfo{author}{\bibfnamefont{I.}~\bibnamefont{Bloch}}, \bibnamefont{and}
  \bibinfo{author}{\bibfnamefont{U.}~\bibnamefont{Schneider}},
  \bibinfo{journal}{Phys. Rev. Lett.} \textbf{\bibinfo{volume}{110}},
  \bibinfo{pages}{205301} (\bibinfo{year}{2013}).

\bibitem[{\citenamefont{Fukuhara
  et~al.}(2013{\natexlab{a}})\citenamefont{Fukuhara, Kantian, Endres, Cheneau,
  Schau\ss, Hild, Gross, Schollw\"ock, Giamarchi, Bloch et~al.}}]{fukuhara13}
\bibinfo{author}{\bibfnamefont{T.}~\bibnamefont{Fukuhara}},
  \bibinfo{author}{\bibfnamefont{A.}~\bibnamefont{Kantian}},
  \bibinfo{author}{\bibfnamefont{M.}~\bibnamefont{Endres}},
  \bibinfo{author}{\bibfnamefont{M.}~\bibnamefont{Cheneau}},
  \bibinfo{author}{\bibfnamefont{P.}~\bibnamefont{Schau\ss}},
  \bibinfo{author}{\bibfnamefont{S.}~\bibnamefont{Hild}},
  \bibinfo{author}{\bibfnamefont{C.}~\bibnamefont{Gross}},
  \bibinfo{author}{\bibfnamefont{U.}~\bibnamefont{Schollw\"ock}},
  \bibinfo{author}{\bibfnamefont{T.}~\bibnamefont{Giamarchi}},
  \bibinfo{author}{\bibfnamefont{I.}~\bibnamefont{Bloch}},
  \bibnamefont{et~al.}, \bibinfo{journal}{Nature Phys.}
  \textbf{\bibinfo{volume}{9}}, \bibinfo{pages}{235}
  (\bibinfo{year}{2013}{\natexlab{a}}).

\bibitem[{\citenamefont{Montagnese et~al.}(2013)\citenamefont{Montagnese,
  Otter, Zotos, Fishman, Hlubek, Mityashkin, Hess, Saint-Martin, Singh,
  Revcolevschi et~al.}}]{montagnese13}
\bibinfo{author}{\bibfnamefont{M.}~\bibnamefont{Montagnese}},
  \bibinfo{author}{\bibfnamefont{M.}~\bibnamefont{Otter}},
  \bibinfo{author}{\bibfnamefont{X.}~\bibnamefont{Zotos}},
  \bibinfo{author}{\bibfnamefont{D.~A.} \bibnamefont{Fishman}},
  \bibinfo{author}{\bibfnamefont{N.}~\bibnamefont{Hlubek}},
  \bibinfo{author}{\bibfnamefont{O.}~\bibnamefont{Mityashkin}},
  \bibinfo{author}{\bibfnamefont{C.}~\bibnamefont{Hess}},
  \bibinfo{author}{\bibfnamefont{R.}~\bibnamefont{Saint-Martin}},
  \bibinfo{author}{\bibfnamefont{S.}~\bibnamefont{Singh}},
  \bibinfo{author}{\bibfnamefont{A.}~\bibnamefont{Revcolevschi}},
  \bibnamefont{et~al.}, \bibinfo{journal}{Phys. Rev. Lett.}
  \textbf{\bibinfo{volume}{110}}, \bibinfo{pages}{147206}
  (\bibinfo{year}{2013}).

\bibitem[{\citenamefont{Wall et~al.}(2010)\citenamefont{Wall, Brida, Clark,
  Ehrke, Jaksch, Ardavan, Bonora, Uemura, Takahashi, Hasegawa et~al.}}]{wall10}
\bibinfo{author}{\bibfnamefont{S.}~\bibnamefont{Wall}},
  \bibinfo{author}{\bibfnamefont{D.}~\bibnamefont{Brida}},
  \bibinfo{author}{\bibfnamefont{S.~R.} \bibnamefont{Clark}},
  \bibinfo{author}{\bibfnamefont{H.~P.} \bibnamefont{Ehrke}},
  \bibinfo{author}{\bibfnamefont{D.}~\bibnamefont{Jaksch}},
  \bibinfo{author}{\bibfnamefont{A.}~\bibnamefont{Ardavan}},
  \bibinfo{author}{\bibfnamefont{S.}~\bibnamefont{Bonora}},
  \bibinfo{author}{\bibfnamefont{H.}~\bibnamefont{Uemura}},
  \bibinfo{author}{\bibfnamefont{Y.}~\bibnamefont{Takahashi}},
  \bibinfo{author}{\bibfnamefont{T.}~\bibnamefont{Hasegawa}},
  \bibnamefont{et~al.}, \bibinfo{journal}{Nature Phys.}
  \textbf{\bibinfo{volume}{7}}, \bibinfo{pages}{114} (\bibinfo{year}{2010}).

\bibitem[{\citenamefont{Rigol et~al.}(2008)\citenamefont{Rigol, Dunjko, and
  Olshanii}}]{rigol08}
\bibinfo{author}{\bibfnamefont{M.}~\bibnamefont{Rigol}},
  \bibinfo{author}{\bibfnamefont{V.}~\bibnamefont{Dunjko}}, \bibnamefont{and}
  \bibinfo{author}{\bibfnamefont{M.}~\bibnamefont{Olshanii}},
  \bibinfo{journal}{Nature} \textbf{\bibinfo{volume}{854}},
  \bibinfo{pages}{858} (\bibinfo{year}{2008}).

\bibitem[{\citenamefont{Polkovnikov et~al.}(2011)\citenamefont{Polkovnikov,
  Sengupta, Silva, and Vengalattore}}]{polkovnikov11}
\bibinfo{author}{\bibfnamefont{A.}~\bibnamefont{Polkovnikov}},
  \bibinfo{author}{\bibfnamefont{K.}~\bibnamefont{Sengupta}},
  \bibinfo{author}{\bibfnamefont{A.}~\bibnamefont{Silva}}, \bibnamefont{and}
  \bibinfo{author}{\bibfnamefont{M.}~\bibnamefont{Vengalattore}},
  \bibinfo{journal}{Rev. Mod. Phys} \textbf{\bibinfo{volume}{83}},
  \bibinfo{pages}{863} (\bibinfo{year}{2011}).

\bibitem[{\citenamefont{Rigol et~al.}(2007)\citenamefont{Rigol, Dunjko,
  Yurovsky, and Olshanii}}]{rigol07}
\bibinfo{author}{\bibfnamefont{M.}~\bibnamefont{Rigol}},
  \bibinfo{author}{\bibfnamefont{V.}~\bibnamefont{Dunjko}},
  \bibinfo{author}{\bibfnamefont{V.}~\bibnamefont{Yurovsky}}, \bibnamefont{and}
  \bibinfo{author}{\bibfnamefont{M.}~\bibnamefont{Olshanii}},
  \bibinfo{journal}{Phys. Rev. Lett.} \textbf{\bibinfo{volume}{98}},
  \bibinfo{pages}{050405} (\bibinfo{year}{2007}).

\bibitem[{\citenamefont{Caux and Mossel}({})}]{caux11}
\bibinfo{author}{\bibfnamefont{J.-S.} \bibnamefont{Caux}} \bibnamefont{and}
  \bibinfo{author}{\bibfnamefont{J.}~\bibnamefont{Mossel}},
  \bibinfo{journal}{J. Stat. Mech.} \textbf{\bibinfo{volume}{{\rm(2011)}}},
  \bibinfo{pages}{P02023} (\bibinfo{year}{{}}).

\bibitem[{\citenamefont{Zotos et~al.}(1997)\citenamefont{Zotos, Naef, and
  Prelov{\v{s}}ek}}]{zotos97}
\bibinfo{author}{\bibfnamefont{X.}~\bibnamefont{Zotos}},
  \bibinfo{author}{\bibfnamefont{F.}~\bibnamefont{Naef}}, \bibnamefont{and}
  \bibinfo{author}{\bibfnamefont{P.}~\bibnamefont{Prelov{\v{s}}ek}},
  \bibinfo{journal}{Phys. Rev. B} \textbf{\bibinfo{volume}{55}},
  \bibinfo{pages}{11029} (\bibinfo{year}{1997}).

\bibitem[{\citenamefont{Heidrich-Meisner
  et~al.}(2007)\citenamefont{Heidrich-Meisner, Honecker, and Brenig}}]{hm07}
\bibinfo{author}{\bibfnamefont{F.}~\bibnamefont{Heidrich-Meisner}},
  \bibinfo{author}{\bibfnamefont{A.}~\bibnamefont{Honecker}}, \bibnamefont{and}
  \bibinfo{author}{\bibfnamefont{W.}~\bibnamefont{Brenig}},
  \bibinfo{journal}{Eur. J. Phys. Special Topics}
  \textbf{\bibinfo{volume}{151}}, \bibinfo{pages}{135} (\bibinfo{year}{2007}).

\bibitem[{\citenamefont{Prosen}(2011)}]{prosen11}
\bibinfo{author}{\bibfnamefont{T.}~\bibnamefont{Prosen}},
  \bibinfo{journal}{Phys. Rev. Lett.} \textbf{\bibinfo{volume}{106}},
  \bibinfo{pages}{217206} (\bibinfo{year}{2011}).

\bibitem[{\citenamefont{Sirker et~al.}(2011)\citenamefont{Sirker, Pereira, and
  Affleck}}]{sirker11}
\bibinfo{author}{\bibfnamefont{J.}~\bibnamefont{Sirker}},
  \bibinfo{author}{\bibfnamefont{R.~G.} \bibnamefont{Pereira}},
  \bibnamefont{and} \bibinfo{author}{\bibfnamefont{I.}~\bibnamefont{Affleck}},
  \bibinfo{journal}{Phys. Rev. B} \textbf{\bibinfo{volume}{83}},
  \bibinfo{pages}{035115} (\bibinfo{year}{2011}).

\bibitem[{\citenamefont{Kl\"umper and Sakai}(2002)}]{kluemper02}
\bibinfo{author}{\bibfnamefont{A.}~\bibnamefont{Kl\"umper}} \bibnamefont{and}
  \bibinfo{author}{\bibfnamefont{K.}~\bibnamefont{Sakai}}, \bibinfo{journal}{J.
  Phys. A} \textbf{\bibinfo{volume}{35}}, \bibinfo{pages}{2173}
  (\bibinfo{year}{2002}).

\bibitem[{\citenamefont{Zotos and Prelov{\v{s}}ek}(1996)}]{zotos96}
\bibinfo{author}{\bibfnamefont{X.}~\bibnamefont{Zotos}} \bibnamefont{and}
  \bibinfo{author}{\bibfnamefont{P.}~\bibnamefont{Prelov{\v{s}}ek}},
  \bibinfo{journal}{Phys. Rev. B} \textbf{\bibinfo{volume}{53}},
  \bibinfo{pages}{983} (\bibinfo{year}{1996}).

\bibitem[{\citenamefont{Zotos}(1999)}]{zotos99}
\bibinfo{author}{\bibfnamefont{X.}~\bibnamefont{Zotos}},
  \bibinfo{journal}{Phys. Rev. Lett.} \textbf{\bibinfo{volume}{82}},
  \bibinfo{pages}{1764} (\bibinfo{year}{1999}).

\bibitem[{\citenamefont{Benz et~al.}(2005)\citenamefont{Benz, Fukui, Kl\"umper,
  and Scheeren}}]{benz05}
\bibinfo{author}{\bibfnamefont{J.}~\bibnamefont{Benz}},
  \bibinfo{author}{\bibfnamefont{T.}~\bibnamefont{Fukui}},
  \bibinfo{author}{\bibfnamefont{A.}~\bibnamefont{Kl\"umper}},
  \bibnamefont{and} \bibinfo{author}{\bibfnamefont{C.}~\bibnamefont{Scheeren}},
  \bibinfo{journal}{J. Phys. Soc. Jpn. Suppl.} \textbf{\bibinfo{volume}{74}},
  \bibinfo{pages}{181} (\bibinfo{year}{2005}).

\bibitem[{\citenamefont{Heidrich-Meisner
  et~al.}(2003)\citenamefont{Heidrich-Meisner, Honecker, Cabra, and
  Brenig}}]{hm03}
\bibinfo{author}{\bibfnamefont{F.}~\bibnamefont{Heidrich-Meisner}},
  \bibinfo{author}{\bibfnamefont{A.}~\bibnamefont{Honecker}},
  \bibinfo{author}{\bibfnamefont{D.~C.} \bibnamefont{Cabra}}, \bibnamefont{and}
  \bibinfo{author}{\bibfnamefont{W.}~\bibnamefont{Brenig}},
  \bibinfo{journal}{Phys. Rev. B} \textbf{\bibinfo{volume}{68}},
  \bibinfo{pages}{134436} (\bibinfo{year}{2003}).

\bibitem[{\citenamefont{Herbrych et~al.}(2011)\citenamefont{Herbrych,
  Prelov{\v{s}}ek, and Zotos}}]{herbrych11}
\bibinfo{author}{\bibfnamefont{J.}~\bibnamefont{Herbrych}},
  \bibinfo{author}{\bibfnamefont{P.}~\bibnamefont{Prelov{\v{s}}ek}},
  \bibnamefont{and} \bibinfo{author}{\bibfnamefont{X.}~\bibnamefont{Zotos}},
  \bibinfo{journal}{Phys. Rev. B} \textbf{\bibinfo{volume}{84}},
  \bibinfo{pages}{155125} (\bibinfo{year}{2011}).

\bibitem[{\citenamefont{Karrasch et~al.}(2012)\citenamefont{Karrasch,
  Bardarson, and Moore}}]{karrasch12}
\bibinfo{author}{\bibfnamefont{C.}~\bibnamefont{Karrasch}},
  \bibinfo{author}{\bibfnamefont{J.}~\bibnamefont{Bardarson}},
  \bibnamefont{and} \bibinfo{author}{\bibfnamefont{J.}~\bibnamefont{Moore}},
  \bibinfo{journal}{Phys. Rev. Lett.} \textbf{\bibinfo{volume}{108}},
  \bibinfo{pages}{227206} (\bibinfo{year}{2012}).

\bibitem[{\citenamefont{Karrasch
  et~al.}(2013{\natexlab{a}})\citenamefont{Karrasch, Hauschild, Langer, and
  Heidrich-Meisner}}]{karrasch13}
\bibinfo{author}{\bibfnamefont{C.}~\bibnamefont{Karrasch}},
  \bibinfo{author}{\bibfnamefont{J.}~\bibnamefont{Hauschild}},
  \bibinfo{author}{\bibfnamefont{S.}~\bibnamefont{Langer}}, \bibnamefont{and}
  \bibinfo{author}{\bibfnamefont{F.}~\bibnamefont{Heidrich-Meisner}},
  \bibinfo{journal}{Phys. Rev. B} \textbf{\bibinfo{volume}{87}},
  \bibinfo{pages}{245128} (\bibinfo{year}{2013}{\natexlab{a}}).

\bibitem[{\citenamefont{Steinigeweg and Gemmer}(2009)}]{steinigeweg09}
\bibinfo{author}{\bibfnamefont{R.}~\bibnamefont{Steinigeweg}} \bibnamefont{and}
  \bibinfo{author}{\bibfnamefont{J.}~\bibnamefont{Gemmer}},
  \bibinfo{journal}{Phys. Rev. B} \textbf{\bibinfo{volume}{80}},
  \bibinfo{pages}{184402} (\bibinfo{year}{2009}).

\bibitem[{\citenamefont{Steinigeweg and Schnalle}(2010)}]{steinigeweg10}
\bibinfo{author}{\bibfnamefont{R.}~\bibnamefont{Steinigeweg}} \bibnamefont{and}
  \bibinfo{author}{\bibfnamefont{R.}~\bibnamefont{Schnalle}},
  \bibinfo{journal}{Phys. Rev. E} \textbf{\bibinfo{volume}{82}},
  \bibinfo{pages}{040103} (\bibinfo{year}{2010}).

\bibitem[{\citenamefont{Steinigeweg and Brenig}(2012)}]{steinigeweg12}
\bibinfo{author}{\bibfnamefont{R.}~\bibnamefont{Steinigeweg}} \bibnamefont{and}
  \bibinfo{author}{\bibfnamefont{W.}~\bibnamefont{Brenig}},
  \bibinfo{journal}{Phys. Rev. Lett.} \textbf{\bibinfo{volume}{107}},
  \bibinfo{pages}{250602} (\bibinfo{year}{2012}).

\bibitem[{\citenamefont{Steinigeweg et~al.}(2012)\citenamefont{Steinigeweg,
  Herbrych, Prelov\ifmmode~\check{s}\else \v{s}\fi{}ek, and
  Mierzejewski}}]{steinigeweg12a}
\bibinfo{author}{\bibfnamefont{R.}~\bibnamefont{Steinigeweg}},
  \bibinfo{author}{\bibfnamefont{J.}~\bibnamefont{Herbrych}},
  \bibinfo{author}{\bibfnamefont{P.}~\bibnamefont{Prelov\ifmmode~\check{s}\else
  \v{s}\fi{}ek}}, \bibnamefont{and}
  \bibinfo{author}{\bibfnamefont{M.}~\bibnamefont{Mierzejewski}},
  \bibinfo{journal}{Phys. Rev. B} \textbf{\bibinfo{volume}{85}},
  \bibinfo{pages}{214409} (\bibinfo{year}{2012}).

\bibitem[{\citenamefont{\v{Z}nidari\v{c}}(2011{\natexlab{a}})}]{znidaric11}
\bibinfo{author}{\bibfnamefont{M.}~\bibnamefont{\v{Z}nidari\v{c}}},
  \bibinfo{journal}{Phys. Rev. Lett.} \textbf{\bibinfo{volume}{106}},
  \bibinfo{pages}{220601} (\bibinfo{year}{2011}{\natexlab{a}}).

\bibitem[{\citenamefont{Steinigeweg}(2012)}]{steinigeweg12b}
\bibinfo{author}{\bibfnamefont{R.}~\bibnamefont{Steinigeweg}},
  \bibinfo{journal}{EPL} \textbf{\bibinfo{volume}{97}}, \bibinfo{pages}{67001}
  (\bibinfo{year}{2012}).

\bibitem[{\citenamefont{Zotos}(2004)}]{zotos04}
\bibinfo{author}{\bibfnamefont{X.}~\bibnamefont{Zotos}},
  \bibinfo{journal}{Phys. Rev. Lett.} \textbf{\bibinfo{volume}{92}},
  \bibinfo{pages}{067202} (\bibinfo{year}{2004}).

\bibitem[{\citenamefont{Jung and Rosch}(2007)}]{jung07}
\bibinfo{author}{\bibfnamefont{P.}~\bibnamefont{Jung}} \bibnamefont{and}
  \bibinfo{author}{\bibfnamefont{A.}~\bibnamefont{Rosch}},
  \bibinfo{journal}{Phys. Rev. B} \textbf{\bibinfo{volume}{76}},
  \bibinfo{pages}{245108} (\bibinfo{year}{2007}).

\bibitem[{\citenamefont{Karrasch
  et~al.}(2013{\natexlab{b}})\citenamefont{Karrasch, Ilan, and
  Moore}}]{karrasch12a}
\bibinfo{author}{\bibfnamefont{C.}~\bibnamefont{Karrasch}},
  \bibinfo{author}{\bibfnamefont{R.}~\bibnamefont{Ilan}}, \bibnamefont{and}
  \bibinfo{author}{\bibfnamefont{J.~E.} \bibnamefont{Moore}},
  \bibinfo{journal}{Phys. Rev. B} \textbf{\bibinfo{volume}{88}},
  \bibinfo{pages}{195129} (\bibinfo{year}{2013}{\natexlab{b}}).

\bibitem[{\citenamefont{\v{Z}nidari\v{c}}(2013)}]{znidaric13}
\bibinfo{author}{\bibfnamefont{M.}~\bibnamefont{\v{Z}nidari\v{c}}},
  \bibinfo{journal}{Phys. Rev. Lett.} \textbf{\bibinfo{volume}{110}},
  \bibinfo{pages}{070602} (\bibinfo{year}{2013}).

\bibitem[{\citenamefont{Prosen and \ifmmode \check{Z}\else
  \v{Z}\fi{}nidari\ifmmode~\check{c}\else \v{c}\fi{}}(2009)}]{prosen09}
\bibinfo{author}{\bibfnamefont{T.}~\bibnamefont{Prosen}} \bibnamefont{and}
  \bibinfo{author}{\bibfnamefont{M.}~\bibnamefont{\ifmmode \check{Z}\else
  \v{Z}\fi{}nidari\ifmmode~\check{c}\else \v{c}\fi{}}}, \bibinfo{journal}{J.
  Stat. Mech} \textbf{\bibinfo{volume}{2009}}, \bibinfo{pages}{P02035}
  (\bibinfo{year}{2009}).

\bibitem[{\citenamefont{Saito}(2003)}]{saito03}
\bibinfo{author}{\bibfnamefont{K.}~\bibnamefont{Saito}}, \bibinfo{journal}{EPL}
  \textbf{\bibinfo{volume}{61}}, \bibinfo{pages}{34} (\bibinfo{year}{2003}).

\bibitem[{\citenamefont{Mendoza-Arenas
  et~al.}(2013{\natexlab{a}})\citenamefont{Mendoza-Arenas, Al-Assam, Clark, and
  Jaksch}}]{mendoza-arenas13}
\bibinfo{author}{\bibfnamefont{J.~J.} \bibnamefont{Mendoza-Arenas}},
  \bibinfo{author}{\bibfnamefont{S.}~\bibnamefont{Al-Assam}},
  \bibinfo{author}{\bibfnamefont{S.~R.} \bibnamefont{Clark}}, \bibnamefont{and}
  \bibinfo{author}{\bibfnamefont{D.}~\bibnamefont{Jaksch}},
  \bibinfo{journal}{J. Stat. Mech.} \textbf{\bibinfo{volume}{(2013)}},
  \bibinfo{pages}{P07007} (\bibinfo{year}{2013}{\natexlab{a}}).

\bibitem[{\citenamefont{Mendoza-Arenas
  et~al.}(2013{\natexlab{b}})\citenamefont{Mendoza-Arenas, Grujic, Jaksch, and
  Clark}}]{mendoza-arenas13a}
\bibinfo{author}{\bibfnamefont{J.~J.} \bibnamefont{Mendoza-Arenas}},
  \bibinfo{author}{\bibfnamefont{T.}~\bibnamefont{Grujic}},
  \bibinfo{author}{\bibfnamefont{D.}~\bibnamefont{Jaksch}}, \bibnamefont{and}
  \bibinfo{author}{\bibfnamefont{S.~R.} \bibnamefont{Clark}},
  \bibinfo{journal}{Phys. Rev. B} \textbf{\bibinfo{volume}{87}},
  \bibinfo{pages}{235130} (\bibinfo{year}{2013}{\natexlab{b}}).

\bibitem[{\citenamefont{Prosen and \ifmmode \check{Z}\else
  \v{Z}\fi{}unkovi\ifmmode~\check{c}\else \v{c}\fi{}}(2013)}]{prosen13b}
\bibinfo{author}{\bibfnamefont{T.}~\bibnamefont{Prosen}} \bibnamefont{and}
  \bibinfo{author}{\bibfnamefont{B.}~\bibnamefont{\ifmmode \check{Z}\else
  \v{Z}\fi{}unkovi\ifmmode~\check{c}\else \v{c}\fi{}}}, \bibinfo{journal}{Phys.
  Rev. Lett.} \textbf{\bibinfo{volume}{111}}, \bibinfo{pages}{040602}
  (\bibinfo{year}{2013}).

\bibitem[{\citenamefont{Langer et~al.}(2009)\citenamefont{Langer,
  Heidrich-Meisner, Gemmer, McCulloch, and Schollw\"ock}}]{langer09}
\bibinfo{author}{\bibfnamefont{S.}~\bibnamefont{Langer}},
  \bibinfo{author}{\bibfnamefont{F.}~\bibnamefont{Heidrich-Meisner}},
  \bibinfo{author}{\bibfnamefont{J.}~\bibnamefont{Gemmer}},
  \bibinfo{author}{\bibfnamefont{I.}~\bibnamefont{McCulloch}},
  \bibnamefont{and}
  \bibinfo{author}{\bibfnamefont{U.}~\bibnamefont{Schollw\"ock}},
  \bibinfo{journal}{Phys. Rev. B} \textbf{\bibinfo{volume}{79}},
  \bibinfo{pages}{214409} (\bibinfo{year}{2009}).

\bibitem[{\citenamefont{Langer et~al.}(2011)\citenamefont{Langer, Heyl,
  McCulloch, and Heidrich-Meisner}}]{langer11}
\bibinfo{author}{\bibfnamefont{S.}~\bibnamefont{Langer}},
  \bibinfo{author}{\bibfnamefont{M.}~\bibnamefont{Heyl}},
  \bibinfo{author}{\bibfnamefont{I.}~\bibnamefont{McCulloch}},
  \bibnamefont{and}
  \bibinfo{author}{\bibfnamefont{F.}~\bibnamefont{Heidrich-Meisner}},
  \bibinfo{journal}{Phys. Rev. B} \textbf{\bibinfo{volume}{84}},
  \bibinfo{pages}{205115} (\bibinfo{year}{2011}).

\bibitem[{\citenamefont{Kim and Huse}(2013)}]{kim13}
\bibinfo{author}{\bibfnamefont{H.}~\bibnamefont{Kim}} \bibnamefont{and}
  \bibinfo{author}{\bibfnamefont{D.~A.} \bibnamefont{Huse}},
  \bibinfo{journal}{Phys. Rev. Lett.} \textbf{\bibinfo{volume}{111}},
  \bibinfo{pages}{127205} (\bibinfo{year}{2013}).

\bibitem[{\citenamefont{Ganahl et~al.}(2012)\citenamefont{Ganahl, Rabel,
  Essler, and Evertz}}]{ganahl12}
\bibinfo{author}{\bibfnamefont{M.}~\bibnamefont{Ganahl}},
  \bibinfo{author}{\bibfnamefont{E.}~\bibnamefont{Rabel}},
  \bibinfo{author}{\bibfnamefont{F.~H.~L.} \bibnamefont{Essler}},
  \bibnamefont{and} \bibinfo{author}{\bibfnamefont{H.~G.}
  \bibnamefont{Evertz}}, \bibinfo{journal}{Phys. Rev. Lett.}
  \textbf{\bibinfo{volume}{108}}, \bibinfo{pages}{077206}
  (\bibinfo{year}{2012}).

\bibitem[{\citenamefont{Liu and Andrei}(unpublished)}]{liu13}
\bibinfo{author}{\bibfnamefont{W.}~\bibnamefont{Liu}} \bibnamefont{and}
  \bibinfo{author}{\bibfnamefont{N.}~\bibnamefont{Andrei}}, p.
  \bibinfo{pages}{arXiv:1311.1118} (\bibinfo{year}{unpublished}).

\bibitem[{\citenamefont{Gobert et~al.}(2005)\citenamefont{Gobert, Kollath,
  Schollw\"ock, and Sch\"utz}}]{gobert05}
\bibinfo{author}{\bibfnamefont{D.}~\bibnamefont{Gobert}},
  \bibinfo{author}{\bibfnamefont{C.}~\bibnamefont{Kollath}},
  \bibinfo{author}{\bibfnamefont{U.}~\bibnamefont{Schollw\"ock}},
  \bibnamefont{and} \bibinfo{author}{\bibfnamefont{G.}~\bibnamefont{Sch\"utz}},
  \bibinfo{journal}{Phys. Rev. E} \textbf{\bibinfo{volume}{71}},
  \bibinfo{pages}{036102} (\bibinfo{year}{2005}).

\bibitem[{\citenamefont{Lancaster and Mitra}(2010)}]{lancaster10a}
\bibinfo{author}{\bibfnamefont{J.}~\bibnamefont{Lancaster}} \bibnamefont{and}
  \bibinfo{author}{\bibfnamefont{A.}~\bibnamefont{Mitra}},
  \bibinfo{journal}{Phys. Rev. E} \textbf{\bibinfo{volume}{81}},
  \bibinfo{pages}{061134} (\bibinfo{year}{2010}).

\bibitem[{\citenamefont{Karrasch
  et~al.}(2013{\natexlab{c}})\citenamefont{Karrasch, Bardarson, and
  Moore}}]{karrasch13a}
\bibinfo{author}{\bibfnamefont{C.}~\bibnamefont{Karrasch}},
  \bibinfo{author}{\bibfnamefont{J.~H.} \bibnamefont{Bardarson}},
  \bibnamefont{and} \bibinfo{author}{\bibfnamefont{J.~E.} \bibnamefont{Moore}},
  \bibinfo{journal}{New J. Phys.} \textbf{\bibinfo{volume}{15}},
  \bibinfo{pages}{083031} (\bibinfo{year}{2013}{\natexlab{c}}).

\bibitem[{\citenamefont{De~Luca et~al.}(2013)\citenamefont{De~Luca, Viti,
  Bernard, and Doyon}}]{luca13}
\bibinfo{author}{\bibfnamefont{A.}~\bibnamefont{De~Luca}},
  \bibinfo{author}{\bibfnamefont{J.}~\bibnamefont{Viti}},
  \bibinfo{author}{\bibfnamefont{D.}~\bibnamefont{Bernard}}, \bibnamefont{and}
  \bibinfo{author}{\bibfnamefont{B.}~\bibnamefont{Doyon}},
  \bibinfo{journal}{Phys. Rev. B} \textbf{\bibinfo{volume}{88}},
  \bibinfo{pages}{134301} (\bibinfo{year}{2013}).

\bibitem[{\citenamefont{Sabetta and Misguich}(2013)}]{sabetta13}
\bibinfo{author}{\bibfnamefont{T.}~\bibnamefont{Sabetta}} \bibnamefont{and}
  \bibinfo{author}{\bibfnamefont{G.}~\bibnamefont{Misguich}},
  \bibinfo{journal}{Phys. Rev. B} \textbf{\bibinfo{volume}{88}},
  \bibinfo{pages}{245114} (\bibinfo{year}{2013}).

\bibitem[{\citenamefont{Vidal}(2004)}]{vidal04}
\bibinfo{author}{\bibfnamefont{G.}~\bibnamefont{Vidal}},
  \bibinfo{journal}{Phys. Rev. Lett.} \textbf{\bibinfo{volume}{93}},
  \bibinfo{pages}{040502} (\bibinfo{year}{2004}).

\bibitem[{\citenamefont{Daley et~al.}(2004)\citenamefont{Daley, Kollath,
  Schollw\"ock, and Vidal}}]{daley04}
\bibinfo{author}{\bibfnamefont{A.}~\bibnamefont{Daley}},
  \bibinfo{author}{\bibfnamefont{C.}~\bibnamefont{Kollath}},
  \bibinfo{author}{\bibfnamefont{U.}~\bibnamefont{Schollw\"ock}},
  \bibnamefont{and} \bibinfo{author}{\bibfnamefont{G.}~\bibnamefont{Vidal}},
  \bibinfo{journal}{J. Stat. Mech.: Theory Exp.}
  \textbf{\bibinfo{volume}{2004}}, \bibinfo{pages}{P04005}
  (\bibinfo{year}{2004}).

\bibitem[{\citenamefont{White and Feiguin}(2004)}]{white04}
\bibinfo{author}{\bibfnamefont{S.~R.} \bibnamefont{White}} \bibnamefont{and}
  \bibinfo{author}{\bibfnamefont{A.~E.} \bibnamefont{Feiguin}},
  \bibinfo{journal}{Phys. Rev. Lett.} \textbf{\bibinfo{volume}{93}},
  \bibinfo{pages}{076401} (\bibinfo{year}{2004}).

\bibitem[{\citenamefont{Jesenko and \v{Z}nidari\v{c}}(2011)}]{jesenko11}
\bibinfo{author}{\bibfnamefont{S.}~\bibnamefont{Jesenko}} \bibnamefont{and}
  \bibinfo{author}{\bibfnamefont{M.}~\bibnamefont{\v{Z}nidari\v{c}}},
  \bibinfo{journal}{Phys. Rev. B} \textbf{\bibinfo{volume}{84}},
  \bibinfo{pages}{174438} (\bibinfo{year}{2011}).

\bibitem[{\citenamefont{Sologubenko et~al.}(2007)\citenamefont{Sologubenko,
  Lorenz, Ott, and Freimuth}}]{sologubenko07}
\bibinfo{author}{\bibfnamefont{A.~V.} \bibnamefont{Sologubenko}},
  \bibinfo{author}{\bibfnamefont{T.}~\bibnamefont{Lorenz}},
  \bibinfo{author}{\bibfnamefont{H.~R.} \bibnamefont{Ott}}, \bibnamefont{and}
  \bibinfo{author}{\bibfnamefont{A.}~\bibnamefont{Freimuth}},
  \bibinfo{journal}{J. Low Temp. Phys.} \textbf{\bibinfo{volume}{147}},
  \bibinfo{pages}{387} (\bibinfo{year}{2007}).

\bibitem[{\citenamefont{Hess}(2007)}]{hess07}
\bibinfo{author}{\bibfnamefont{C.}~\bibnamefont{Hess}}, \bibinfo{journal}{Eur.
  Phys. J. Spec. Topics} \textbf{\bibinfo{volume}{151}}, \bibinfo{pages}{73}
  (\bibinfo{year}{2007}).

\bibitem[{\citenamefont{Sologubenko
  et~al.}(2000{\natexlab{a}})\citenamefont{Sologubenko, Felder, Gianno, Ott,
  Vietkine, and Revcolevschi}}]{sologubenko00a}
\bibinfo{author}{\bibfnamefont{A.~V.} \bibnamefont{Sologubenko}},
  \bibinfo{author}{\bibfnamefont{E.}~\bibnamefont{Felder}},
  \bibinfo{author}{\bibfnamefont{K.}~\bibnamefont{Gianno}},
  \bibinfo{author}{\bibfnamefont{H.~R.} \bibnamefont{Ott}},
  \bibinfo{author}{\bibfnamefont{A.}~\bibnamefont{Vietkine}}, \bibnamefont{and}
  \bibinfo{author}{\bibfnamefont{A.}~\bibnamefont{Revcolevschi}},
  \bibinfo{journal}{Phys. Rev. B} \textbf{\bibinfo{volume}{62}},
  \bibinfo{pages}{R6108} (\bibinfo{year}{2000}{\natexlab{a}}).

\bibitem[{\citenamefont{Sologubenko et~al.}(2001)\citenamefont{Sologubenko,
  Gianno, Ott, Vietkine, and Revcolevschi}}]{sologubenko01}
\bibinfo{author}{\bibfnamefont{A.~V.} \bibnamefont{Sologubenko}},
  \bibinfo{author}{\bibfnamefont{K.}~\bibnamefont{Gianno}},
  \bibinfo{author}{\bibfnamefont{H.~R.} \bibnamefont{Ott}},
  \bibinfo{author}{\bibfnamefont{A.}~\bibnamefont{Vietkine}}, \bibnamefont{and}
  \bibinfo{author}{\bibfnamefont{A.}~\bibnamefont{Revcolevschi}},
  \bibinfo{journal}{Phys. Rev. B} \textbf{\bibinfo{volume}{64}},
  \bibinfo{pages}{054412} (\bibinfo{year}{2001}).

\bibitem[{\citenamefont{Hlubek et~al.}(2010)\citenamefont{Hlubek, Ribeiro,
  Saint-Martin, Revcolevschi, Roth, Behr, B\"uchner, and Hess}}]{hlubek10}
\bibinfo{author}{\bibfnamefont{N.}~\bibnamefont{Hlubek}},
  \bibinfo{author}{\bibfnamefont{P.}~\bibnamefont{Ribeiro}},
  \bibinfo{author}{\bibfnamefont{R.}~\bibnamefont{Saint-Martin}},
  \bibinfo{author}{\bibfnamefont{A.}~\bibnamefont{Revcolevschi}},
  \bibinfo{author}{\bibfnamefont{G.}~\bibnamefont{Roth}},
  \bibinfo{author}{\bibfnamefont{G.}~\bibnamefont{Behr}},
  \bibinfo{author}{\bibfnamefont{B.}~\bibnamefont{B\"uchner}},
  \bibnamefont{and} \bibinfo{author}{\bibfnamefont{C.}~\bibnamefont{Hess}},
  \bibinfo{journal}{Phys. Rev. B} \textbf{\bibinfo{volume}{81}},
  \bibinfo{pages}{020405} (\bibinfo{year}{2010}).

\bibitem[{\citenamefont{Sologubenko
  et~al.}(2000{\natexlab{b}})\citenamefont{Sologubenko, Gianno, Ott, Ammerahl,
  and Revcolevschi}}]{sologubenko00}
\bibinfo{author}{\bibfnamefont{A.~V.} \bibnamefont{Sologubenko}},
  \bibinfo{author}{\bibfnamefont{K.}~\bibnamefont{Gianno}},
  \bibinfo{author}{\bibfnamefont{H.~R.} \bibnamefont{Ott}},
  \bibinfo{author}{\bibfnamefont{U.}~\bibnamefont{Ammerahl}}, \bibnamefont{and}
  \bibinfo{author}{\bibfnamefont{A.}~\bibnamefont{Revcolevschi}},
  \bibinfo{journal}{Phys. Rev. Lett.} \textbf{\bibinfo{volume}{84}},
  \bibinfo{pages}{2714} (\bibinfo{year}{2000}{\natexlab{b}}).

\bibitem[{\citenamefont{Hess et~al.}(2001)\citenamefont{Hess, Baumann,
  Ammerahl, B\"uchner, Heidrich-Meisner, Brenig, and Revcolevschi}}]{hess01}
\bibinfo{author}{\bibfnamefont{C.}~\bibnamefont{Hess}},
  \bibinfo{author}{\bibfnamefont{C.}~\bibnamefont{Baumann}},
  \bibinfo{author}{\bibfnamefont{U.}~\bibnamefont{Ammerahl}},
  \bibinfo{author}{\bibfnamefont{B.}~\bibnamefont{B\"uchner}},
  \bibinfo{author}{\bibfnamefont{F.}~\bibnamefont{Heidrich-Meisner}},
  \bibinfo{author}{\bibfnamefont{W.}~\bibnamefont{Brenig}}, \bibnamefont{and}
  \bibinfo{author}{\bibfnamefont{A.}~\bibnamefont{Revcolevschi}},
  \bibinfo{journal}{Phys. Rev. B} \textbf{\bibinfo{volume}{64}},
  \bibinfo{pages}{184305} (\bibinfo{year}{2001}).

\bibitem[{\citenamefont{Otter et~al.}(2009)\citenamefont{Otter, Krasnikov,
  Fishman, Pshenichnikov, Saint-Martin, Revcolevschi, and van
  Loodsrecht}}]{otter09}
\bibinfo{author}{\bibfnamefont{M.}~\bibnamefont{Otter}},
  \bibinfo{author}{\bibfnamefont{V.}~\bibnamefont{Krasnikov}},
  \bibinfo{author}{\bibfnamefont{D.}~\bibnamefont{Fishman}},
  \bibinfo{author}{\bibfnamefont{M.}~\bibnamefont{Pshenichnikov}},
  \bibinfo{author}{\bibfnamefont{R.}~\bibnamefont{Saint-Martin}},
  \bibinfo{author}{\bibfnamefont{A.}~\bibnamefont{Revcolevschi}},
  \bibnamefont{and} \bibinfo{author}{\bibfnamefont{P.}~\bibnamefont{van
  Loodsrecht}}, \bibinfo{journal}{J. Mag. Mag. Mat.}
  \textbf{\bibinfo{volume}{321}}, \bibinfo{pages}{796} (\bibinfo{year}{2009}).

\bibitem[{\citenamefont{Otter et~al.}(2012)\citenamefont{Otter, Athanasopoulos,
  Hlubek, Montagnese, Labois, Fishman, de~Haan, Singh, Lakehal, Giapintzakis
  et~al.}}]{otter12}
\bibinfo{author}{\bibfnamefont{M.}~\bibnamefont{Otter}},
  \bibinfo{author}{\bibfnamefont{G.}~\bibnamefont{Athanasopoulos}},
  \bibinfo{author}{\bibfnamefont{N.}~\bibnamefont{Hlubek}},
  \bibinfo{author}{\bibfnamefont{M.}~\bibnamefont{Montagnese}},
  \bibinfo{author}{\bibfnamefont{M.}~\bibnamefont{Labois}},
  \bibinfo{author}{\bibfnamefont{D.}~\bibnamefont{Fishman}},
  \bibinfo{author}{\bibfnamefont{F.}~\bibnamefont{de~Haan}},
  \bibinfo{author}{\bibfnamefont{S.}~\bibnamefont{Singh}},
  \bibinfo{author}{\bibfnamefont{D.}~\bibnamefont{Lakehal}},
  \bibinfo{author}{\bibfnamefont{J.}~\bibnamefont{Giapintzakis}},
  \bibnamefont{et~al.}, \bibinfo{journal}{Int. J. of Heat and Mass Transfer}
  \textbf{\bibinfo{volume}{55}}, \bibinfo{pages}{2531} (\bibinfo{year}{2012}).

\bibitem[{\citenamefont{Rozhkov and Chernyshev}(2005)}]{rozhkov05}
\bibinfo{author}{\bibfnamefont{A.~V.} \bibnamefont{Rozhkov}} \bibnamefont{and}
  \bibinfo{author}{\bibfnamefont{A.~L.} \bibnamefont{Chernyshev}},
  \bibinfo{journal}{Phys. Rev. Lett.} \textbf{\bibinfo{volume}{94}},
  \bibinfo{pages}{087201} (\bibinfo{year}{2005}).

\bibitem[{\citenamefont{Chernyshev and Rozhkov}(2005)}]{chernychev05}
\bibinfo{author}{\bibfnamefont{A.~L.} \bibnamefont{Chernyshev}}
  \bibnamefont{and} \bibinfo{author}{\bibfnamefont{A.~V.}
  \bibnamefont{Rozhkov}}, \bibinfo{journal}{Phys. Rev. B}
  \textbf{\bibinfo{volume}{72}}, \bibinfo{pages}{104423}
  (\bibinfo{year}{2005}).

\bibitem[{\citenamefont{Shimshoni et~al.}(2003)\citenamefont{Shimshoni, Andrei,
  and Rosch}}]{shimshoni03}
\bibinfo{author}{\bibfnamefont{E.}~\bibnamefont{Shimshoni}},
  \bibinfo{author}{\bibfnamefont{N.}~\bibnamefont{Andrei}}, \bibnamefont{and}
  \bibinfo{author}{\bibfnamefont{A.}~\bibnamefont{Rosch}},
  \bibinfo{journal}{Phys. Rev. B} \textbf{\bibinfo{volume}{68}},
  \bibinfo{pages}{104401} (\bibinfo{year}{2003}).

\bibitem[{\citenamefont{Boulat et~al.}(2007)\citenamefont{Boulat, Mehta,
  Andrei, Shimshoni, and Rosch}}]{boulat07}
\bibinfo{author}{\bibfnamefont{E.}~\bibnamefont{Boulat}},
  \bibinfo{author}{\bibfnamefont{P.}~\bibnamefont{Mehta}},
  \bibinfo{author}{\bibfnamefont{N.}~\bibnamefont{Andrei}},
  \bibinfo{author}{\bibfnamefont{E.}~\bibnamefont{Shimshoni}},
  \bibnamefont{and} \bibinfo{author}{\bibfnamefont{A.}~\bibnamefont{Rosch}},
  \bibinfo{journal}{Phys. Rev. B} \textbf{\bibinfo{volume}{76}},
  \bibinfo{pages}{214411} (\bibinfo{year}{2007}).

\bibitem[{\citenamefont{Metavitsiadis et~al.}(2010)\citenamefont{Metavitsiadis,
  Zotos, Bari\ifmmode \check{s}\else \v{s}\fi{}i\ifmmode~\acute{c}\else
  \'{c}\fi{}, and Prelov\ifmmode~\check{s}\else
  \v{s}\fi{}ek}}]{metavitsiadis10}
\bibinfo{author}{\bibfnamefont{A.}~\bibnamefont{Metavitsiadis}},
  \bibinfo{author}{\bibfnamefont{X.}~\bibnamefont{Zotos}},
  \bibinfo{author}{\bibfnamefont{O.~S.} \bibnamefont{Bari\ifmmode
  \check{s}\else \v{s}\fi{}i\ifmmode~\acute{c}\else \'{c}\fi{}}},
  \bibnamefont{and}
  \bibinfo{author}{\bibfnamefont{P.}~\bibnamefont{Prelov\ifmmode~\check{s}\else
  \v{s}\fi{}ek}}, \bibinfo{journal}{Phys. Rev. B}
  \textbf{\bibinfo{volume}{81}}, \bibinfo{pages}{205101}
  (\bibinfo{year}{2010}).

\bibitem[{\citenamefont{Bari\ifmmode \check{s}\else
  \v{s}\fi{}i\ifmmode~\acute{c}\else \'{c}\fi{}
  et~al.}(2009)\citenamefont{Bari\ifmmode \check{s}\else
  \v{s}\fi{}i\ifmmode~\acute{c}\else \'{c}\fi{}, Prelov\ifmmode~\check{s}\else
  \v{s}\fi{}ek, Metavitsiadis, and Zotos}}]{barisic09}
\bibinfo{author}{\bibfnamefont{O.~S.} \bibnamefont{Bari\ifmmode \check{s}\else
  \v{s}\fi{}i\ifmmode~\acute{c}\else \'{c}\fi{}}},
  \bibinfo{author}{\bibfnamefont{P.}~\bibnamefont{Prelov\ifmmode~\check{s}\else
  \v{s}\fi{}ek}},
  \bibinfo{author}{\bibfnamefont{A.}~\bibnamefont{Metavitsiadis}},
  \bibnamefont{and} \bibinfo{author}{\bibfnamefont{X.}~\bibnamefont{Zotos}},
  \bibinfo{journal}{Phys. Rev. B} \textbf{\bibinfo{volume}{80}},
  \bibinfo{pages}{125118} (\bibinfo{year}{2009}).

\bibitem[{\citenamefont{Karahalios et~al.}(2009)\citenamefont{Karahalios,
  Metavitsiadis, Zotos, Gorczyca, and Prelov\ifmmode~\check{s}\else
  \v{s}\fi{}ek}}]{karahalios09}
\bibinfo{author}{\bibfnamefont{A.}~\bibnamefont{Karahalios}},
  \bibinfo{author}{\bibfnamefont{A.}~\bibnamefont{Metavitsiadis}},
  \bibinfo{author}{\bibfnamefont{X.}~\bibnamefont{Zotos}},
  \bibinfo{author}{\bibfnamefont{A.}~\bibnamefont{Gorczyca}}, \bibnamefont{and}
  \bibinfo{author}{\bibfnamefont{P.}~\bibnamefont{Prelov\ifmmode~\check{s}\else
  \v{s}\fi{}ek}}, \bibinfo{journal}{Phys. Rev. B}
  \textbf{\bibinfo{volume}{79}}, \bibinfo{pages}{024425}
  (\bibinfo{year}{2009}).

\bibitem[{\citenamefont{Bartsch and Brenig}(2013)}]{bartsch13}
\bibinfo{author}{\bibfnamefont{C.}~\bibnamefont{Bartsch}} \bibnamefont{and}
  \bibinfo{author}{\bibfnamefont{W.}~\bibnamefont{Brenig}},
  \bibinfo{journal}{Phys. Rev. B} \textbf{\bibinfo{volume}{88}},
  \bibinfo{pages}{214412} (\bibinfo{year}{2013}).

\bibitem[{\citenamefont{J\"ordens et~al.}(2008)\citenamefont{J\"ordens,
  Strohmaier, G\"unter, Moritz, and Esslinger}}]{joerdens08}
\bibinfo{author}{\bibfnamefont{R.}~\bibnamefont{J\"ordens}},
  \bibinfo{author}{\bibfnamefont{N.}~\bibnamefont{Strohmaier}},
  \bibinfo{author}{\bibfnamefont{K.}~\bibnamefont{G\"unter}},
  \bibinfo{author}{\bibfnamefont{H.}~\bibnamefont{Moritz}}, \bibnamefont{and}
  \bibinfo{author}{\bibfnamefont{T.}~\bibnamefont{Esslinger}},
  \bibinfo{journal}{Nature (London)} \textbf{\bibinfo{volume}{455}},
  \bibinfo{pages}{204} (\bibinfo{year}{2008}).

\bibitem[{\citenamefont{Schneider et~al.}(2008)\citenamefont{Schneider,
  Hackerm\"uller, Will, Best, Bloch, Costi, Helmes, Rasch, and
  Rosch}}]{schneider08}
\bibinfo{author}{\bibfnamefont{U.}~\bibnamefont{Schneider}},
  \bibinfo{author}{\bibfnamefont{L.}~\bibnamefont{Hackerm\"uller}},
  \bibinfo{author}{\bibfnamefont{S.}~\bibnamefont{Will}},
  \bibinfo{author}{\bibfnamefont{T.}~\bibnamefont{Best}},
  \bibinfo{author}{\bibfnamefont{I.}~\bibnamefont{Bloch}},
  \bibinfo{author}{\bibfnamefont{T.~A.} \bibnamefont{Costi}},
  \bibinfo{author}{\bibfnamefont{R.~W.} \bibnamefont{Helmes}},
  \bibinfo{author}{\bibfnamefont{D.}~\bibnamefont{Rasch}}, \bibnamefont{and}
  \bibinfo{author}{\bibfnamefont{A.}~\bibnamefont{Rosch}},
  \bibinfo{journal}{Science} \textbf{\bibinfo{volume}{322}},
  \bibinfo{pages}{1520} (\bibinfo{year}{2008}).

\bibitem[{\citenamefont{Bloch et~al.}(2008)\citenamefont{Bloch, Dalibard, and
  Zwerger}}]{bloch08}
\bibinfo{author}{\bibfnamefont{I.}~\bibnamefont{Bloch}},
  \bibinfo{author}{\bibfnamefont{J.}~\bibnamefont{Dalibard}}, \bibnamefont{and}
  \bibinfo{author}{\bibfnamefont{W.}~\bibnamefont{Zwerger}},
  \bibinfo{journal}{Rev. Mod. Phys.} \textbf{\bibinfo{volume}{80}},
  \bibinfo{pages}{885} (\bibinfo{year}{2008}).

\bibitem[{\citenamefont{Cazalilla et~al.}(2011)\citenamefont{Cazalilla, Citro,
  Giamarchi, Orignac, and Rigol}}]{cazalilla11}
\bibinfo{author}{\bibfnamefont{M.~A.} \bibnamefont{Cazalilla}},
  \bibinfo{author}{\bibfnamefont{R.}~\bibnamefont{Citro}},
  \bibinfo{author}{\bibfnamefont{T.}~\bibnamefont{Giamarchi}},
  \bibinfo{author}{\bibfnamefont{E.}~\bibnamefont{Orignac}}, \bibnamefont{and}
  \bibinfo{author}{\bibfnamefont{M.}~\bibnamefont{Rigol}},
  \bibinfo{journal}{Rev. Mod. Phys.} \textbf{\bibinfo{volume}{83}},
  \bibinfo{pages}{1405} (\bibinfo{year}{2011}).

\bibitem[{\citenamefont{Reinhard et~al.}(2013)\citenamefont{Reinhard, Riou,
  Zundel, Weiss, Li, Rey, and Hipolito}}]{reinhard13}
\bibinfo{author}{\bibfnamefont{A.}~\bibnamefont{Reinhard}},
  \bibinfo{author}{\bibfnamefont{J.-F.} \bibnamefont{Riou}},
  \bibinfo{author}{\bibfnamefont{L.~A.} \bibnamefont{Zundel}},
  \bibinfo{author}{\bibfnamefont{D.~S.} \bibnamefont{Weiss}},
  \bibinfo{author}{\bibfnamefont{S.}~\bibnamefont{Li}},
  \bibinfo{author}{\bibfnamefont{A.~M.} \bibnamefont{Rey}}, \bibnamefont{and}
  \bibinfo{author}{\bibfnamefont{R.}~\bibnamefont{Hipolito}},
  \bibinfo{journal}{Phys. Rev. Lett.} \textbf{\bibinfo{volume}{110}},
  \bibinfo{pages}{033001} (\bibinfo{year}{2013}).

\bibitem[{\citenamefont{Vidmar et~al.}(2013)\citenamefont{Vidmar, Langer,
  McCulloch, Schneider, Schollw\"ock, and Heidrich-Meisner}}]{vidmar13}
\bibinfo{author}{\bibfnamefont{L.}~\bibnamefont{Vidmar}},
  \bibinfo{author}{\bibfnamefont{S.}~\bibnamefont{Langer}},
  \bibinfo{author}{\bibfnamefont{I.~P.} \bibnamefont{McCulloch}},
  \bibinfo{author}{\bibfnamefont{U.}~\bibnamefont{Schneider}},
  \bibinfo{author}{\bibfnamefont{U.}~\bibnamefont{Schollw\"ock}},
  \bibnamefont{and}
  \bibinfo{author}{\bibfnamefont{F.}~\bibnamefont{Heidrich-Meisner}},
  \bibinfo{journal}{Phys. Rev. B} \textbf{\bibinfo{volume}{88}},
  \bibinfo{pages}{235117} (\bibinfo{year}{2013}).

\bibitem[{\citenamefont{Jreissaty et~al.}(2013)\citenamefont{Jreissaty,
  Carrasquilla, and Rigol}}]{jreissaty13}
\bibinfo{author}{\bibfnamefont{A.}~\bibnamefont{Jreissaty}},
  \bibinfo{author}{\bibfnamefont{J.}~\bibnamefont{Carrasquilla}},
  \bibnamefont{and} \bibinfo{author}{\bibfnamefont{M.}~\bibnamefont{Rigol}},
  \bibinfo{journal}{Phys. Rev. A} \textbf{\bibinfo{volume}{88}},
  \bibinfo{pages}{031606} (\bibinfo{year}{2013}).

\bibitem[{\citenamefont{Sch\"onmeier-Kromer and Pollet}(2014)}]{schonmeier13}
\bibinfo{author}{\bibfnamefont{J.}~\bibnamefont{Sch\"onmeier-Kromer}}
  \bibnamefont{and} \bibinfo{author}{\bibfnamefont{L.}~\bibnamefont{Pollet}},
  \bibinfo{journal}{Phys. Rev. A} \textbf{\bibinfo{volume}{89}},
  \bibinfo{pages}{023605} (\bibinfo{year}{2014}).

\bibitem[{\citenamefont{Barthel}(2013)}]{barthel13b}
\bibinfo{author}{\bibfnamefont{T.}~\bibnamefont{Barthel}},
  \bibinfo{journal}{New J. Phys.} \textbf{\bibinfo{volume}{15}},
  \bibinfo{pages}{073010} (\bibinfo{year}{2013}).

\bibitem[{\citenamefont{Sachdev and Damle}(1997)}]{sachdev97}
\bibinfo{author}{\bibfnamefont{S.}~\bibnamefont{Sachdev}} \bibnamefont{and}
  \bibinfo{author}{\bibfnamefont{K.}~\bibnamefont{Damle}},
  \bibinfo{journal}{Phys. Rev. Lett.} \textbf{\bibinfo{volume}{78}},
  \bibinfo{pages}{943} (\bibinfo{year}{1997}).

\bibitem[{\citenamefont{Damle and Sachdev}(2005)}]{damle05}
\bibinfo{author}{\bibfnamefont{K.}~\bibnamefont{Damle}} \bibnamefont{and}
  \bibinfo{author}{\bibfnamefont{S.}~\bibnamefont{Sachdev}},
  \bibinfo{journal}{Phys. Rev. Lett.} \textbf{\bibinfo{volume}{95}},
  \bibinfo{pages}{187201} (\bibinfo{year}{2005}).

\bibitem[{\citenamefont{Damle and Sachdev}(1998)}]{damle98}
\bibinfo{author}{\bibfnamefont{K.}~\bibnamefont{Damle}} \bibnamefont{and}
  \bibinfo{author}{\bibfnamefont{S.}~\bibnamefont{Sachdev}},
  \bibinfo{journal}{Phys. Rev. B} \textbf{\bibinfo{volume}{57}},
  \bibinfo{pages}{8307} (\bibinfo{year}{1998}).

\bibitem[{\citenamefont{Kolezhuk and Mikeska}(2004)}]{kolezhuk-review}
\bibinfo{author}{\bibfnamefont{A.}~\bibnamefont{Kolezhuk}} \bibnamefont{and}
  \bibinfo{author}{\bibfnamefont{H.}~\bibnamefont{Mikeska}},
  \bibinfo{journal}{Lect. Not. Phys.} \textbf{\bibinfo{volume}{645}},
  \bibinfo{pages}{1} (\bibinfo{year}{2004}).

\bibitem[{\citenamefont{Schmitteckert}(2004)}]{schmitteckert04}
\bibinfo{author}{\bibfnamefont{P.}~\bibnamefont{Schmitteckert}},
  \bibinfo{journal}{Phys. Rev. B.} \textbf{\bibinfo{volume}{70}},
  \bibinfo{pages}{121302(R)} (\bibinfo{year}{2004}).

\bibitem[{\citenamefont{Verstraete et~al.}(2004)\citenamefont{Verstraete,
  Garc\'ia-Ripoll, and Cirac}}]{verstraete04}
\bibinfo{author}{\bibfnamefont{F.}~\bibnamefont{Verstraete}},
  \bibinfo{author}{\bibfnamefont{J.~J.} \bibnamefont{Garc\'ia-Ripoll}},
  \bibnamefont{and} \bibinfo{author}{\bibfnamefont{J.~I.} \bibnamefont{Cirac}},
  \bibinfo{journal}{Phys. Rev. Lett.} \textbf{\bibinfo{volume}{93}},
  \bibinfo{pages}{207204} (\bibinfo{year}{2004}).

\bibitem[{\citenamefont{Feiguin and White}(2005)}]{feiguin05}
\bibinfo{author}{\bibfnamefont{A.~E.} \bibnamefont{Feiguin}} \bibnamefont{and}
  \bibinfo{author}{\bibfnamefont{S.~R.} \bibnamefont{White}},
  \bibinfo{journal}{Phys. Rev. B} \textbf{\bibinfo{volume}{72}},
  \bibinfo{pages}{220401} (\bibinfo{year}{2005}).

\bibitem[{\citenamefont{White}(1992)}]{white92}
\bibinfo{author}{\bibfnamefont{S.~R.} \bibnamefont{White}},
  \bibinfo{journal}{Phys. Rev. Lett.} \textbf{\bibinfo{volume}{69}},
  \bibinfo{pages}{2863} (\bibinfo{year}{1992}).

\bibitem[{\citenamefont{White}(1993)}]{white93}
\bibinfo{author}{\bibfnamefont{S.~R.} \bibnamefont{White}},
  \bibinfo{journal}{Phys. Rev. B} \textbf{\bibinfo{volume}{48}},
  \bibinfo{pages}{10345} (\bibinfo{year}{1993}).

\bibitem[{\citenamefont{Schollw\"ock}(2005)}]{schollwoeck05}
\bibinfo{author}{\bibfnamefont{U.}~\bibnamefont{Schollw\"ock}},
  \bibinfo{journal}{Rev. Mod. Phys.} \textbf{\bibinfo{volume}{77}},
  \bibinfo{pages}{259} (\bibinfo{year}{2005}).

\bibitem[{\citenamefont{Schollw\"ock}(2011)}]{schollwoeck11}
\bibinfo{author}{\bibfnamefont{U.}~\bibnamefont{Schollw\"ock}},
  \bibinfo{journal}{Ann. Phys. (NY)} \textbf{\bibinfo{volume}{326}},
  \bibinfo{pages}{96} (\bibinfo{year}{2011}).

\bibitem[{\citenamefont{Eisler and R\'acz}(2013)}]{eisler13}
\bibinfo{author}{\bibfnamefont{V.}~\bibnamefont{Eisler}} \bibnamefont{and}
  \bibinfo{author}{\bibfnamefont{Z.}~\bibnamefont{R\'acz}},
  \bibinfo{journal}{Phys. Rev. Lett.} \textbf{\bibinfo{volume}{110}},
  \bibinfo{pages}{060602} (\bibinfo{year}{2013}).

\bibitem[{\citenamefont{Huber}(2012)}]{huber12}
\bibinfo{author}{\bibfnamefont{D.}~\bibnamefont{Huber}},
  \bibinfo{journal}{Physica B} \textbf{\bibinfo{volume}{407}},
  \bibinfo{pages}{4274} (\bibinfo{year}{2012}).

\bibitem[{\citenamefont{\v{Z}nidari\v{c}}(2011{\natexlab{b}})}]{znidaric11a}
\bibinfo{author}{\bibfnamefont{M.}~\bibnamefont{\v{Z}nidari\v{c}}},
  \bibinfo{journal}{J. Stat. Mech.} p. \bibinfo{pages}{P12008}
  (\bibinfo{year}{2011}{\natexlab{b}}).

\bibitem[{\citenamefont{Prosen}(unpublished)}]{prosen13a}
\bibinfo{author}{\bibfnamefont{T.}~\bibnamefont{Prosen}}, p.
  \bibinfo{pages}{arXiv:1310.8629} (\bibinfo{year}{unpublished}).

\bibitem[{\citenamefont{Schulz}(1996)}]{schulz96}
\bibinfo{author}{\bibfnamefont{H.}~\bibnamefont{Schulz}},
  \emph{\bibinfo{title}{in: Correlated Fermions and Transport in Mesoscopic
  Systems, edited by T. Martin, G. Montambaux, and J. Tran Thanh Van}}
  (\bibinfo{publisher}{Editions Fronti\`eres, Gif-sur-Yvette},
  \bibinfo{year}{1996}).

\bibitem[{\citenamefont{Shastry and Sutherland}(1990)}]{shastry90}
\bibinfo{author}{\bibfnamefont{B.}~\bibnamefont{Shastry}} \bibnamefont{and}
  \bibinfo{author}{\bibfnamefont{B.}~\bibnamefont{Sutherland}},
  \bibinfo{journal}{Phys. Rev. Lett.} \textbf{\bibinfo{volume}{65}},
  \bibinfo{pages}{243} (\bibinfo{year}{1990}).

\bibitem[{\citenamefont{Prosen and Ilievski}(2013)}]{prosen13}
\bibinfo{author}{\bibfnamefont{T.}~\bibnamefont{Prosen}} \bibnamefont{and}
  \bibinfo{author}{\bibfnamefont{E.}~\bibnamefont{Ilievski}},
  \bibinfo{journal}{Phys. Rev. Lett.} \textbf{\bibinfo{volume}{111}},
  \bibinfo{pages}{057203} (\bibinfo{year}{2013}).

\bibitem[{\citenamefont{Sakai and Kl\"umper}(2003)}]{sakai03}
\bibinfo{author}{\bibfnamefont{K.}~\bibnamefont{Sakai}} \bibnamefont{and}
  \bibinfo{author}{\bibfnamefont{A.}~\bibnamefont{Kl\"umper}},
  \bibinfo{journal}{J. Phys. A} \textbf{\bibinfo{volume}{36}},
  \bibinfo{pages}{11617} (\bibinfo{year}{2003}).

\bibitem[{\citenamefont{Huber and Semura}(1969)}]{huber69}
\bibinfo{author}{\bibfnamefont{D.~L.} \bibnamefont{Huber}} \bibnamefont{and}
  \bibinfo{author}{\bibfnamefont{J.~S.} \bibnamefont{Semura}},
  \bibinfo{journal}{Phys. Rev.} \textbf{\bibinfo{volume}{182}},
  \bibinfo{pages}{602} (\bibinfo{year}{1969}).

\bibitem[{\citenamefont{Krueger}(1971)}]{krueger71}
\bibinfo{author}{\bibfnamefont{D.~A.} \bibnamefont{Krueger}},
  \bibinfo{journal}{Phys. Rev. B} \textbf{\bibinfo{volume}{3}},
  \bibinfo{pages}{2348} (\bibinfo{year}{1971}).

\bibitem[{\citenamefont{Fukuhara
  et~al.}(2013{\natexlab{b}})\citenamefont{Fukuhara, Schau\ss, Endres, Hild,
  Cheneau, Bloch, and Gross}}]{fukuhara13a}
\bibinfo{author}{\bibfnamefont{T.}~\bibnamefont{Fukuhara}},
  \bibinfo{author}{\bibfnamefont{P.}~\bibnamefont{Schau\ss}},
  \bibinfo{author}{\bibfnamefont{M.}~\bibnamefont{Endres}},
  \bibinfo{author}{\bibfnamefont{S.}~\bibnamefont{Hild}},
  \bibinfo{author}{\bibfnamefont{M.}~\bibnamefont{Cheneau}},
  \bibinfo{author}{\bibfnamefont{I.}~\bibnamefont{Bloch}}, \bibnamefont{and}
  \bibinfo{author}{\bibfnamefont{C.}~\bibnamefont{Gross}},
  \bibinfo{journal}{Nature} \textbf{\bibinfo{volume}{506}}, \bibinfo{pages}{76}
  (\bibinfo{year}{2013}{\natexlab{b}}).

\bibitem[{\citenamefont{Bakr et~al.}(2009)\citenamefont{Bakr, Gillen, Peng,
  Tai, F\"olling, and Greiner}}]{bakr09}
\bibinfo{author}{\bibfnamefont{W.~S.} \bibnamefont{Bakr}},
  \bibinfo{author}{\bibfnamefont{J.~I.} \bibnamefont{Gillen}},
  \bibinfo{author}{\bibfnamefont{A.}~\bibnamefont{Peng}},
  \bibinfo{author}{\bibfnamefont{M.~E.} \bibnamefont{Tai}},
  \bibinfo{author}{\bibfnamefont{S.}~\bibnamefont{F\"olling}},
  \bibnamefont{and} \bibinfo{author}{\bibfnamefont{M.}~\bibnamefont{Greiner}},
  \bibinfo{journal}{Nature (London)} \textbf{\bibinfo{volume}{462}},
  \bibinfo{pages}{74} (\bibinfo{year}{2009}).

\bibitem[{\citenamefont{Sherson et~al.}(2010)\citenamefont{Sherson, Weitenberg,
  Endres, Cheneau, Bloch, and Kuhr}}]{sherson10}
\bibinfo{author}{\bibfnamefont{J.~F.} \bibnamefont{Sherson}},
  \bibinfo{author}{\bibfnamefont{C.}~\bibnamefont{Weitenberg}},
  \bibinfo{author}{\bibfnamefont{M.}~\bibnamefont{Endres}},
  \bibinfo{author}{\bibfnamefont{M.}~\bibnamefont{Cheneau}},
  \bibinfo{author}{\bibfnamefont{I.}~\bibnamefont{Bloch}}, \bibnamefont{and}
  \bibinfo{author}{\bibfnamefont{S.}~\bibnamefont{Kuhr}},
  \bibinfo{journal}{Nature (London)} \textbf{\bibinfo{volume}{467}},
  \bibinfo{pages}{68} (\bibinfo{year}{2010}).

\bibitem[{\citenamefont{Weitenberg et~al.}(2011)\citenamefont{Weitenberg,
  Endres, Sherson, Cheneau, Schau\ss, Fukuhara, Bloch, and
  Kuhr}}]{weitenberg11}
\bibinfo{author}{\bibfnamefont{C.}~\bibnamefont{Weitenberg}},
  \bibinfo{author}{\bibfnamefont{M.}~\bibnamefont{Endres}},
  \bibinfo{author}{\bibfnamefont{J.~F.} \bibnamefont{Sherson}},
  \bibinfo{author}{\bibfnamefont{M.}~\bibnamefont{Cheneau}},
  \bibinfo{author}{\bibfnamefont{P.}~\bibnamefont{Schau\ss}},
  \bibinfo{author}{\bibfnamefont{T.}~\bibnamefont{Fukuhara}},
  \bibinfo{author}{\bibfnamefont{I.}~\bibnamefont{Bloch}}, \bibnamefont{and}
  \bibinfo{author}{\bibfnamefont{S.}~\bibnamefont{Kuhr}},
  \bibinfo{journal}{Nature (London)} \textbf{\bibinfo{volume}{471}},
  \bibinfo{pages}{319} (\bibinfo{year}{2011}).

\bibitem[{\citenamefont{Steinigeweg
  et~al.}(unpublished)\citenamefont{Steinigeweg, Gemmer, and
  Brenig}}]{steinigeweg13}
\bibinfo{author}{\bibfnamefont{R.}~\bibnamefont{Steinigeweg}},
  \bibinfo{author}{\bibfnamefont{J.}~\bibnamefont{Gemmer}}, \bibnamefont{and}
  \bibinfo{author}{\bibfnamefont{W.}~\bibnamefont{Brenig}}, p.
  \bibinfo{pages}{arXiv:1312.5319} (\bibinfo{year}{unpublished}).

\end{thebibliography}

\end{document}